\documentclass[%
 reprint,
 amsmath,amssymb,
 aps,
]{revtex4-1}

\usepackage{hyperref}
\usepackage{footnote}
\usepackage{tablefootnote}
\usepackage{color}
\usepackage{graphicx}
\usepackage{dcolumn}
\usepackage{bm}
\usepackage[caption=false]{subfig}
\newcommand*\Eval[2]{\left.#1\right\rvert_{#2}}

\begin{document}

\preprint{APS/123-QED}

\title{Clustering, intermittency and scaling for passive particles on fluctuating surfaces}

\author{Tapas Singha} 
\email{tapas134@gmail.com}
\author{Mustansir Barma}
\email{barma23@gmail.com}
\affiliation{TIFR Centre for Interdisciplinary Sciences, Tata Institute of Fundamental Research, Gopanpally, Hyderabad 500107, India}

\date{\today}
\begin{abstract}
We show that a scaling approach successfully characterizes clustering and intermittency in space and time, in systems of noninteracting particles driven by fluctuating surfaces. We study both the steady state and the approach to it, for passive particles sliding on one-dimensional Edwards-Wilkinson or Kardar-Parisi-Zhang (KPZ) surfaces, with particles moving either along (advection) or against (antiadvection) the growth direction in the latter case. Extensive numerical simulations are supplemented by analytical results for a sticky slider model in which particles coalesce when they meet.  Results for single particle displacement versus time show to what extent particle dynamics is slaved to the surface, while scaling properties of the probability distribution of the separation of two particles determine the scaling form of average overlap of a pair of trajectories. For the many-particle system, clustering in steady state is studied via moments of particle number fluctuations in a single stretch, revealing different degrees of spatial multiscaling with different drivings. Temporal intermittency in steady state is established by showing that the scaled flatness diverges 
as the stretch size scaled by system size approaches zero for all the three drivings, but with different exponents, reflecting strongest clustering for KPZ advection and weakest for KPZ antiadvection. Finally we consider the approach to the steady state, study both the  flatness and the evolution of equal-time correlation functions as in coarsening of phase ordering systems. Our studies give clear evidence for a simple scaling description of the approach to steady state, with the scale set by a length which grows in time. An investigation of aging properties reveals that flatness is nonmonotonic in time with two distinct branches, and that a scaling description holds for each one.

\end{abstract}
\maketitle

\section{Introduction}
The motion of passive particles advected by a fluid field has long been of interest in nonequilibrium statistical mechanics \cite{Falkovich2001,Shraiman2000}. More generally, when subjected to a random force field with long-ranged correlations, passive particles exhibit strong clustering \cite{Deutsch1985, Wilkinson2003, Kraichnan1994}, characterized by pronounced inhomogeneities in density. Clustering has a strong effect on both static and dynamic correlations. This is because the formation of a large cluster increases the density in its local neighborhood, with a concomitant depletion of density in an extended region around it. The effect is thus that a fixed point in space experiences long periods of stasis, punctuated by infrequent, strong bursts of activity whenever a cluster visits its immediate neighborhood. This phenomenon is known as intermittency. In this paper we are primarily concerned with a quantification of intermittency, both after a steady state with clustering has formed, and, importantly, also during the approach to such a state.

\begin{figure*}[ht!]
\begin{minipage}{0.285\textwidth}
\includegraphics[width=\textwidth,height=0.125\textheight]{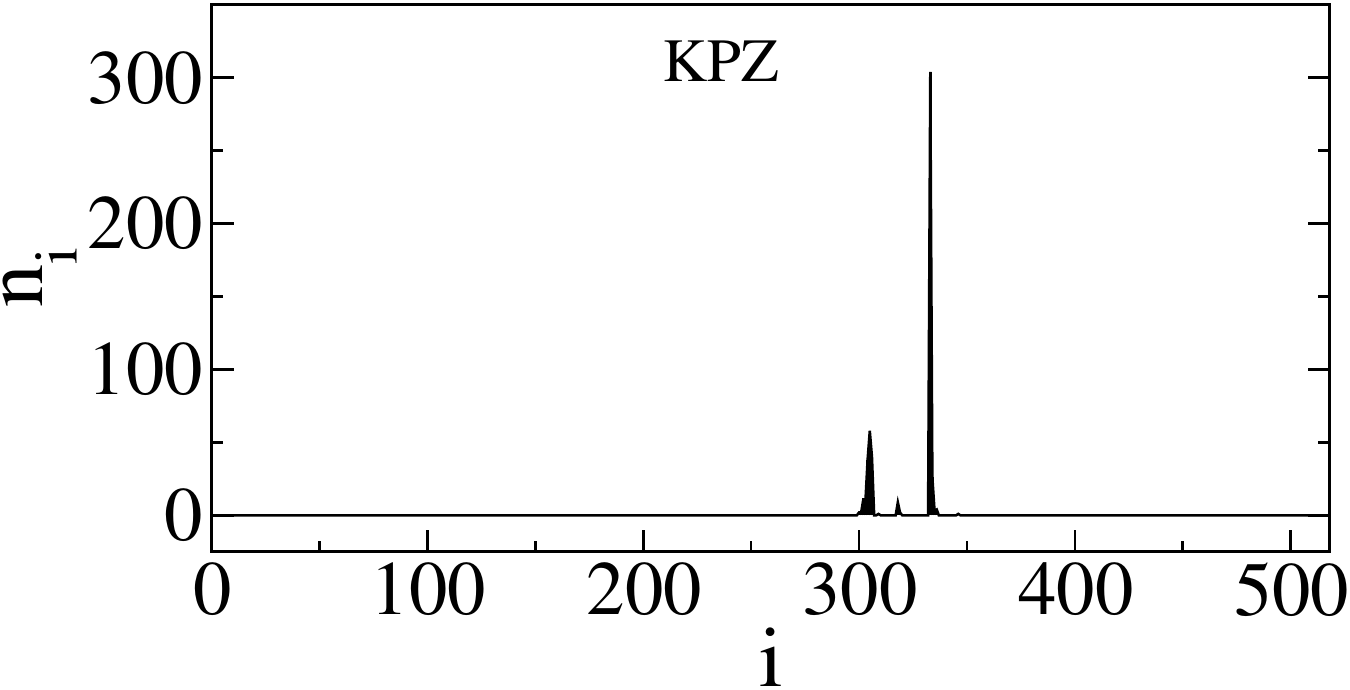}
\end{minipage}%
\hspace{1.1mm}
\begin{minipage}{0.28\textwidth}
\includegraphics[width=\textwidth,height=0.13\textheight]{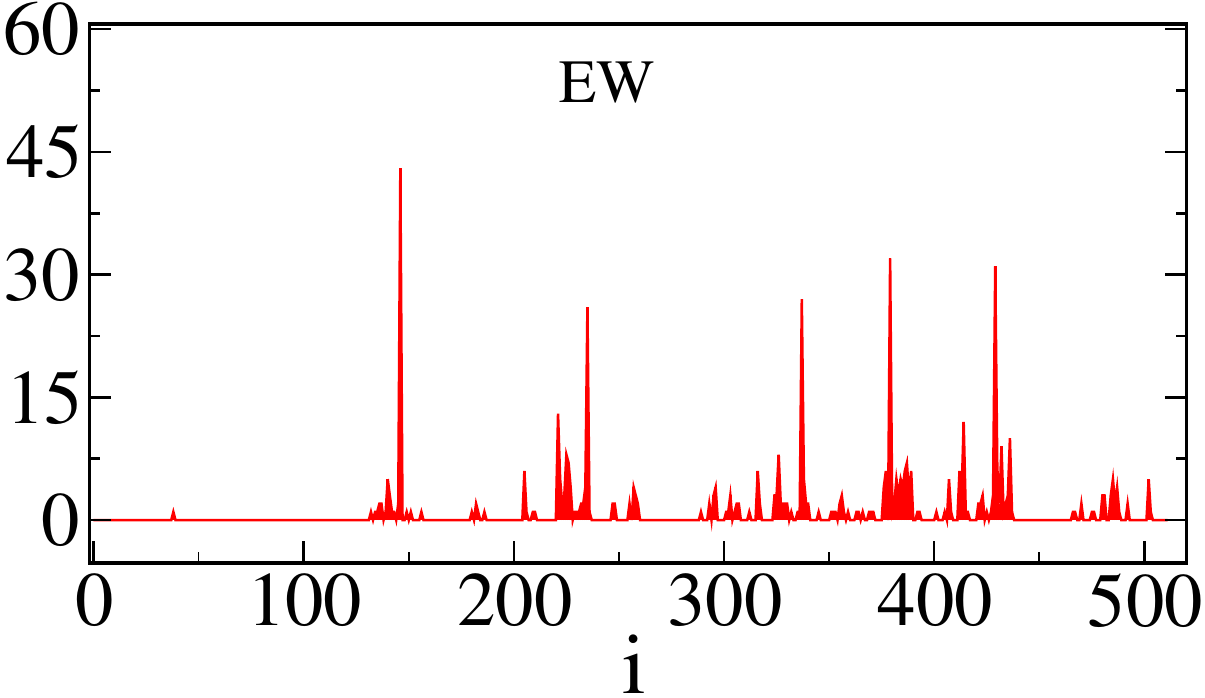}
\end{minipage}%
\hspace{1.13mm}
\begin{minipage}{0.275\textwidth}
\includegraphics[width=\textwidth,height=0.125\textheight]{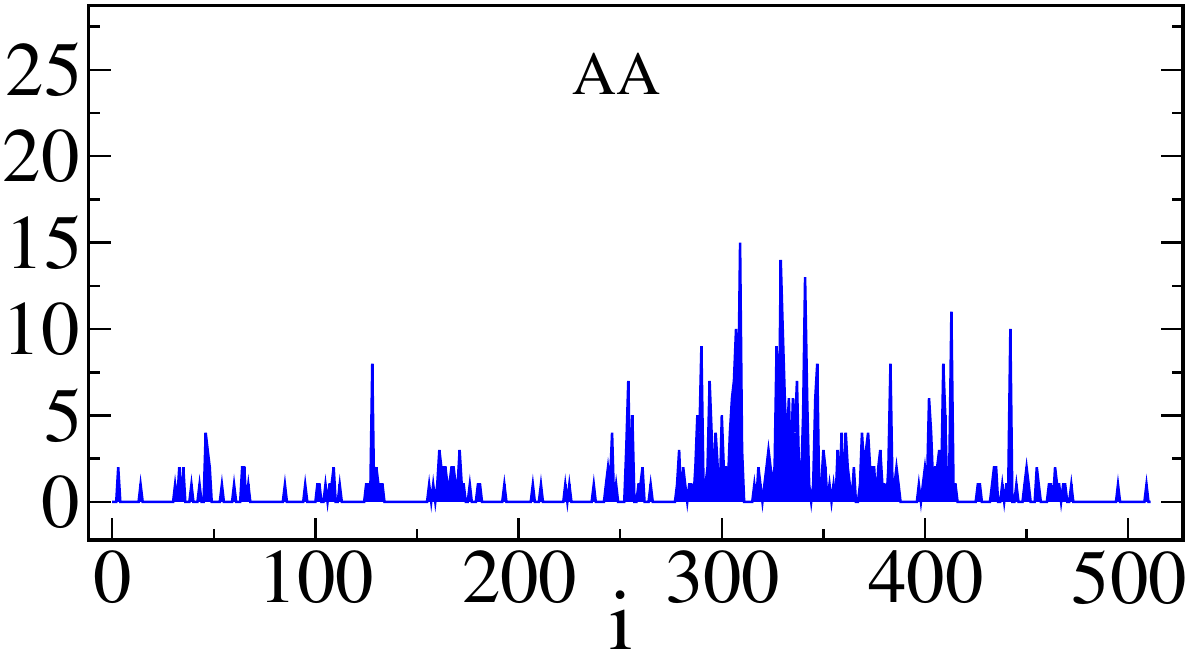}
\end{minipage}%
\caption{Typical configurations of passive particle density over space in steady state for KPZ, EW and KPZ-AA driving.} 
\label{Density_profile}
\end{figure*}

We study a family of simply defined models involving passive particles, which show different degrees of clustering. The particles slide on a fluctuating surface, stochastically following surface slopes without affecting the surface dynamics. The degree of particle clustering and intermittency depend strongly on the nature of surface driving; the main purpose of this paper is to quantify this dependence and study associated scaling properties of the resulting clustered state. We study driving by a one-dimensional Kardar-Parisi-Zhang (KPZ) surface \cite{KPZ1986, Medina1989} with particles moving either along (advection) or against (anti-advection, KPZ-AA) the direction of surface growth \cite{DK2000, DK2000}, and also by an Edwards-Wilkinson (EW) surface \cite{EW1982}. The density profile in typical steady state configurations shown in Fig.\ \ref{Density_profile} illustrates the different degrees of clustering for the three different drivings.

To quantify this, we study particle number fluctuations in an extended stretch, using measures defined in terms of structure functions; a simple diagnostic is the flatness, the ratio of the fourth moment to the square of the second moment \cite{Frisch1995}. The unifying feature of our results is the occurrence of scaling, with a divergence of flatness for small scaling argument being the hallmark of intermittency in both space and time. Also, values of exponents characterizing scaling  divergences allow quantification of different degrees of clustering. These methods are of broad applicability and may fruitfully be used in a number of settings. For instance, in the context of biological systems which exhibit clustering, temporal intermittency has been studied for both  molecular transport through Golgi\cite{SachdevaPRE2011, SachdevaSciRep2016}, and protein aggregation on cell membranes \cite{DasPRL2016}. 

\vspace{2mm}

\vspace{5mm}

Earlier studies of passive particles on fluctuating surfaces have considered two limiting cases, (i) noninteracting passive particles, in which case any number of particles is allowed to reside at a site \cite{BP1993,DK2000,DrosselPRB2002,Chin2002,Manoj2004,NBM_2005,NMB_2006,Singha2018,Huveneers} (ii) passive particles with hard core interactions where at most one particle can reside at a site \cite{DasPRE2001,DasPRL2000,Chatterjee2006,KapriPRE2016}. The first case is the one of interest in this paper. In the steady state of both systems, the correlation function is a scaling function which depends on the ratio of separation to system size. But at small argument, the scaling function has a cusp singularity in case (ii), while it shows a divergence in case (i), pointing to strong clustering states \cite{NBM_2005,NMB_2006}. A consequence is that intermittency in space and time is very pronounced in such states, and this is the primary aspect studied in this paper. A brief account, primarily with KPZ advection, appeared in \cite{Singha2018}, while in this work we study several types of driving, and characterize the different degrees of clustering and intermittency that result, often with interesting differences.  Another aspect we take up concerns the approach to the steady state, including coarsening and aging, which has been studied in detail in case (ii) \cite{DasPRE2001,DasPRL2000,Chatterjee2006,KapriPRE2016}, but not in (i). Finally, as each particle moves independently in our case with correlations arising only from shared histories of the driving, the behaviour of one and two particles has important repurcussions for the statistical properties of the noninteracting many-particle system. Consequently, we also add to earlier studies of single-particle displacements as a function of time \cite{BP1993,DK2000,DrosselPRB2002,Chin2002,Manoj2004,Huveneers}, focusing on the occurrence of multiplicative logarithms with power laws, and on the time-dependence of two-particle correlations, focusing on  consequences of scaling \cite{SBComment}.

Because of the stochastic element in passive particle motion, particles which start at the same point do not follow identical trajectories. However, they have strong correlations, as they are subject to the same history of driving by the fluctuating surface. This leads ultimately to anomalously large fluctuations of particle density, which are the central concern of this work. In order to develop this theme, it is useful to ask a number of questions. 
\begin{itemize}
\item How far does a single particle move in time $t$? If the typical distance moved is $r \sim t^{1/z}$, how is $z$ related to the dynamic exponent $z_s$ of surface fluctuations? 

\item How does the separation of two particles which start together, vary with time? In the long time limit, what fraction of the time would they be found within a  specified finite range?

\item For the many-particle system, is there spatial intermittency in steady state and does the density profile show multiscaling? Do fluctuations in steady state exhibit temporal intermittency? 

\item Starting from a random distribution, how do clustering and intermittency build up during the coarsening regime describing the approach to steady state? Is there a growing length scale, and if so how do scaling functions differ from those in phase ordering systems, for both coarsening and aging? 
\end{itemize}

In the remainder of this section, we attempt to provide a coherent account of the answers to the questions posed above. We refer to the results of earlier work as well as results obtained in this paper, highlighting the differences brought in by the three different drivings we have considered.



Our interest in single particle motion stems from the fact that it has strong implications for the many particle system: the displacement of a single particle in time $t$ governs the size of the basin $\mathcal L(t) \sim t^{1/z}$ within which particles cluster in the many-particle system, as explained in Section VII. Thus the exponent $z$ enters a scaling description of coarsening and aging.
For KPZ advection, earlier work has shown that the particle dynamics is slaved to the surface, implying $z=z_s=3/2$ \cite{BP1993,DrosselPRB2002,Chin2002}, whereas for KPZ-AA, $z \simeq 1.74$ which is distinct from $z_s$ \cite{DK2000,DrosselPRB2002}. For EW  driving, the growth of the mean squared displacement is proportional to the time, with multiplicative logarithmic corrections \cite{BP1993, Huveneers}. In this paper, we provide a numerical estimate of the power of the logarithm.

To monitor correlations between two particles which start at the same spot, it is useful to study the time evolution of the probability distribution of the separation \cite{Ueda_Sasa2015}. A scaling approach \cite{SBComment} reveals an important distinction between KPZ advection on the one hand, and EW and KPZ-AA dynamics on the other. For KPZ driving, interestingly, the fraction of time of the trajectory pair would be within a finite range is finite even in the $t\rightarrow \infty$ limit, whereas the fraction goes to zero for EW and KPZ-AA driving. 

\begin{figure*}[ht!]
\begin{minipage}{0.27\textwidth}
\includegraphics[width=\textwidth,height=0.15\textheight]{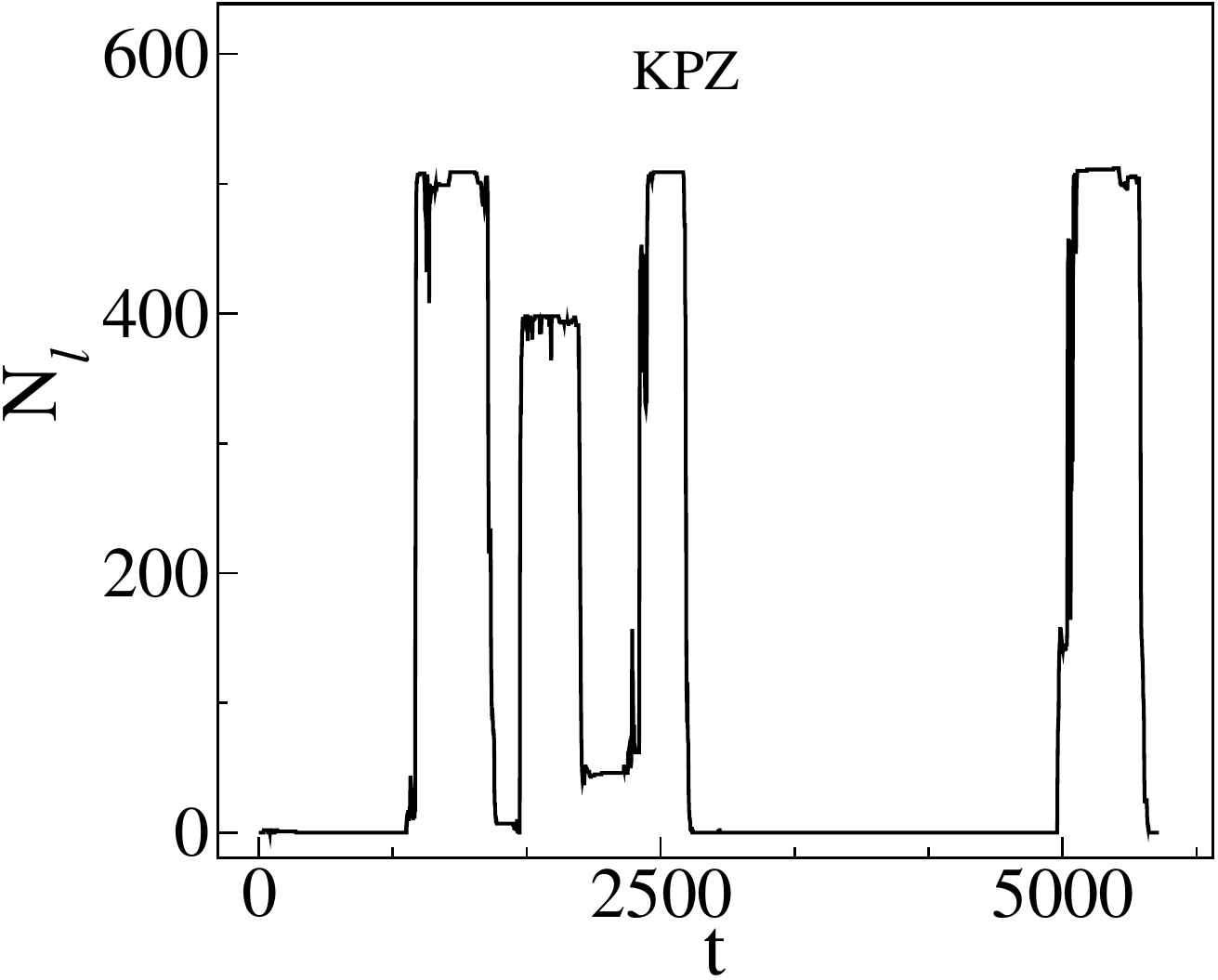}
\end{minipage}%
\hspace{1.1mm}
\begin{minipage}{0.27\textwidth}
\includegraphics[width=\textwidth,height=0.145\textheight]{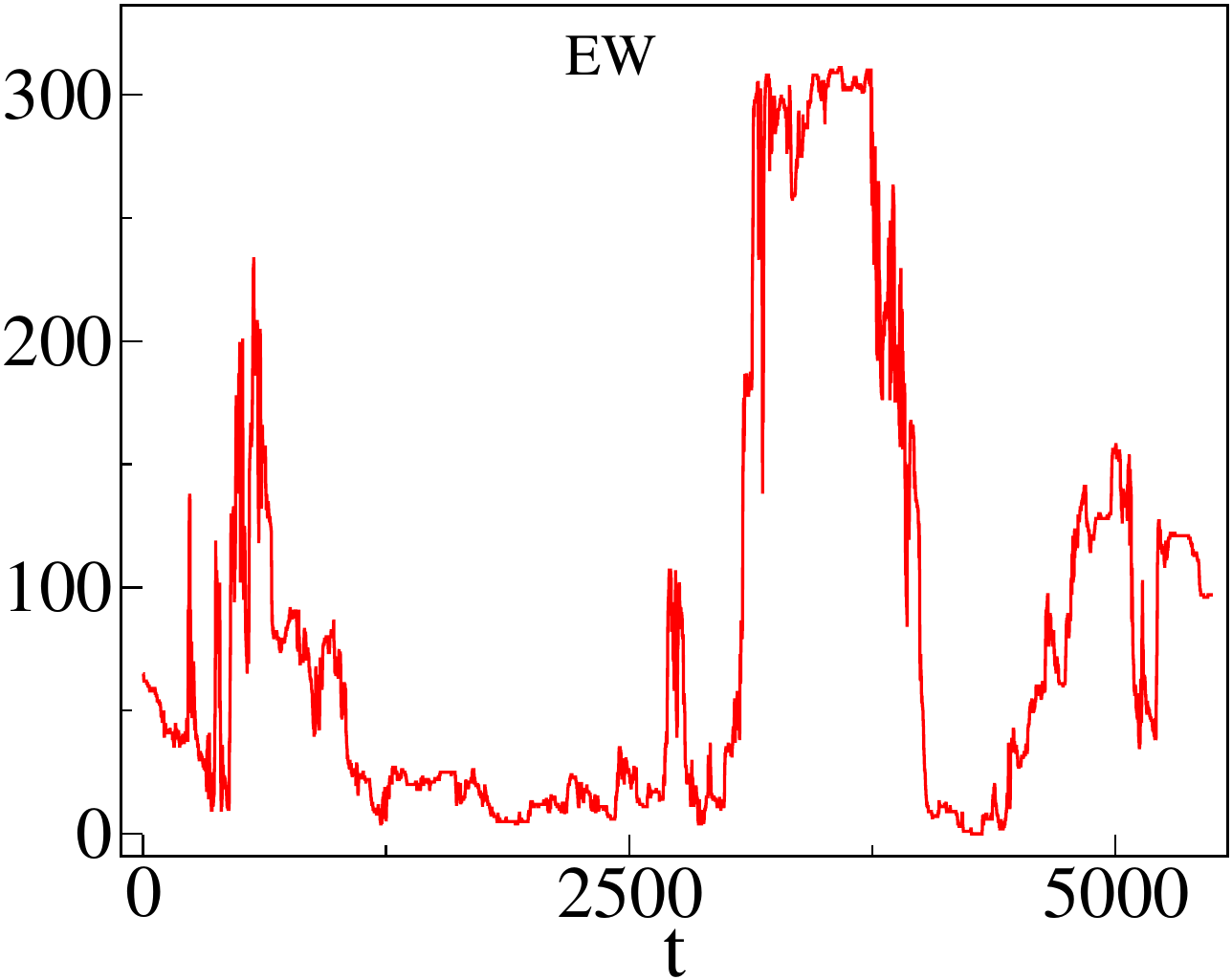}
\end{minipage}%
\hspace{1.13mm}
\begin{minipage}{0.27\textwidth}
\includegraphics[width=\textwidth,height=0.145\textheight]{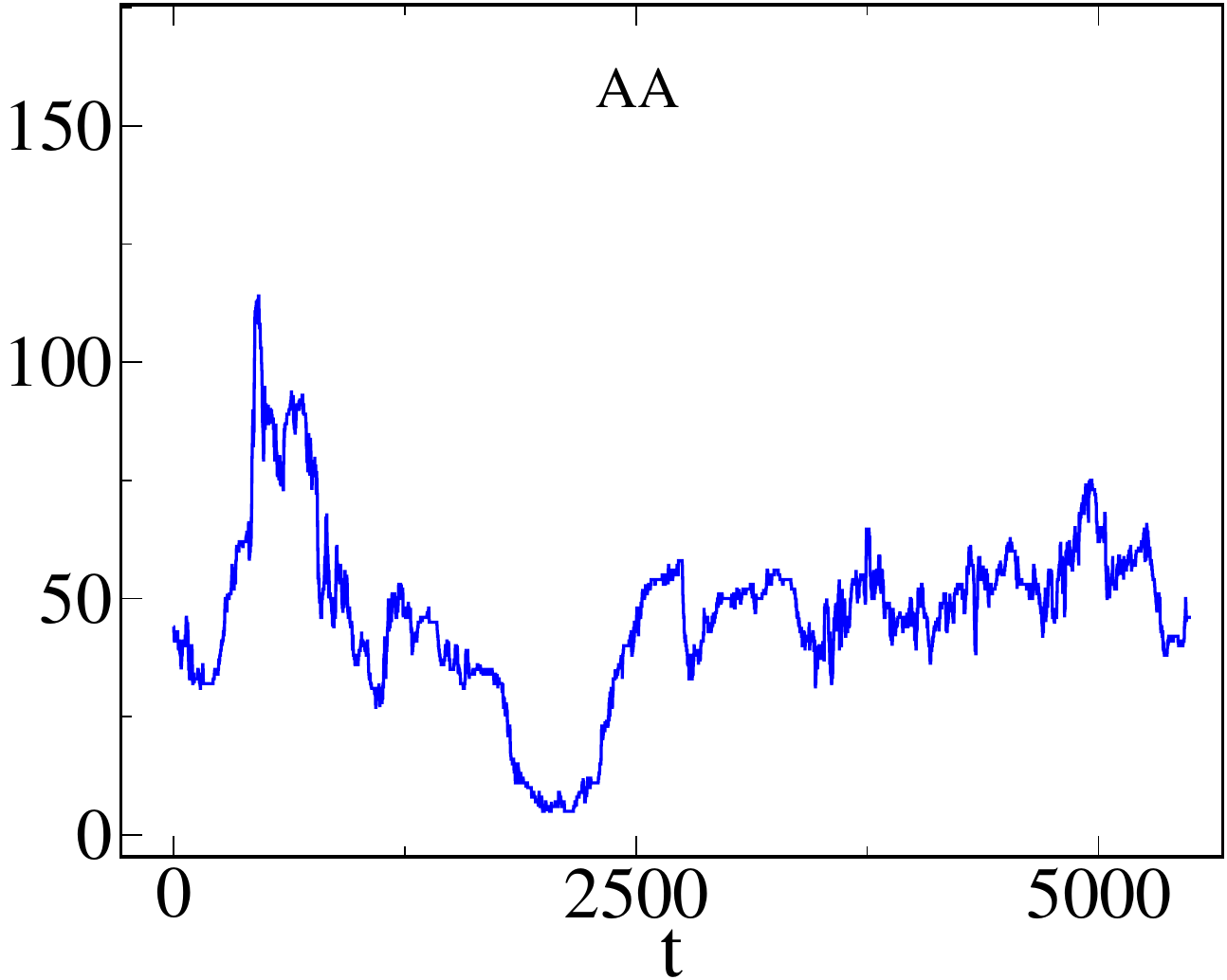}
\end{minipage}%
\caption{Time series of particle number in a stretch $l=L/8$ for KPZ, EW and KPZ-AA driving.}
\label{Time_Series_inl} 
\end{figure*}

To study intermittency in steady state, we analyze the high-order structure functions of particle number in a stretch of $l$ successive sites. We find evidence of spatial multiscaling in all the cases; in the case of KPZ advection, it takes on a particularly extreme but simple form while it leads to a whole spectrum of exponents associated with EW and KPZ-AA driving, revealing that both self-similarity and intermittency coexist in the spatial structures. Further, temporal fluctuations also exhibit intermittency, quantified by showing that the flatness diverges as the scaled time $t/\tau_l$ approaches zero where $\tau_l \sim l^{z}$.

 In the coarsening regime the two-point correlation is found to be a scaling function of scaled separation $r/\mathcal{L}(t)$, reminiscent of phase ordering dynamics \cite{Bray1994}. As pointed out above,  $\mathcal L(t) $ is determined by a single particle property. Moreover, the flatness increases indefinitely, in proportion to $\mathcal{L}(t)/l$, showing how intermittency sets in. Physically, $\mathcal{L}(t)$ is the size of the basin from which particles are drawn and form clusters near the bottom of a valley. In the aging regime, the flatness shows an interesting nonmonotonicity as a function of time provided that the waiting time $t_0$ exceeds $\tau_l$; 
both left and right branches of the aging curve for flatness then diverge, showing distinct scaling functions, but both involve the same scaling variable $l/\mathcal{L}(t)$.

We also consider a simpler model of `sticky sliders' which do not dissociate once they meet, as it provides considerable insight into the behavior in various regimes. In particular, it predicts scaling forms which are found to hold also for the passive particle systems, although exponents differ in the case of KPZ-AA driving.

The paper is organized as follows. In Section II, we discuss the model. In Section III, we study the motion of a single passive particle for different fluctuating surfaces. Section IV presents the probability distribution of the separation between two particles for different drivings and the average fraction of the time they are together. In Section V, we discuss static properties in the steady state, including the two-point density-density correlation function and structure functions which provide evidence for multiscaling. Steady-state dynamics and characterization of temporal intermittency are presented in Section VI. Section VII is devoted to the study of scaling in the coarsening regime, during the approach to steady state. In Section VIII, we study aging, which involves studying correlations after a certain waiting time. Finally, Section IX is the conclusion.

\vspace{4mm}
\begin{figure*}[ht!]
\begin{minipage}{0.19\textwidth}
\includegraphics[width=\textwidth,height=0.25\textheight]{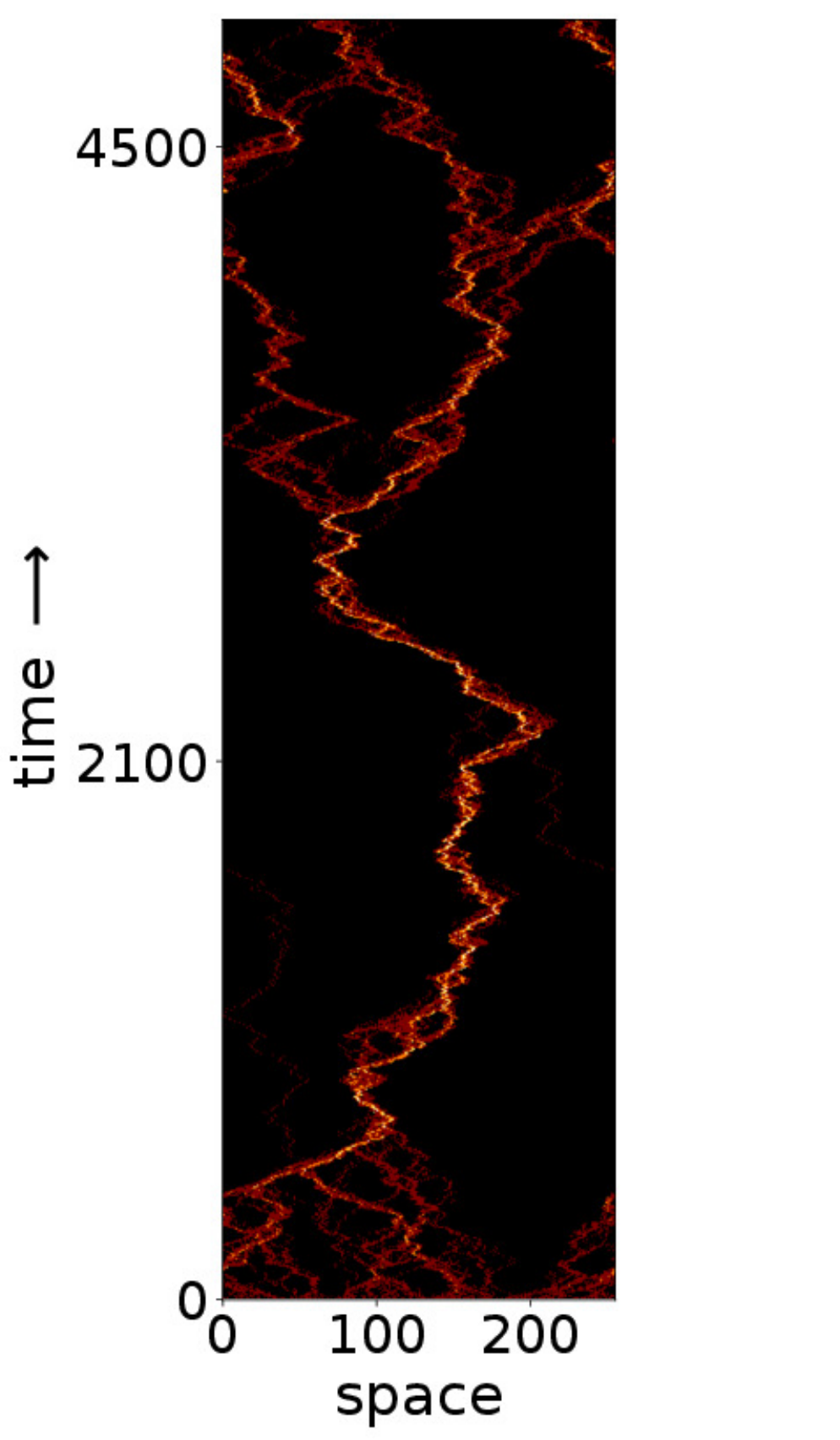}
\end{minipage}%
\hspace{-9mm}
\begin{minipage}{0.155\textwidth}
\includegraphics[width=\textwidth,height=0.25\textheight]{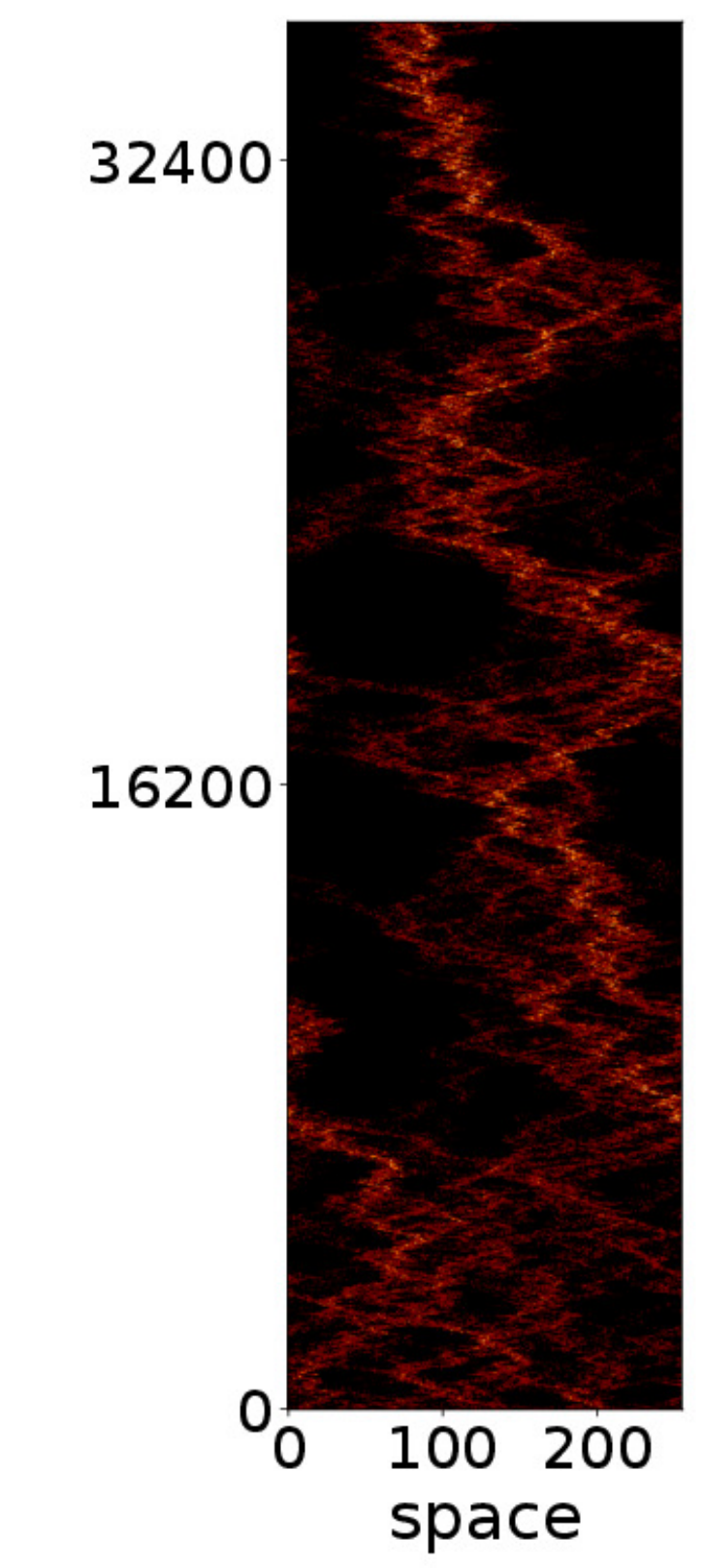}
\end{minipage}%
\hspace{-1mm}
\begin{minipage}{0.20\textwidth}
\includegraphics[width=\textwidth,height=0.265\textheight]{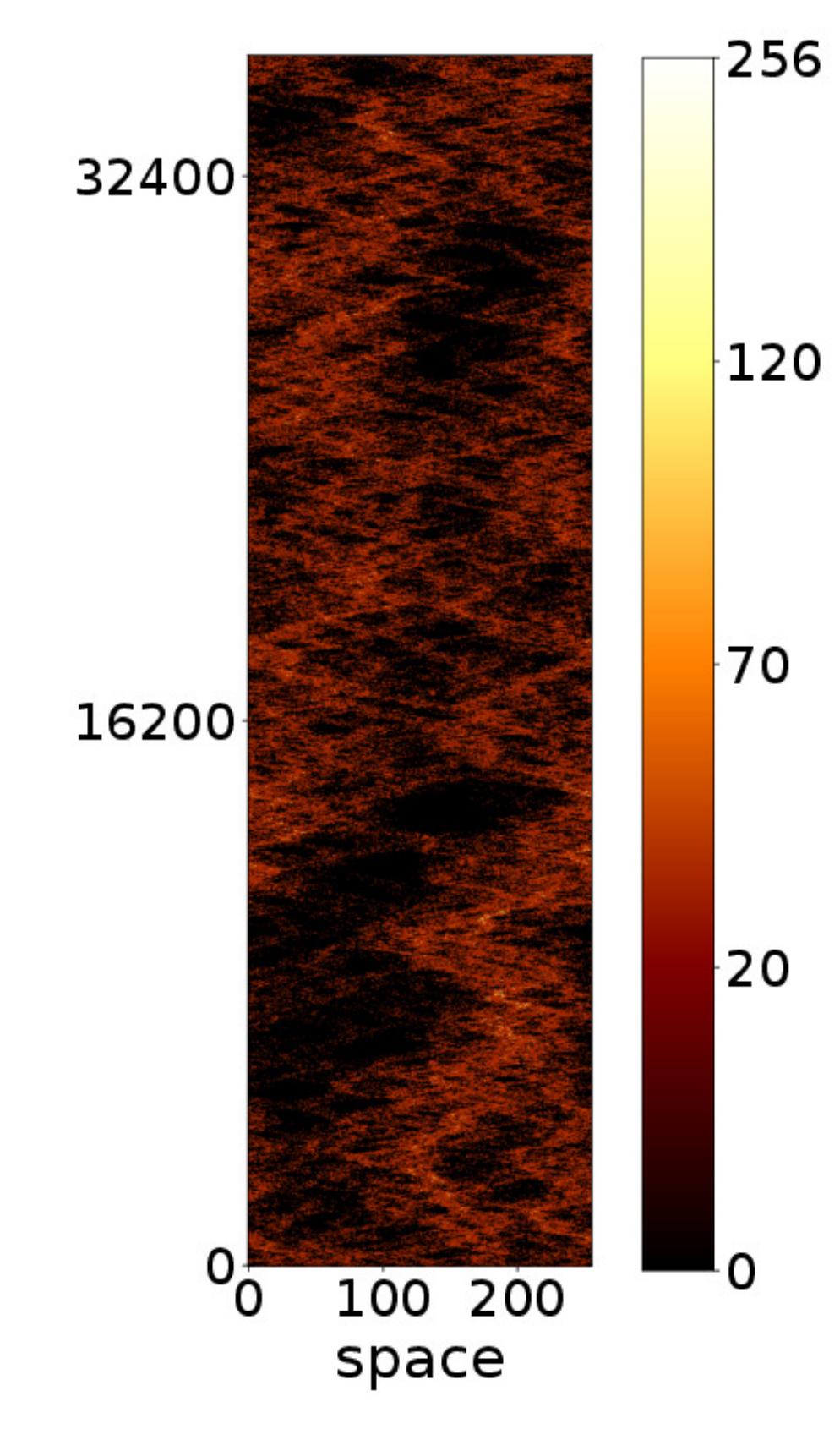}
\end{minipage}%
\caption{World lines of passive particles with KPZ, EW and KPZ-AA drivings are shown from left to right, respectively.}
\label{Space-Time}
\end{figure*}
\vspace{4mm}

\section{Models}
 In a continuum description, the surface evolution is taken to follow the KPZ equation   
\begin{equation}  
\frac{\partial h}{\partial t} = \nu_0 \nabla^2 h + \frac{\lambda_0}{2} (\nabla h)^2 + \eta(\mathbf{x},t)
\label{equation_KPZ}
\end{equation}
which describes a growing, fluctuating interface, where $\nu_0$ and $\eta(\mathbf{x},t)$ are the surface tension and spatio-temporal uncorrelated Gaussian noise, respectively. $\lambda_0$ is the strength of the nonlinearity which arises for a growing, fluctuating KPZ interface. On setting $\lambda_0=0$, Eq. (\ref{equation_KPZ}) reduces to the Edwards-Wilkinson (EW) equation, in which the average surface height does not change in time. On substituting $\nabla h(\mathbf{x},t) = - \mathbf {u} (\mathbf{x},t)$ and setting $\lambda_0 =1$ in Eq.\ \ref{equation_KPZ}, one obtains the vorticity-free noisy Burgers equation \cite{Medina1989}, where $\mathbf {u} (\mathbf{x},t)$ is the velocity of Burgers fluid.  

The equation of motion of a passive particle is
\begin{equation}
\frac{d r(t)}{dt} = - a \ \Eval{\dfrac{\partial h }{\partial x}}{x=r(t)} + \eta_r(t)  
\label{EOM_Single_Particle}
\end{equation}
where $r(t)$ is the position of the particle and the slope $\partial h/\partial x$ of the surface is evaluated at $r(t)$. The noise $\eta_r(t)$ is Gaussian with zero average and $\langle \eta_r(t) \eta_r(t')\rangle = 2 D \delta(t-t')$ where $D$ is the strength of the noise. The particle stochastically moves either along (when $a$ is positive) or opposite to (when $a$ is negative) the growth direction of the surface. 

\subsubsection{Passive Slider Model on a Lattice}

In this work, we consider a discrete lattice model in one dimension, with bonds inclined upward ($/$) or downward ($\backslash$). The two ends of a finite lattice of length $L$ are connected via periodic boundary conditions so that $h(0)=h(L)$ and we have an equal number of upward and downward slopes. The dynamics involves local hills ($/$$\backslash$) stochastically transforming into local valleys ($\backslash$$/$) at rate $u_1$ and local valleys ($\backslash$$/$) into local hills ($/$$\backslash$) at rate $u_2$. This is the single step model for a discrete surface \cite{Liggett1985}, which can be mapped to the asymmetric simple exclusion process (ASEP) by associating an upward-tilted bond ($/$) with a particle and a downward-tilted bond ($\backslash$) with a hole. On large enough length and time scales, the surface is described by the KPZ equation if $u_1 \neq u_2$ and by the EW equation if $u_1 = u_2$. 

\vspace{4mm}

\begin{figure}[ht!]
\begin{minipage}{0.46\textwidth}
\includegraphics[width=\textwidth,height=0.04\textheight]{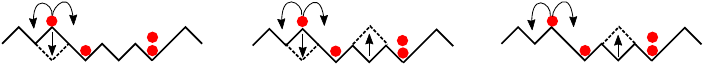}
\label{Fig_Aging_Monotonic}
\end{minipage}%
\caption{A schematic diagram to show the elementary moves of the surface and particles for KPZ, EW and KPZ-AA drivings, shown from left to right, respectively.}
\end{figure}

\vspace{2mm}

Initially, the passive particles are distributed randomly over the surface sites between successive bonds. Particles are labeled and move independently and there is no restriction on the number of particles on each site. The total number of particles $N$ is taken to be equal to the total number of sites $L$ which makes the average particle density $\langle \rho_i \rangle = 1$. Each particle moves stochastically down the fluctuating surface, one step at a time, following the local surface slope. In our numerical simulations, one Monte Carlo step comprises $L$ microstep for particles and $L$ microstep for slopes, with an alternation between particle and surface microstep. In each particle microstep, we choose a particle at random, while in each slope micro step, we choose a bond at random. For $u_1 \neq u_2$, if the slope is positive (negative), we examine the bonds on the right (left) of the selected bond. For $u_1 = u_2$, we only examine the bond on the right of the selected bond. In order to mimic Eq.\ \ref{EOM_Single_Particle}, the rules for particle update are the following. A randomly chosen particle slides down a bond; if atop a local hill $/$$\backslash$, it slides down one of the two bonds with equal likelihood, and if in a valley $\backslash$$/$, it does not move.  Antiadvection corresponds to the case $u_1 < u_2$, in which case the surface grows upwards, opposite to the direction of particle sliding.

Stochasticity, while it is tantamount to the effect of noise, enters into the dynamics of the particles in the following ways. The random selection of a tagged particle implies that a given particle may not be selected in a Monte Carlo time step even if the particle resides on a downward slope. On the other hand, if a selected particle resides on top of a hill, it moves either towards left or right with the equal probability, which is also a source of noise in the particle dynamics. 

In short, particles stochasically slide down the surface slopes, whereas the surface itself grows downward for KPZ advection, fluctuates around the mean in the EW case, and grows upwards for KPZ-AA. In Fig.\ \ref{Space-Time}, we show the world-lines of passive particles with KPZ, EW, and KPZ-AA dynamics to illustrate the typical evolution. Significant differences in the amount of clustering are apparent in Fig. \ref{Time_Series_inl}; the quantification of this feature and how it develops in time is the major concern of this paper.

\subsubsection{Sticky Slider Model}

We also introduce a simpler model, namely a sticky slider model (SSM): once the particles come to the same site, they stick together and then move on the surface as a single entity. Updation rules for the numerical simulation of the SSM are the following. Instead of randomly choosing an individual particle as in the passive slider model (PSM), for the SSM a randomly chosen individual cluster slides on the surface and eventually, in the steady state, they merge to form a single cluster which then moves on the fluctuating surfaces.

\section{Single particle on a fluctuating surface}
Particle displacements are characterized by
\begin{equation}
\mathcal{R}(t)=\langle [r(t+t')-r(t')]^2\rangle^{1/2} 
\end{equation}
where $r(t)$ is the location of a single passive particle. $\mathcal{R}(t)$ 
grows as $t^{1/z}$, and the question is whether $z=z_s$. For KPZ advection, earlier work has established $z=z_s=3/2$ \cite{BP1993,DrosselPRB2002} while for KPZ-AA, $z$ was found to be very different from $z_s$ \cite{DrosselPRB2002,DK2000,NMB_2006}. With EW driving, there appears to be a marginal difference, in that $z=z_s=2$, but there is a multiplicative logarithm $\mathcal{R}(t)\sim t^{1/2} [\ln(t/t_0)]^{\alpha/2}$. The numerical work reported in this section suggests $\alpha$ is close to $1/2$. 
  
The single-particle displacement $\mathcal{R}(t)$ is important for the many-particle system, as the growing length scale $\mathcal{L}(t)$,  which governs scaling during coarsening  (Section VII) coincides with $\mathcal{R}(t)$.

\subsection{KPZ driving}
We summarize earlier predictions for the growth of $\mathcal{R}(t)$. 

Bohr and Pikovsky \cite{BP1993} studied the root mean-squared displacement of passive particles advected by the noisy Burgers fluid, within a mean-field approach to the scaling form of the two-point velocity. They obtained a self-consistent asymptotic solution $\mathcal{R}(t) \sim t^{1/z}$ with $z=3/2$. 

Drossel and Kardar performed a numerical simulation of a restricted solid-on-solid model belonging to the KPZ universality class, coupled with the passive particle dynamics; they found $z=3/2$ \cite{DrosselPRB2002}. 

Later, the same value of the dynamic exponent was obtained by modeling the surface dynamics through the Kim-Kosterlitz  model \cite{Chin2002} and the single step model \cite{NMB_2006}. As emphasized in \cite{Chin2002}, with KPZ driving the particle motion becomes slaved to the fluctuations of the surface so that $z=z_s$. Our results from numerical simulations are consistent with $z=z_s=3/2$.

\subsection{EW driving}

Let us turn to EW surface dynamics, in which case the motion of the particle is less strongly coupled to the surface fluctuations.

 A numerical simulation of particles driven by an unbiased single step model was carried out by Manoj \cite{Manoj2004}. He found $\mathcal{R}(t) \sim t^{1/z}$ where $z$ shows an apparent dependence on the ratio $\omega$ of update rates for the surface and particle evolution, varying from $1/z \simeq 0.67$ for $\omega \gg 1$ (rapid surface motion) to $ 1/z \simeq 0.56$ for $\omega =1$, and to $1/z \simeq 0.50$ for $\omega \ll 1$ (rapid particle motion) \cite{Manoj2004}.
 
 However, the apparent dependence on $\omega$ may result from crossover effects, with the true form involving multiplicative logarithms:
\begin{equation}
\mathcal{R}^2(t) \sim t \ f\left(\frac{t}{t_0}\right)
\label{EW_log_Correction}
\end{equation}
where $f(t/t_0)$ indicates a multiplicative logarithmic correction. In fact, such logarithmic corrections have been proposed earlier. Bohr and Pikovsky \cite{BP1993} studied a linear version of the noisy Burgers equation (tantamount to EW dynamics) for which the velocity-velocity correlation function is known. Within a mean-field approach, they concluded $f(t/t_0) \sim \ln[1+ (t/t_0)^{1/2}]$ where $t_0$ depends on the model parameters of the system. This would imply $\mathcal{R}^2(t) \sim t^{3/2}$ for $t \ll t_0$, and $\mathcal{R}^2(t) \sim t \ln(t)$ as $t \rightarrow \infty$.
In recent work, Huveneers suggests that $f(t)$ may follow $ \sim [ \ln (t) ]^{\alpha}$ with $ 0 \leq \alpha < 1$ but raises the possibility that $\alpha > 0$ may be a transient effect \cite{Huveneers}.  

\vspace{4mm}

\begin{figure}[ht!]
\centering
\includegraphics[width=0.45\textwidth,height=0.25\textheight]{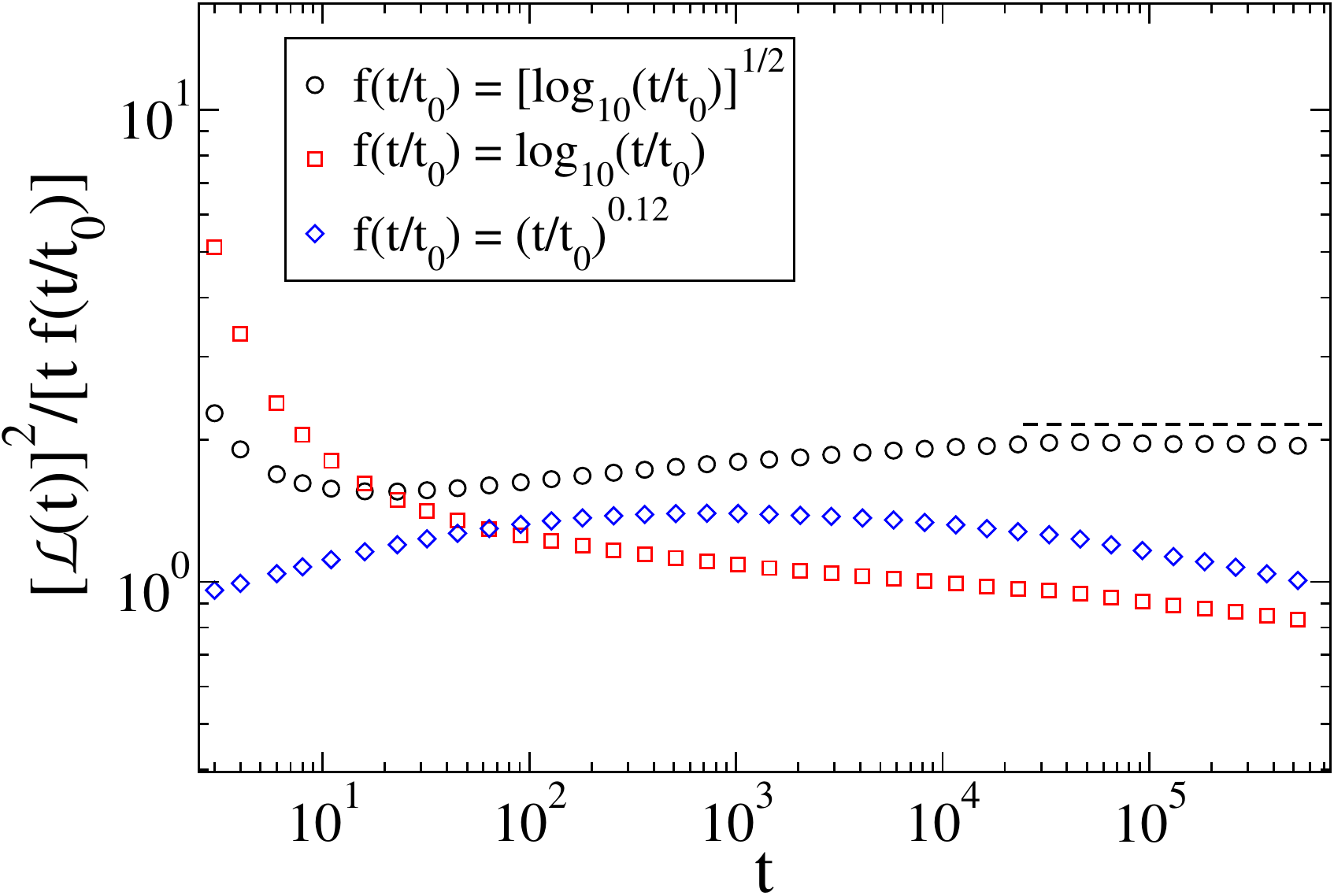}
\caption{Mean-squared displacement of a passive particle on an EW surface driving where the system size $L=2^{16}$. We chose $t_0=1.9$ in $f(t/t_0)$.}
\label{EW_MSD_Log_Correct}
\end{figure}

\vspace{2mm}

The results of our Monte Carlo simulations are shown in Fig.\ \ref{EW_MSD_Log_Correct}. To distinguish between competing predictions  \cite{Manoj2004,BP1993,Huveneers}, we have plotted $\mathcal{R}^2(t)/(t f(t/t_0))$ with (a) $f(t/t_0) \sim t^{\theta_{eff}}$ with $\theta_{eff} \simeq 0.12$ as in \cite{Manoj2004} for $\omega =1$ (b) $f(t/t_0) = \ln (t/t_0)$ as in \cite{BP1993} (c) $f(t/t_0) = [\ln (t/t_0)]^{\alpha}$ with $\alpha = 1/2$. The results in Fig.\ \ref{EW_MSD_Log_Correct} seem to indicate $\alpha = 1/2$,  but there is a degree of uncertainty.  In the remainder of this work, for convenience we use an effective dynamic exponent $1/z \simeq 0.56$ for EW driving as proposed by Manoj \cite{Manoj2004}, recognizing that the different estimates of $f(t)$ do not lead to substantial differences in the ranges to be considered. 

\subsection{Antiadvection}

For the case of antiadvection (AA), the exponent $z$ of the particles was estimated numerically by Drossel and Kardar \cite{DK2000} and found to be nonuniversal, changing continuously with $a$  \cite{DrosselPRB2002,DK2000}. For $a=1$ they obtained $z \simeq 1.74$ consistent with the numerical findings in \cite{NMB_2006,Singha2018} as also in the current work. As the coupling constant $a$ decreases from $1$ \text{to} $0$, the exponent $z$ increases from approximately $1.74$ \text{to} $2$, with $a=0$ corresponding to a simple random walk. 

Note that single particle dynamics with different sorts of driving also determines the dynamics of SSM in the steady state, as in that case there is a single cluster. Evidently, the motion of this cluster is exactly that of a single particle.

\section{Two particles}

The relative motion of two passive particles is studied through the time evolution of the probability distribution of their separation, and the overlap, that is the fraction of time in which the separation falls within a certain range. A scaling form for the probability distribution \cite{SBComment} is shown in this section to hold for all three types of surface driving.  Results for EW and KPZ-AA driving are found to differ strongly from those for KPZ advection. The average of the overlap function is also shown to follow a scaling form by relating it to the probability distribution of separation.



\subsection{PDF of separation between two particles}

\begin{figure}[ht!]
\hspace{-8mm}
\begin{minipage}{0.45\textwidth}
\includegraphics[width=\textwidth,height=.25\textheight]{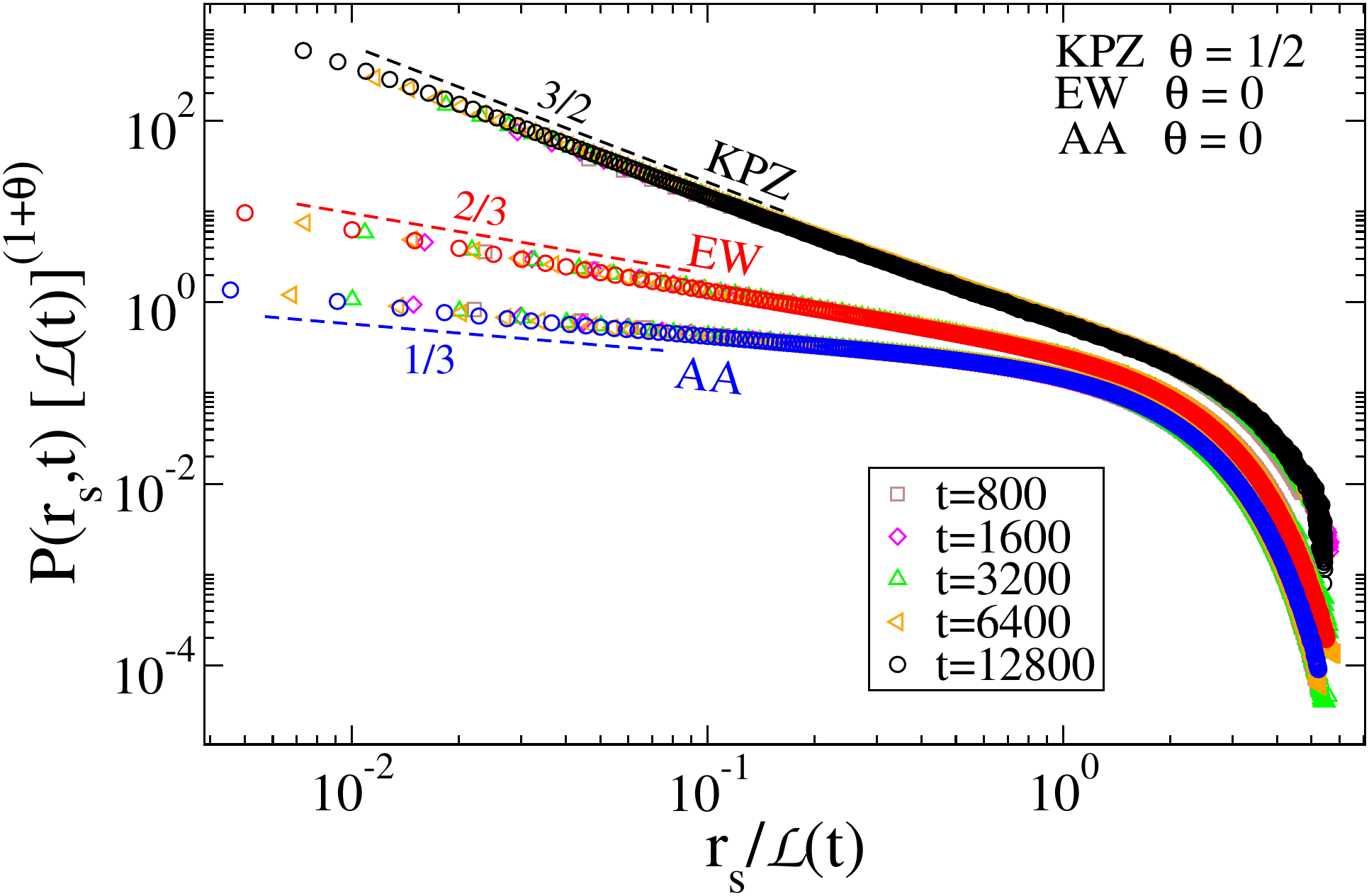}
\caption{Probability distribution of separation $r_s$ between two particles with KPZ, EW, and KPZ-AA driving. The PDFs for different times collapse when the separation $r_s$ is scaled by the corresponding $\mathcal{L}(t)$ and $P(r_s,t)$ with $[\mathcal{L}(t)]^{1+\theta}$.}
\label{PDF_time_Fixed_separation}
\end{minipage}%
\end{figure}

 The interplay of advection and independent noise cause a pair of trajectories to overlap during one part of the evolution and deviate from each other during other parts. A measure of the closeness of the trajectories is the PDF of the interparticle separation 
$$r_s(t) \equiv |x^{(1)}(t)-x^{(2)}(t)|,$$ 
which has been studied in \cite{Ueda_Sasa2015}, where an interesting coexistence of a high-overlap and low-overlap regimes in space-time were found. We studied the PDF of $r_s$ for KPZ \cite{SBComment}, EW and KPZ-AA drivings and found power-law decays with different exponents for different drivings. Interestingly, the PDFs turn out to be \emph{scaling functions} of $r_s$ and $t$. Data for different times can be collapsed for each of the drivings, as shown in Fig.\  \ref{PDF_time_Fixed_separation}, when $r_s$ is scaled by $\mathcal{L}(t)$ and $P(r_s,t)$ is scaled with $[\mathcal{L}(t)]^{1+\theta}$ with $\theta \simeq \frac{1}{2}$ for KPZ and $\theta \simeq 0$ for EW and KPZ-AA drivings. The scaled PDF can then be expressed as

\vspace{4mm}

\begin{figure}[ht!]
\hspace{-4mm}
\begin{minipage}{0.45\textwidth}
\includegraphics[width=\textwidth,height=.25\textheight]{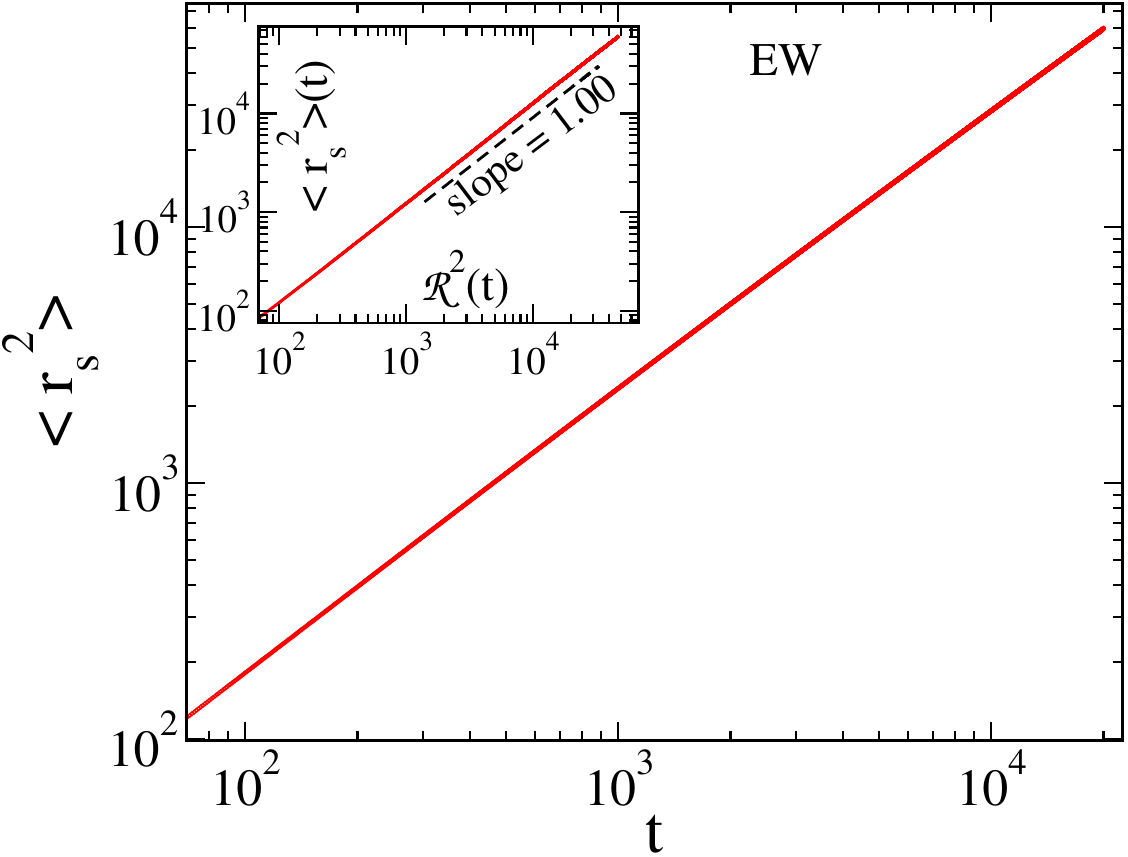} 
\end{minipage}%
\hspace{0.9mm}
\begin{minipage}{0.45\textwidth}
\includegraphics[width=\textwidth,height=.25\textheight]{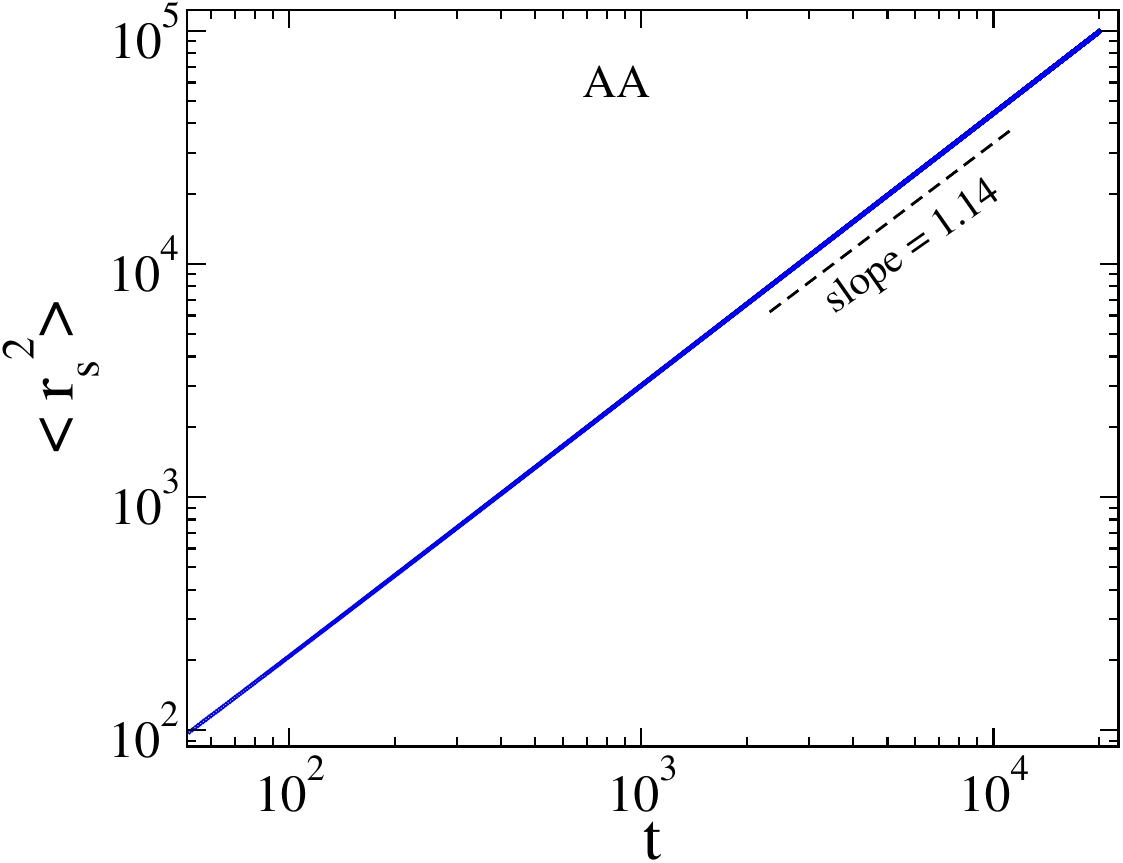} 
\end{minipage}%
\caption{Mean-squared separation between two particles as a function of time for EW and KPZ-AA driving. For EW driving, the inset shows $\langle r_s^2\rangle$ is proportional to $\mathcal{R}^2(t)$. For KPZ-AA driving as well, the growth of $\langle r_s^2\rangle$ is similar to that of $\mathcal{R}^2(t)$.}
\label{Fig_MSS_EW_AA}
\end{figure}

\vspace{4mm}

\begin{equation}
P(r_s,t) \approx  \frac{1}{[\mathcal{L}(t)]^{(1+\theta)}} \  Y \left( \frac{r_s}{\mathcal{L}(t)} \right)
\label{Separation_PDF}
\end{equation}
where the scaling function follows $Y(y)\sim y^{-\nu}$ as $y \rightarrow 0$, and falls exponentially as $y \rightarrow \infty$. The exponent values corresponding to the power-law decay are estimated as $\nu \simeq \frac{3}{2}$, $\simeq \frac{2}{3}$, and $\simeq \frac{1}{3}$ for KPZ, EW and KPZ-AA drivings, respectively. 


For KPZ driving, Ueda and Sasa \cite{Ueda_Sasa2015} had numerically found that the mean-squared separation follows 
\begin{equation}
\langle r_s^2 \rangle \sim t,  
\label{MSD_Pair}
\end{equation}
while at the same time the probability distribution of separation approaches a constant value for $r_s < r^{*}_s$ where 
$r^{*}_s$ is fixed. It should be noted that, for KPZ driving, both these properties are an immediate consequence of the scaling form of Eq. \ref{Separation_PDF} which is proposed in Ref. \cite{SBComment} and confirmed in Ref. \cite{USReply}. 

 On the other hand, in this work, for EW and KPZ-AA drivings, we see that the time dependence of $\langle r^2_s \rangle $ follows $\mathcal{R}^2(t)$ which is numerically verified as shown in Fig. \ref{Fig_MSS_EW_AA}. The scaling form (given by Eq. \ref{Separation_PDF}) leads to estimate $q$-th moment of the separation which grows as 
\begin{equation}
\langle r_s^q \rangle \sim t^{\frac{q-\theta}{z}}.
\label{qth_mom_multi}  
\end{equation}

\vspace{4mm}

\begin{figure*}[ht!]
\begin{minipage}{0.315\textwidth}
\includegraphics[width=\textwidth,height=0.175\textheight]{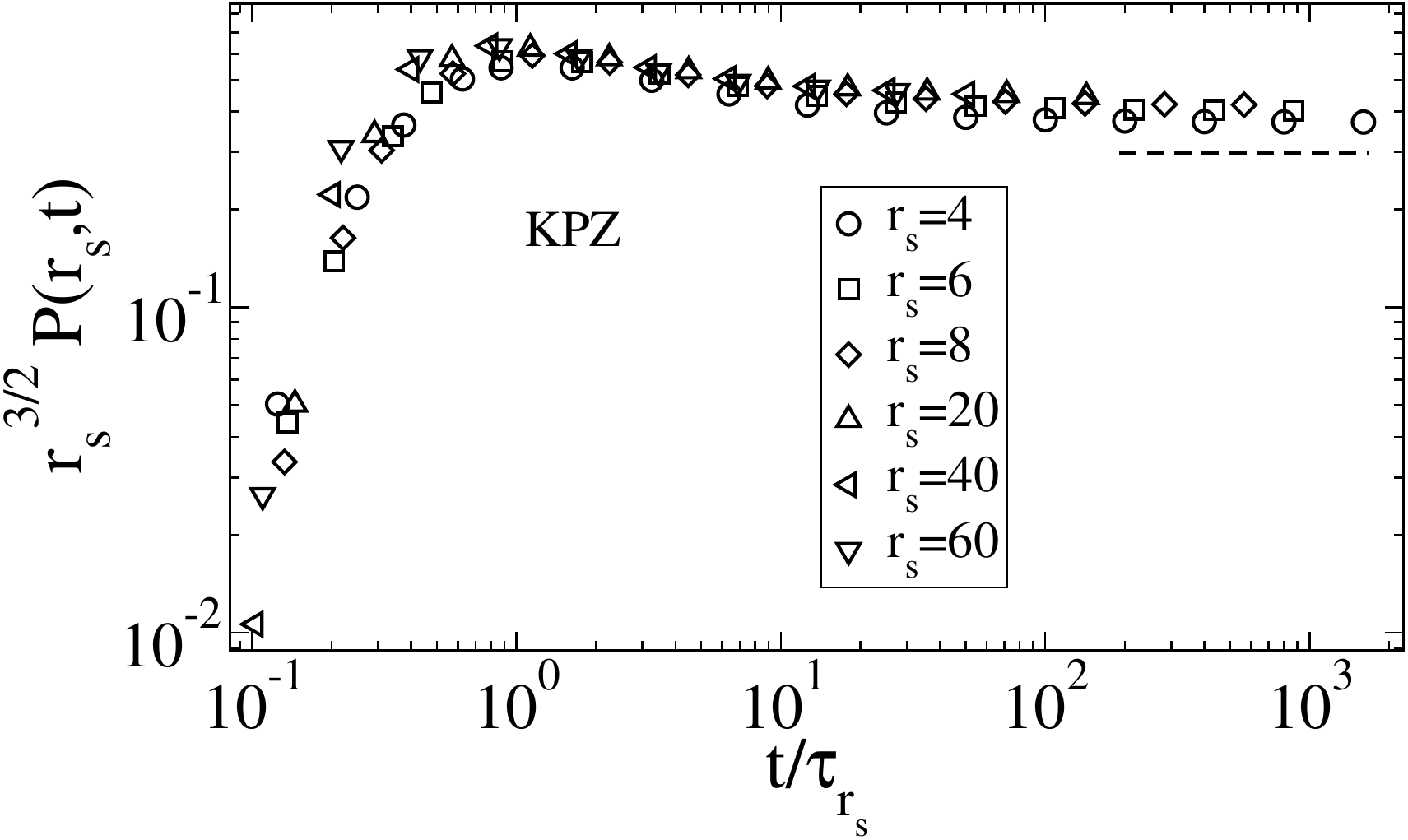}
\end{minipage}%
\hspace{2mm}
\begin{minipage}{0.315\textwidth}
\includegraphics[width=\textwidth,height=0.175\textheight]{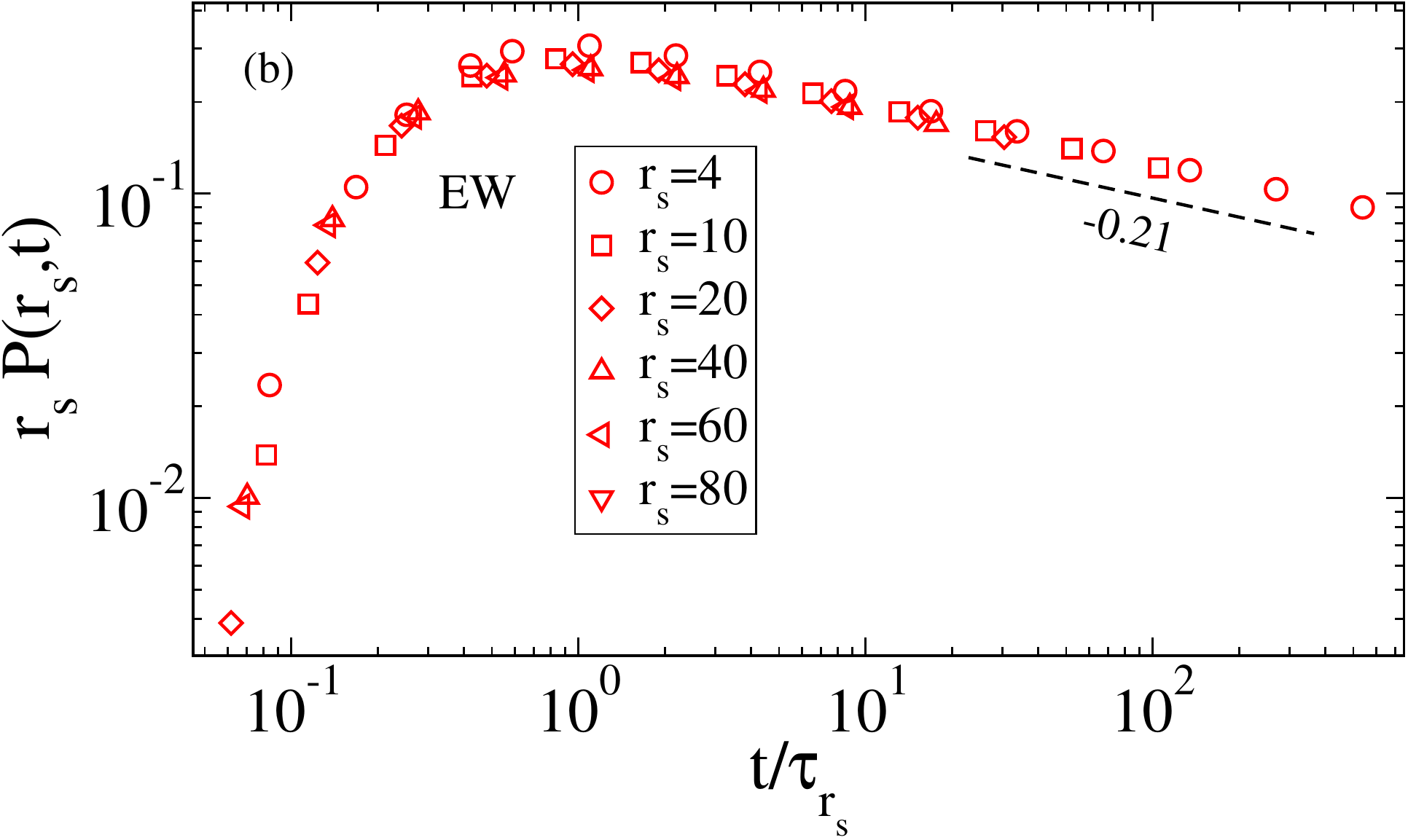}
\end{minipage}%
\hspace{2mm}
\begin{minipage}{0.315\textwidth}
\includegraphics[width=\textwidth,height=0.175\textheight]{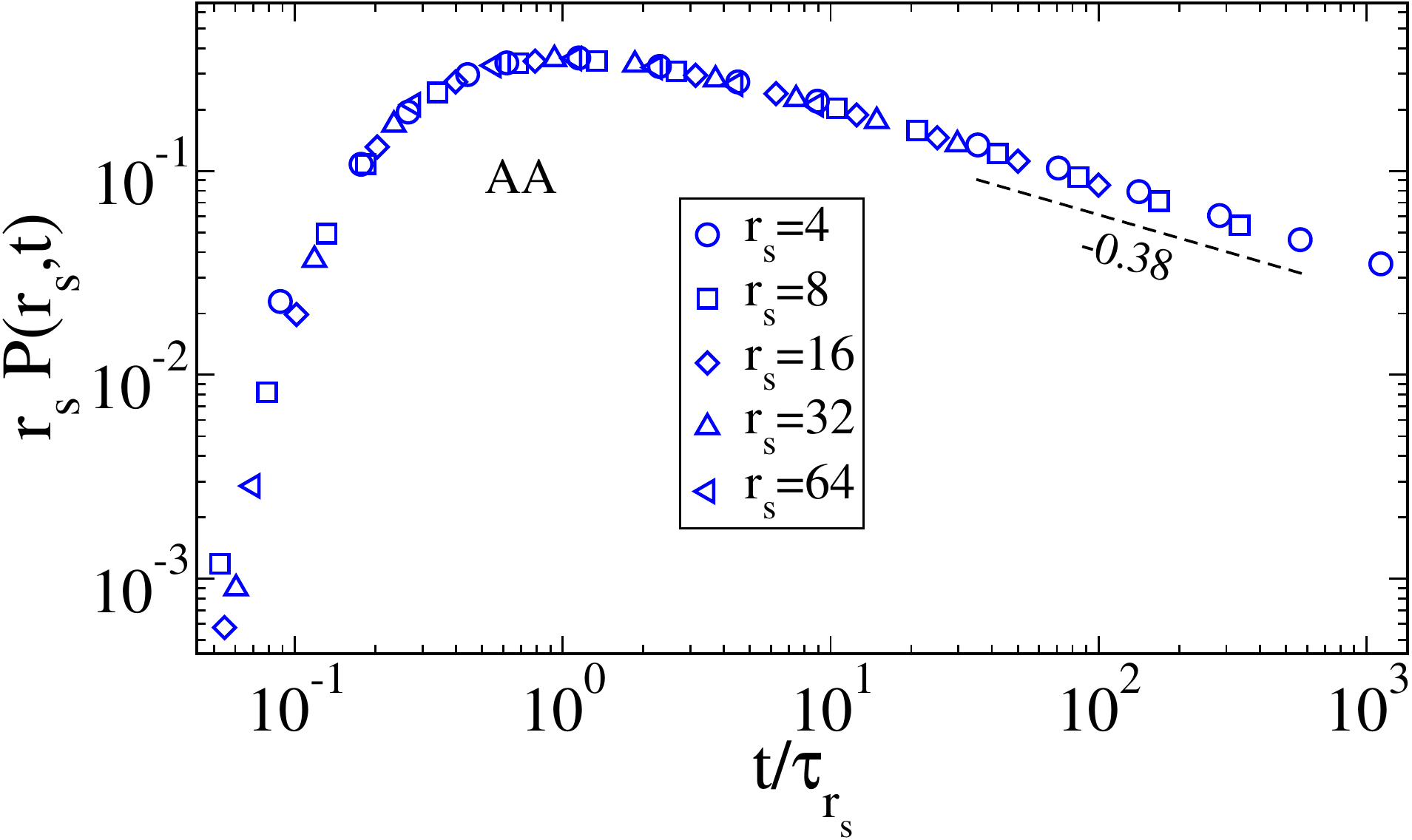}
\end{minipage}%
\caption{Variation of the probability $P(r_s,t)$ as a function of time. For KPZ driving, for every value of $r_s$, the probability $P(r_s,t)$ approaches a nonzero value as $t \rightarrow \infty$. For EW and KPZ-AA driving, however, $P(r_s,t)$ approaches zero as $t \rightarrow \infty$.}
\label{PDF_TimeFixedR}
\end{figure*}

\vspace{2mm}

\begin{figure}[ht!]
\begin{minipage}{0.45\textwidth}
\includegraphics[width=\textwidth,height=.25\textheight]{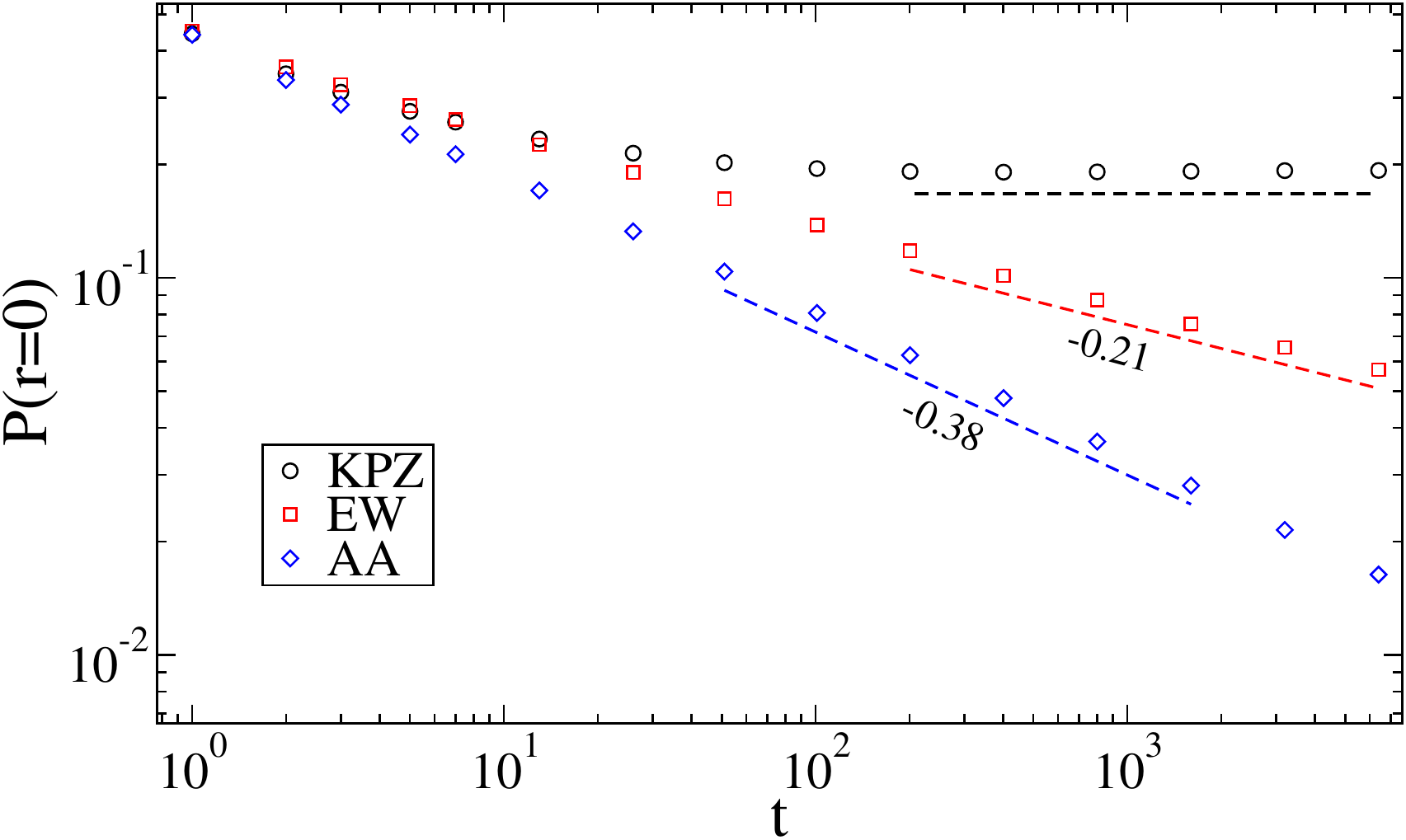}
\end{minipage}%
\caption{Probability that both particles are at the same site at time $t$. As $t \rightarrow \infty $, it approaches a constant value for KPZ driving whereas it falls as a power law for EW and KPZ-AA driving.}
\label{TwoParticleZero}
\end{figure}

From the scaling form of Eq. \ref{Separation_PDF}, along with the corresponding values of $\nu$ and $\theta$, we conclude that there exists a limiting form for the PDF for KPZ driving for large $t$, whereas PDFs decay with time and eventually vanish for EW and KPZ-AA drivings (Fig. \ref{PDF_TimeFixedR}). Thus, for KPZ driving, given any value of separation $r_s$, the distribution $P(r_s,t)$ approaches a time-independent value $P_{ss}(r_s) \sim r^{-3/2}_s$ for times $t \gg r_s^z$. For instance, we consider PDFs $P(r_s=0,t)$, shown in Fig. \ref{TwoParticleZero}, which quickly approach a constant value for KPZ driving and decay as a power with different exponents for EW and KPZ-AA driving. 
 
\vspace{2mm}

\vspace{2mm}


We conclude this sub section with a discussion of two-particle correlations in finite systems of size $L$. We expect to recover the correct scaling with $r_s$ on replacing $\mathcal{L}(t)$ by $L$ in Eq.\ref{Separation_PDF}. We find $P_{ss}(r_s,L) \sim r^{-3/2}_s$ for KPZ driving, and $P_{ss}(r_s,L) \sim \frac{1}{L^{1/3}} \, r_s^{-2/3}$  and $\sim \frac{1}{L^{2/3}} \, r_s^{-1/3}$ for EW and KPZ-AA driving, respectively. This is confirmed on noting that the two-point correlation function $G(r_s,L)$ is related to $P_{ss}(r_s,L)$ through $\frac{L}{N^2}G_{ss}(r_s,L) = P_{ss}(r_s,L)$. The scaling form for the steady state correlations in Ref. \cite{NMB_2006} is fully consistent with Eq.\ \ref{Separation_PDF}, if we make the replacement $\mathcal{L}(t)=L$

In a similar vein, the problem of two passive particles on a fluctuating KPZ surface is closely related to the problem of two second-class particles in the ASEP. An exact solution \cite{Derrida1993}, shows that the probability of finding two second-class particles at distance $r_s$ apart follows $P(r_s) \sim \frac{1}{r_s^{3/2}}$ for large $r_s$ as $L \rightarrow \infty$.  There are similarities and differences in the two cases. The rules of hopping of a single second class particle in the usual ASEP are the same as the rules of advection of a passive particle on a KPZ surface (on mapping particle and hole occupancies in the ASEP to uphills and downhills in the KPZ dynamics). However, two second-class particles cannot overlap unlike our noninteracting passive particles. Nevertheless, the large distance behavior is similar in the two cases.


\subsection{Overlap function}
A good way to quantify the closeness of the trajectories of two particles, is to follow the `overlap' of the trajectories up to time $t$. To this end, we follow \cite{Ueda_Sasa2015} and consider an overlap function 
\begin{equation}
q_o(t) = \frac{1}{t} \int^{t}_{0} \mathrm{d}t' \  \Theta(l-|r_s(t')|) 
\end{equation}
where $\Theta(x)$ is a theta function, $r_s(t')$ is the separation of two particles at time $t'$ \cite{Ueda_Sasa2015} and $l$ is a length which is used to qualify whether or not the trajectories do overlap. 

\vspace{4mm}
\begin{figure*}[ht!]
\begin{minipage}{0.315\textwidth}
\includegraphics[width=\textwidth,height=0.19\textheight]{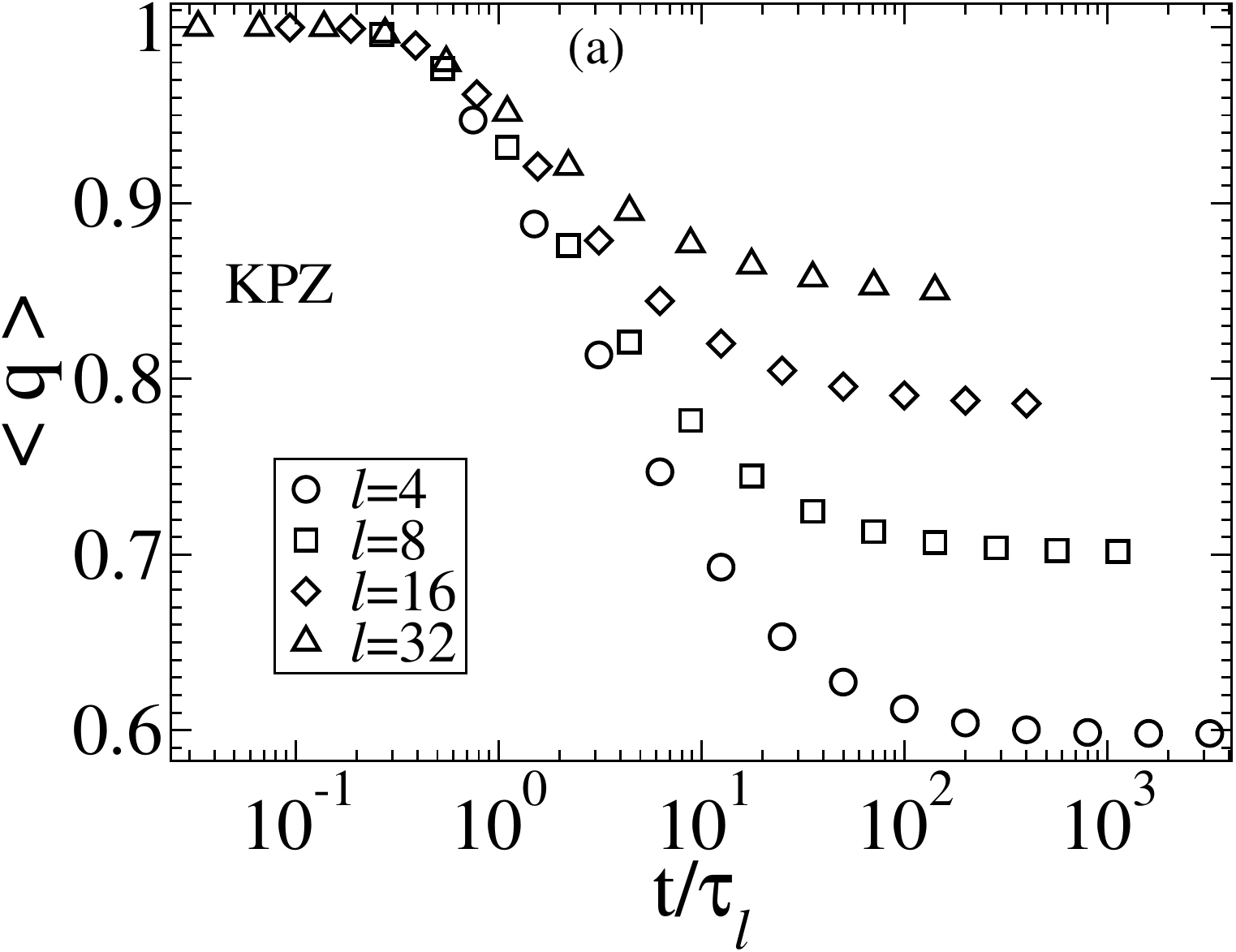}
\end{minipage}%
\hspace{1mm}
\begin{minipage}{0.315\textwidth}
\includegraphics[width=\textwidth,height=0.19\textheight]{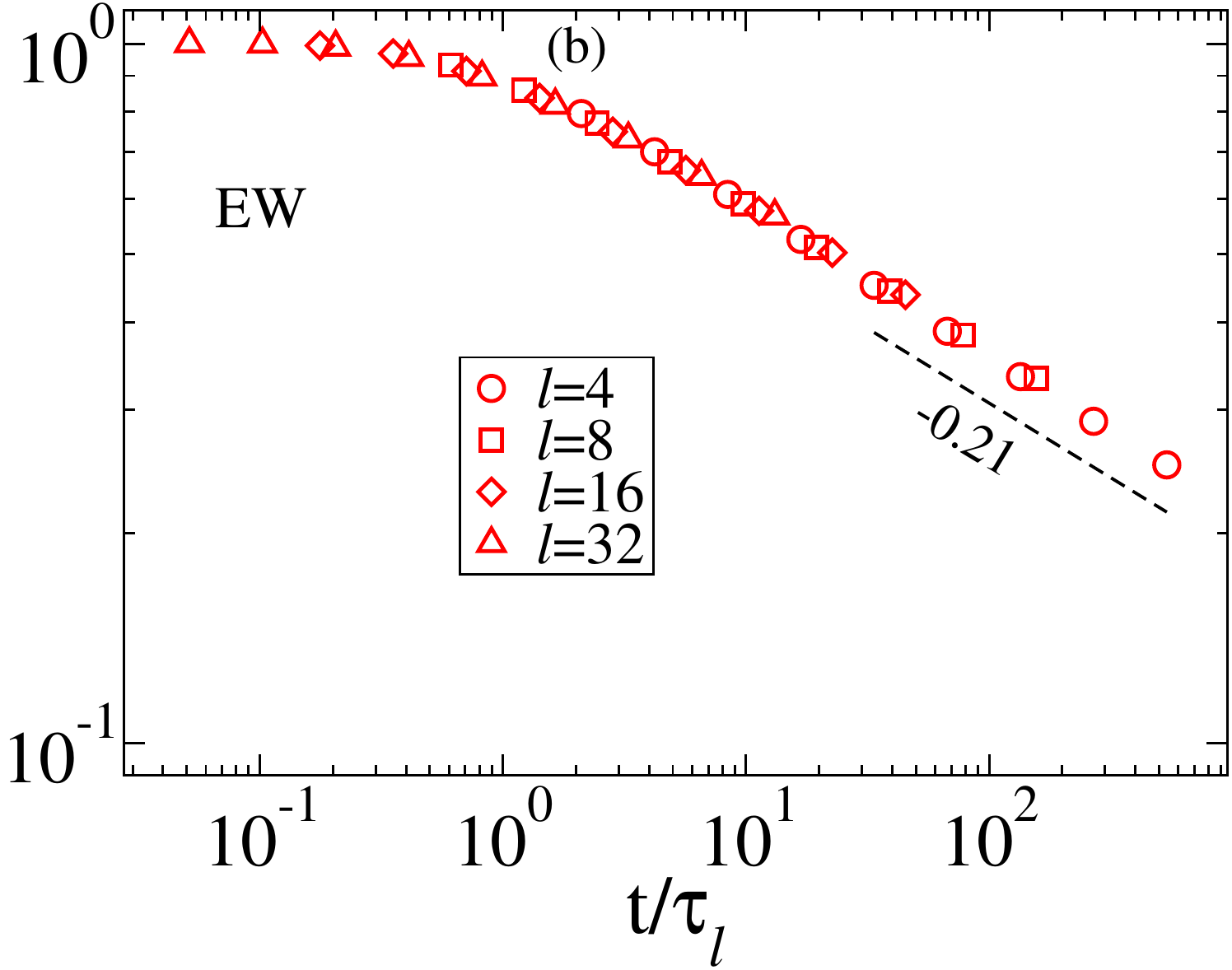}
\end{minipage}%
\hspace{1mm}
\begin{minipage}{0.315\textwidth}
\includegraphics[width=\textwidth,height=0.19\textheight]{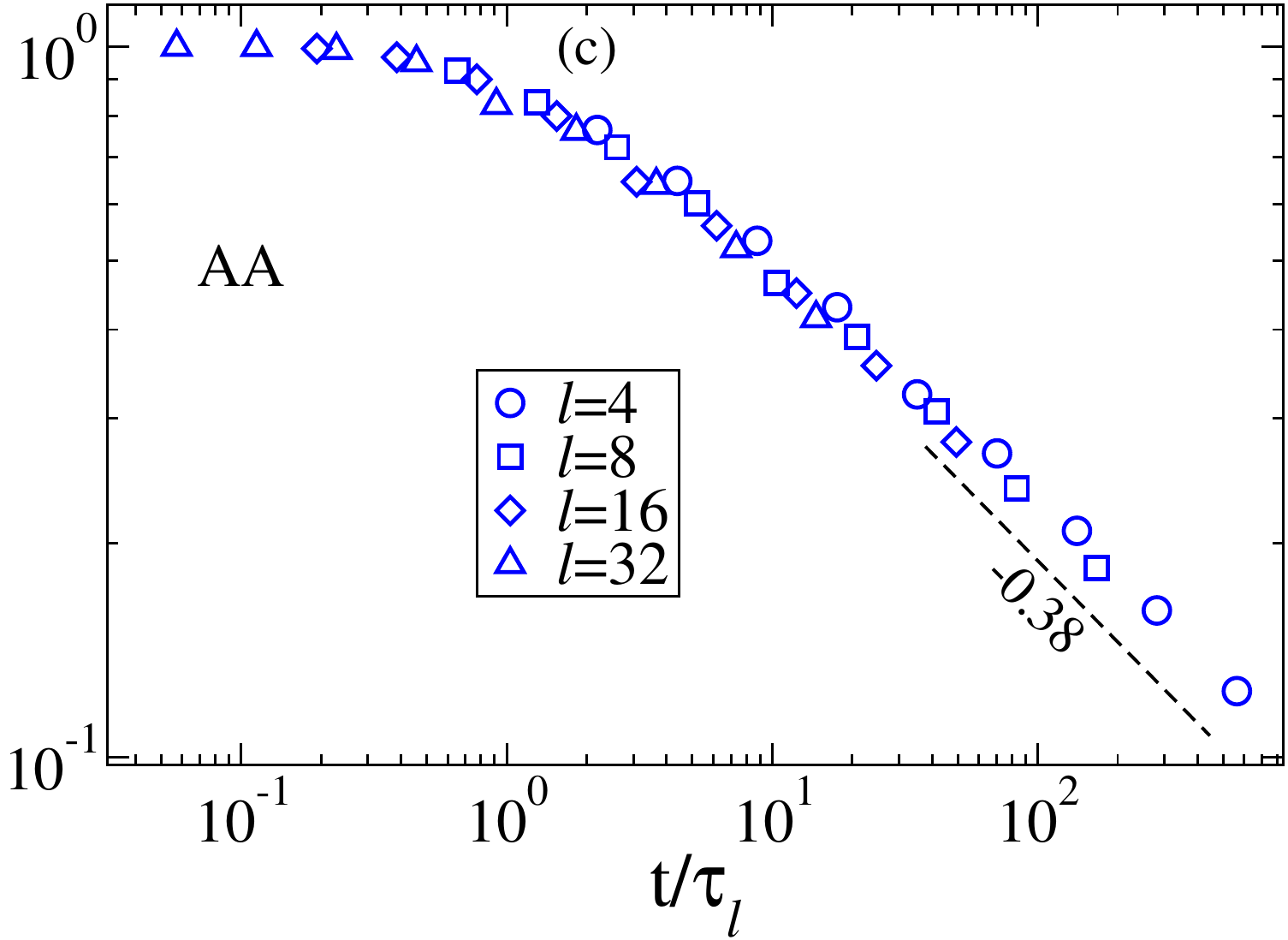}
\end{minipage}%
\caption{Average overlap as a function of scaled time for KPZ, EW, and KPZ-AA driving.}
\label{Fig_Average_Overlap}
\end{figure*}
Evidently, $q_o(t)$ measures the fraction of time during which trajectories overlap.
 
 Now consider the average overlap up to time $t$
\begin{equation}
\langle q_o(t) \rangle = \frac{1}{t} \int^{t}_{0} \mathrm{d}t' \ \left\langle \Theta(l-|r_s(t')|) \right\rangle
\label{Average_Overlap1}
\end{equation}
where $\langle \Theta \rangle$ indicates average over independent realizations. Rewriting in terms of $P(r_s,t)$, we obtain
\begin{equation}
\langle q_o(t) \rangle = \frac{1}{t} \int^{t}_{0} \mathrm{d}t' \int^{l}_{0} \mathrm{d}r_s \ P(r_s,t') .
\label{Average_Overlap}
\end{equation}
Substituting the scaling form of Eq.\ \ref{Separation_PDF} in  Eq.\ \ref{Average_Overlap}, we obtain 
\begin{equation}
\langle q_o(t) \rangle \sim \left(\frac{t}{l^z}\right)^{\frac{\nu-1}{z}}  t^{-\frac{\theta}{z}}  . 
\label{Eq_Average_Overlap}
\end{equation}

For KPZ driving, substituting $\nu \simeq 3/2$ and $\theta \simeq 1/2$ in Eq. \ref{Eq_Average_Overlap}, we obtain 
\begin{equation}
\langle q_o(t) \rangle \approx const.
\end{equation}
which is independent of time in the large-distance and long-time limit.

For EW and KPZ-AA drivings, we have $\theta\simeq 0$. Therefore Eq. \ref{Eq_Average_Overlap} reduces  to 
\begin{equation}
\langle q_o(t)\rangle \sim \left(\frac{t}{l^z}\right)^{\frac{\nu-1}{z}} 
\label{EW_AA_AverageOverlap}
\end{equation}
in the asymptotic limit of time. For EW driving, on substituting $\nu = 0.67$ and $z=2$ (omitting the logarithmic correction to $z$  given by Eq. \ref{EW_log_Correction}), we obtain $\langle q_o(t) \rangle \sim (t/l^z)^{-\phi_{EW}}$ with $\phi_{EW} \simeq 0.17$. Similarly, for KPZ-AA driving, on substituting $\nu = 0.33$ and $z=1.75$ in Eq.\ref{EW_AA_AverageOverlap}, we obtain $\langle q_o(t) \rangle \sim (t/l^z)^{-\phi_{AA}}$ with $\phi_{AA} \simeq 0.38$. 

In order to verify the dependence of time and localization length $l$ on $\langle q_o(t) \rangle$, we carried out a numerical simulation, and find fair agreement with Eq.\ \ref{Eq_Average_Overlap} for all three drivings, as shown in Fig.\ \ref{Fig_Average_Overlap}.  


\section{Steady state : Statics}

In order to study spatial fluctuations, we focus on the higher order structure functions of the particle number in a given stretch of sites. We find a nontrivial spectrum of exponents which implies multiscaling. We also study the flatness, a measure of spatial intermittency, for the three types of drivings.

\subsection{Static Correlation Function}
The steady states obtained for the three types of driving show interesting similarities and differences as illustrated in Fig.\ \ref{Density_profile}. These are reflected in the two-point density-density correlation function 
\begin{equation}
G_{ss}(r,L) = \langle n_i \ n_{i+r} \rangle
\end{equation} 
where $n_i$ denotes the total number of particles at $i$-th site. Numerical simulations \cite{NMB_2006,NBM_2005} reveal 
\begin{equation}
G_{ss}(r,L) = \frac{1}{L^{\theta}} Y_{ss}(r/L),
\label{correlation_SS}
\end{equation}
i.e. $G_{ss}$ is a scaling function of separation $r$ scaled by system size $L$. This unusual behavior is reminiscent of phase ordering, but the new point is that the scaling function $Y_{ss}(y)$ is \emph{divergent} in this case. In Eq.\ref{correlation_SS}, $Y_{ss}(y) \sim y^{-\nu}$ as $y \rightarrow 0$ where the exponent $\nu$ is estimated to be $\simeq \frac{3}{2}$,  $\simeq \frac{2}{3}$, and $\simeq \frac{1}{3}$ for KPZ, EW, and KPZ-AA drivings, respectively. The exponent $\theta$ is $\simeq \frac{1}{2}$ for KPZ driving ensuring normalizability, while $\theta \simeq 0$ for EW and KPZ-AA drivings. 

\subsection{Static structure functions and flatness}
As is evident from the density profiles shown in Fig.\ \ref{Density_profile}, there is a good deal of clustering for all three types of surface driving, though the degree of surface clustering seems to vary substantially from one case to the other. In order to quantify this, we study the moments of particle numbers $N_l$ in a stretch of $l$ successive sites in steady state. A good idea of clustering is obtained by studying the dependence  of the $q$'th order moment on the stretch length $l$:

\begin{equation}
R_q^{ss}(l) = \langle N_l^q \rangle \sim l^{\zeta(q)},
\end{equation}
where $\langle ... \rangle$ indicates the average over steady state configurations. We choose the stretch length to be a finite fraction of system size with $l/L=1/2^7$, $1/2^6$, $1/2^5$, $1/2^4$ for $L=4096$ and $8192.$ 

\vspace{4mm}

\begin{figure*}[ht!]
\begin{minipage}{0.315\textwidth}
\includegraphics[width=\textwidth,height=0.19\textheight]{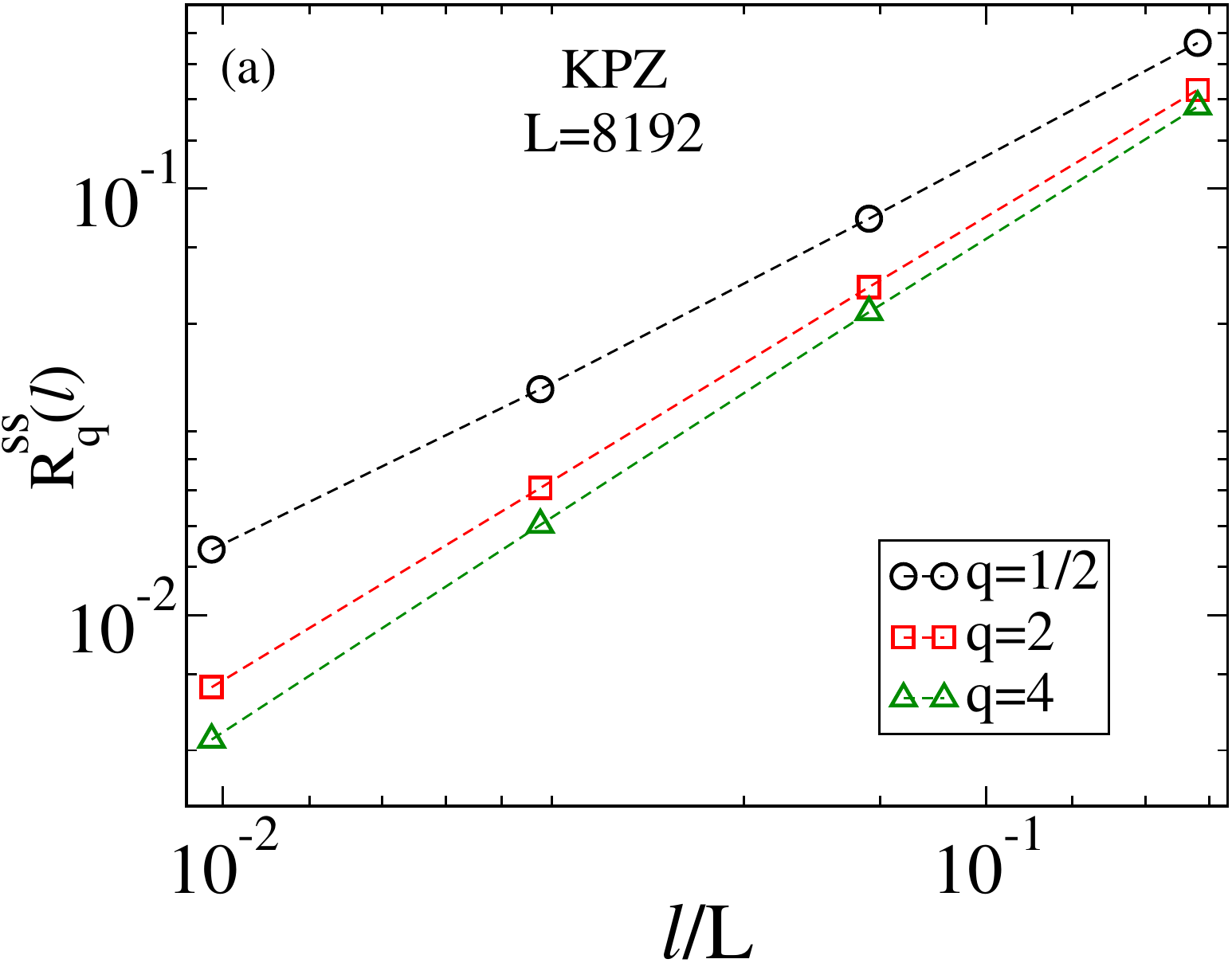}
\end{minipage}%
\hspace{1.2mm}
\begin{minipage}{0.315\textwidth}
\includegraphics[width=\textwidth,height=0.19\textheight]{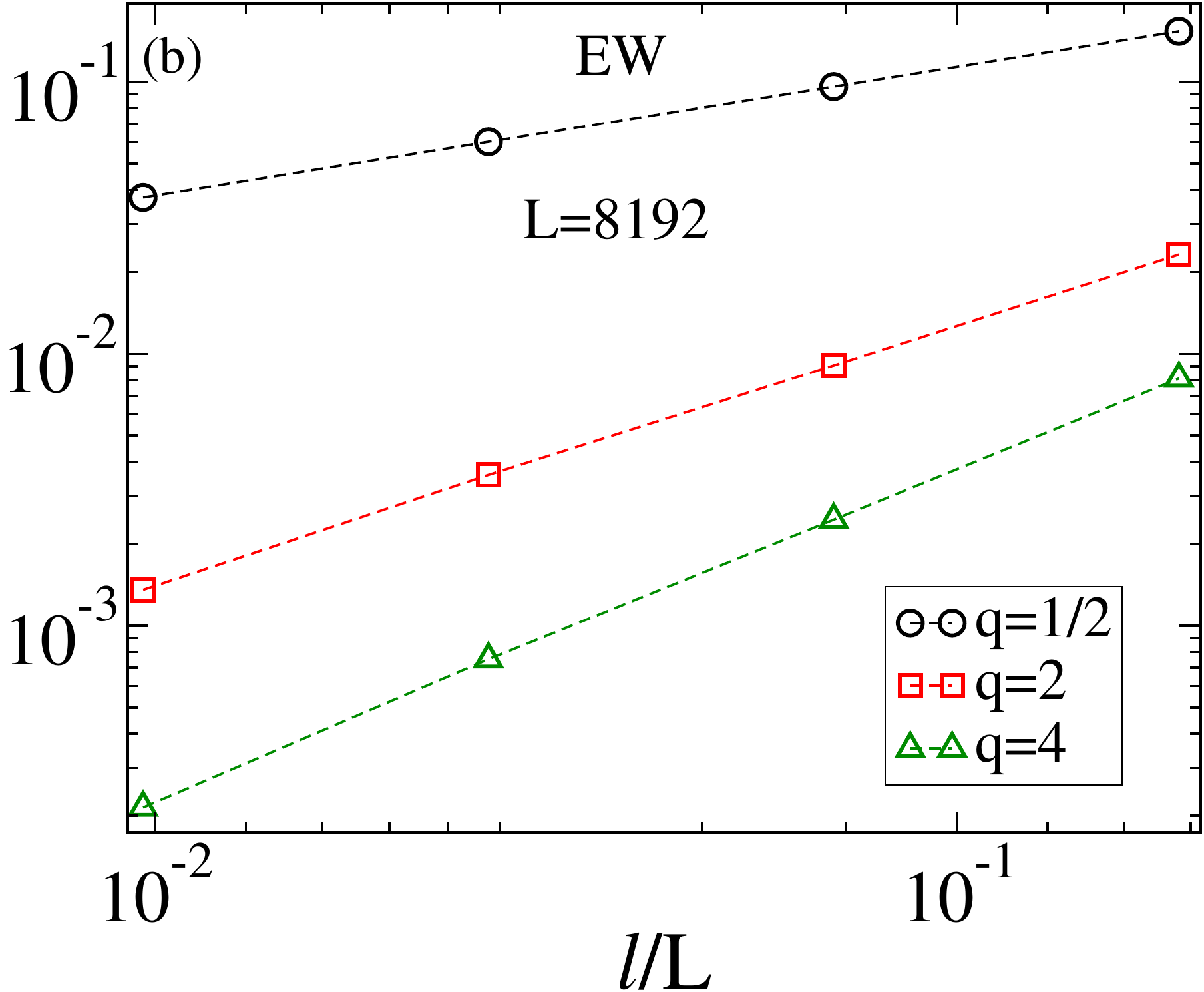}
\end{minipage}%
\hspace{1.2mm}
\begin{minipage}{0.315\textwidth}
\includegraphics[width=\textwidth,height=0.19\textheight]{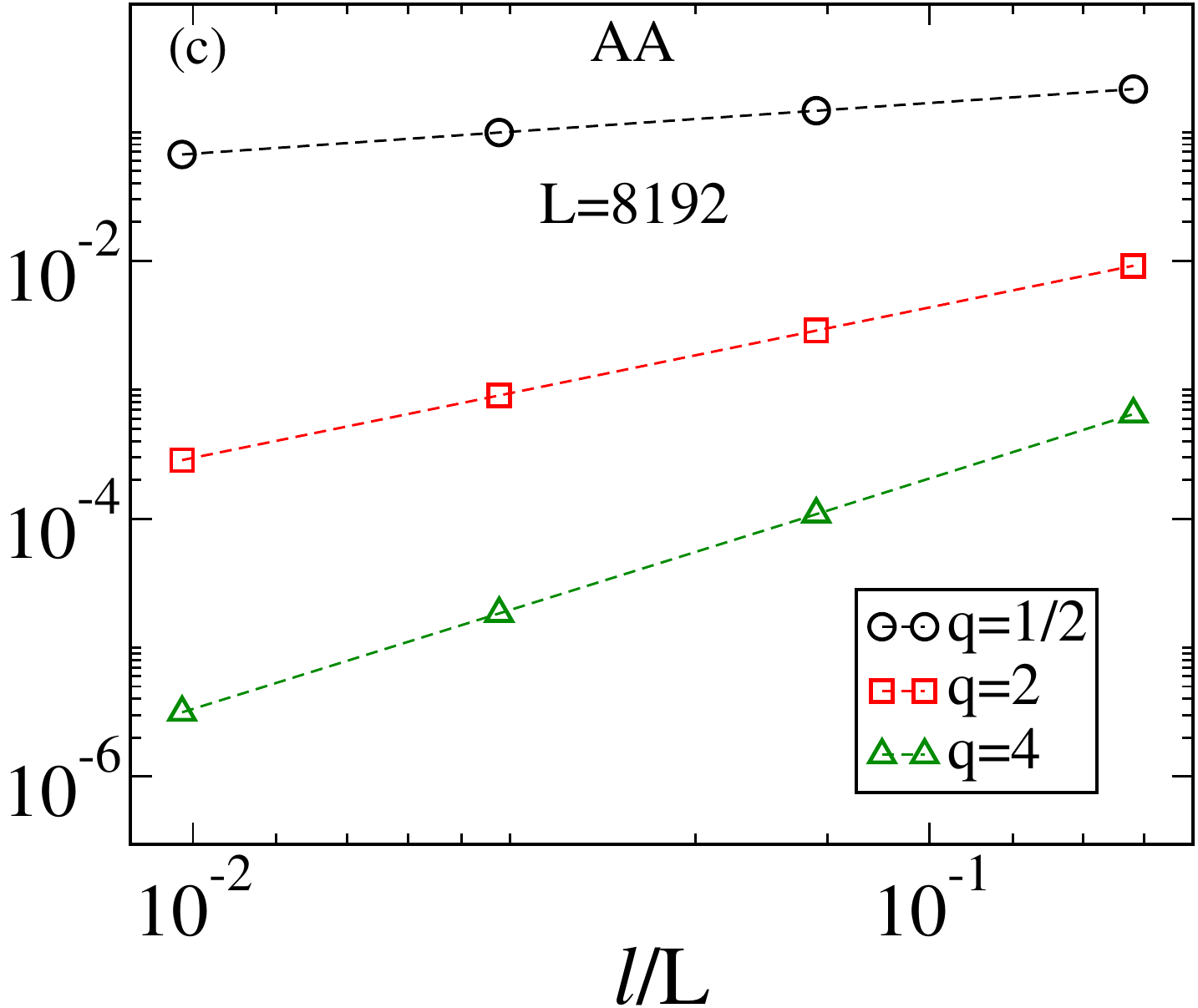}
\end{minipage}%
\caption{Variation of $q$-th order structure functions $R^{ss}_q(l)$ with stretch length $l$. For the purpose of presentation, we rescaled $R^{ss}_q(l)$ by a factor of $(8192)^{1/2}$, $(8192)^{2}$ and $(8192)^4$ for $q=1/2$, 2 and $4$, respectively.} 
\label{Moments}
\end{figure*}

\vspace{4mm}

In Fig.\ \ref{Moments}, we plot $R_q^{ss}$ versus $l/L$ for $q=\frac{1}{5}$, $\frac{1}{4}$ $\frac{1}{2}$, $\frac{3}{4}$, $2$ and $4$ for PSMs with KPZ, EW and KPZ-AA drivings. Measuring the slope of $R^{ss}_q(l)$ with $l/L$, we numerically determine $\zeta(q)$ for PSMs.

\vspace{4mm}

\begin{figure}[ht!]  
\begin{minipage}{0.45\textwidth}
\includegraphics[width=\textwidth,height=.25\textheight]{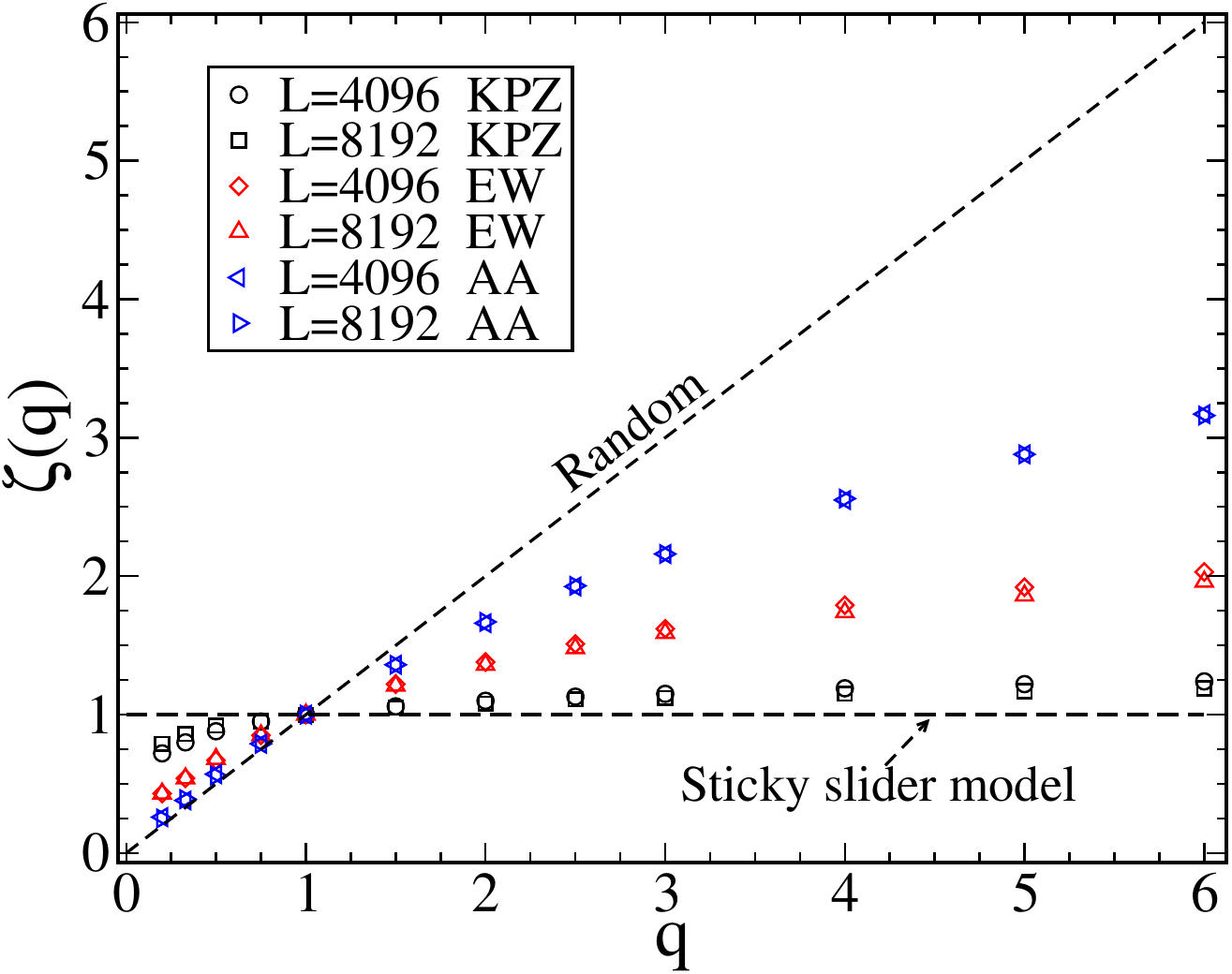}
\end{minipage}%
\caption{$q$-dependence of exponents $\zeta(q)$ which define the growth of structure functions. Particles show a larger degree of clustering with KPZ driving than in the EW case. The two dashed lines depict the limits of the no clustering (random) and extreme clustering (sticky slider model). Error-bars are smaller than the size of the symbols.}
\label{Exponents_Structure_factors}
\end{figure}
\vspace{4mm}

Figure\ \ref{Exponents_Structure_factors} shows the $q$-dependence of the exponents $\zeta(q)$ for several values of $q$ starting from $q=1/5$ up to $q=6$. For reference, we have included the curve for random placement of particles (no clustering) and SSM defined in Section 2, which shows intense clustering. Large values ($>1$) of $q$ amplify clustering, while the small values of $q$ ($<1$) bring out the background small signals.  A nonlinear dependence of $\zeta(q)$ on $q$ indicates multiscaling. The marked difference between the curves for KPZ, EW, and KPZ-AA driving quantifies  the degree of clustering in the three cases, evident in a qualitative sense in Fig.\ \ref{Density_profile}. 

\emph{KPZ driving:} As $L$ increases, $\zeta(q)$ seems to approach the SSM value unity   
for all $q \neq 0$, indicating extremely strong clustering. 

\emph{EW and KPZ-AA driving:} For $q < 1$, $\zeta(q)$ varies linearly with $q$ which implies that the smaller signals are self-similar while $\zeta(q)$ for $q>1$ is not linear with $q$ indicating multiscaling. 

It is interesting that though the $\zeta(q)$ versus $q$ plot shows a significant difference between KPZ and EW drivings, the flatness varies in a similar way for both. If we define the exponent $\sigma$ through  
\begin{equation}
\kappa_4=R^{ss}_4/(R_2^{ss})^2 \sim (l/L)^{-\sigma}
\label{SS_static_flatness}
\end{equation}
the value of $\sigma$ is $ \simeq 1$ for KPZ and EW driving whereas  $\sigma \simeq 0.75$ in the for KPZ-AA case.  
\vspace{4mm}

\begin{figure}[ht!]
\hspace{-5mm}
\begin{minipage}{0.45\textwidth}
\includegraphics[width=\textwidth,height=.25\textheight]{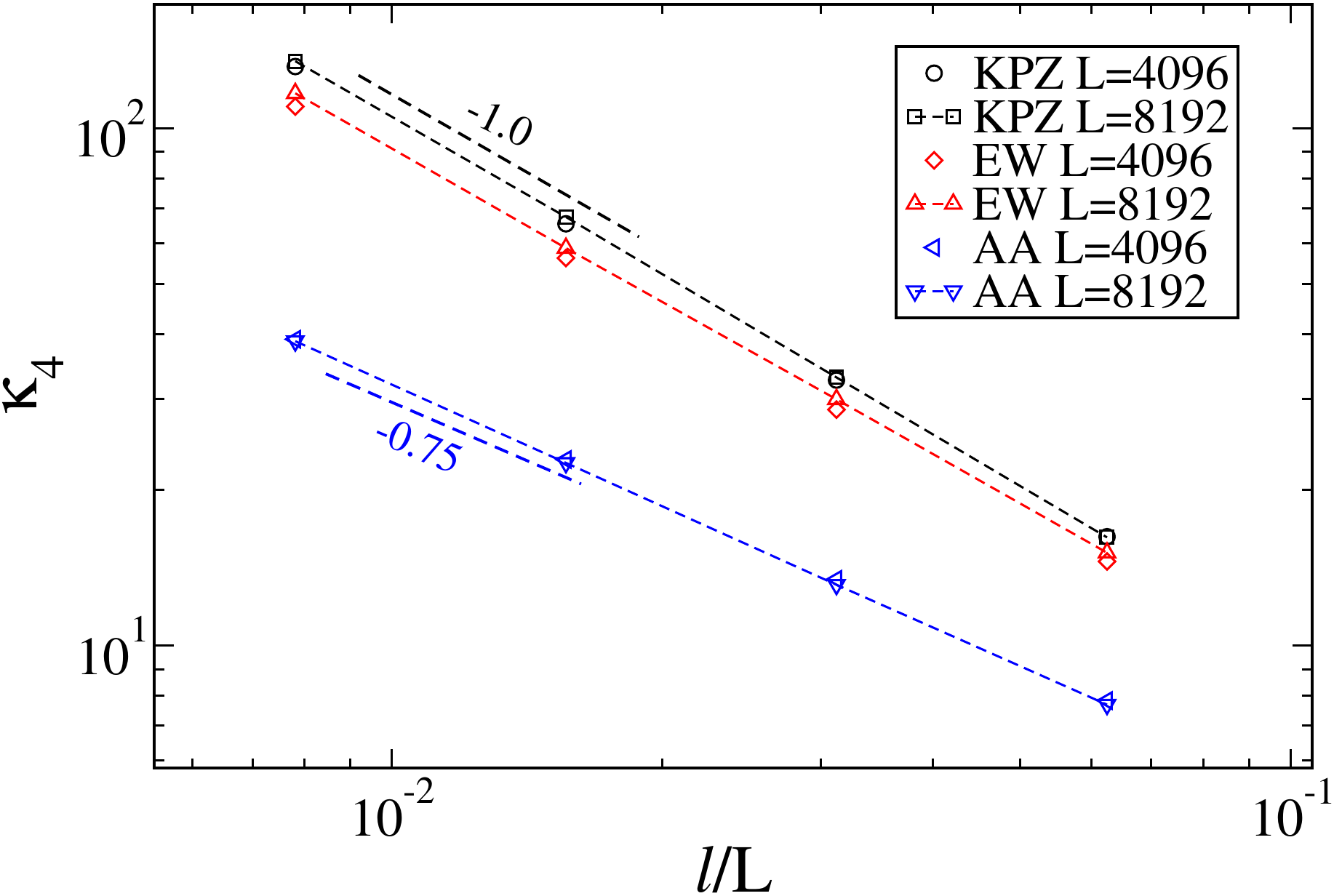}
\end{minipage}%
\caption{Flatness $\kappa_4$ in steady state as a function of $l/L$. $\kappa_4$ diverges as 
a power law with exponent $\sigma \simeq 1$ for KPZ and EW driving, and $\sigma \simeq 0.75$ for KPZ-AA driving.}
\label{Spatial_flatness}
\end{figure}
\vspace{4mm}

\section{Steady State : Dynamics}
In this section we compare results for temporal intermittency, with KPZ advection, EW and KPZ-AA driving. The flatness has a scaling form which shows a divergence as the scaled time approaches zero; the exponent characterizing the divergence gives a useful quantification of the degree of intermittency.  

We also study the problem in the adiabatic limit for the three drivings, and find that while the flatness diverges, indicating intermittency, the scaling exponents do not agree with those found for the particle model. 

 Finally, we show that numerical results for sticky slider models verify the scaling form of flatness predicted by analytical arguments.

\subsection{Dynamic correlation functions}

The time dependent density-density auto-correlation function 
\begin{equation}
G_s(t,L)=\langle n_i(0)n_i(t) \rangle 
\label{Eq_Defi_AutoCorreTime}
\end{equation}
has been studied \cite{NBM_2005}, and found to follow the scaling form 
\begin{equation}
G_s(t,L) \sim \tilde{Y}(t/L^z)
\label{Eq_Scaling_AutoCorreTime}
\end{equation}
where $\tilde{Y}(y) \sim \tilde{y}^{-\tilde{\nu}}$ as $\tilde{y} \rightarrow 0$. The estimated values of the exponent $\tilde{\nu}$ are $ \simeq \frac{2}{3}$, $\simeq \frac{1}{3}$ and $\simeq 0.19$ for 
KPZ, EW, and KPZ-AA drivings, respectively  \cite{NBM_2005}.

\subsection{Dynamic structure functions and intermittency}

We now present numerical results for particle number fluctuations in a stretch of the lattice for PSMs and SSMs. We show that a scaling description holds for structure functions, and support this through analytical arguments for the SSM. We also compare our numerical results for PSMs with KPZ, EW and KPZ-AA dynamics with analytical and numerical results for the corresponding SSMs. 

 The time-dependent $q$-th order structure function of particle number fluctuations in steady state is given by
\begin{equation}
S^{ss}_q(t_0,t,l) = \langle [N_l(t_0+t)-N_l(t_0)]^{q} \rangle
\label{struc}
\end{equation}
where the condition $t_0 \gg L^z$ is imposed to gurantee for steady state. Here $N_l(t)$ is the total number of particles at time $t$ in a stretch length $l$, which we take to be a finite fraction of the system size. We consider a large value of $l$ as clusters may be spread out, both in the steady state \cite{NBM_2005}  and coarsening regime (which is discussed in the next section). Associated with the stretch length $l$, there is a time scale $\tau_l$ beyond which particle number fluctuations are uncorrelated. Consequently the structure function $S^{ss}_q(t_0,t,l)$ saturates for $t>\tau_l$.

 Representative time series of particle number $N_l$ for the three types of dynamics are shown in Fig. \ref{Time_Series_inl}. We monitor the ratio of fourth ($S_4$) and square of second moment ($S_2$), namely flatness $\kappa_4^{ss} \equiv S^{ss}_4/(S^{ss}_2)^2$. Intermittency is indicated by the divergence of $\kappa_4^{ss}$ in the limit $t/l^z \rightarrow 0$ \cite{Frisch1995}.

\begin{table*}
\centering
\caption{ Values of the exponents of $\kappa_4$ for PSMs and SSMs \footnote{For SSM drivings, we have analytical arguments to obtain the exponents in different regimes.}  in different regimes.} 
\centering
\begin{tabular}{ |p{2.1cm}||p{2.8cm}|p{1.8cm}|p{2.8cm}|p{2.8cm}| }
 \hline
 System & Steady-state statics & \multicolumn{2}{c|}{Steady-state dynamics } & Coarsening \\
 \hline
Exponents &   $\sigma$ \hspace{0.4cm} (Eq.\ \ref{SS_static_flatness})	&   $\phi$  &  $\gamma$ \hspace{0.4cm} (Eq.\ \ref{eq:SS_scaling}) & $\psi$ \hspace{0.4cm} (Eq.\ \ref{Eq_Coarsening_scaling})  \\
\hline
PSM-KPZ  &	$1.00 \pm 0.02$ 	& $ 1$ & $  0.70 \pm 0.01 $ 
 & $1.02 \pm 0.01$   \\
\hline
PSM-EW  & $0.98 \pm 0.01$	 & $ 1$  & $  0.43 \pm 0.03 $  & $0.94 \pm 0.01$  \\
\hline
PSM-KPZ-AA  & $0.75 \pm 0.01$ & $  0.75$  & $  0.41 \pm 0.02 $ & $0.75 \pm 0.02$   \\
\hline
\hline
\text{SSM-KPZ}   & $1$	& $  1$ & $  0.67 $ & $1$    \\
\hline
\text{SSM-EW}  & $1$	& $  1$ & $  0.56 $ & $1$  \\
\hline
\text{SSM-KPZ-AA}  & $1$	& $  1$ & $  0.57 $ & $1$   \\
\hline
\end{tabular}
\label{Table_flatness_inter}

\end{table*}
\vspace{4mm}

\begin{figure}[ht!]
\hspace{-2mm}
\begin{minipage}{0.45\textwidth}
\includegraphics[width=\textwidth,height=.25\textheight]{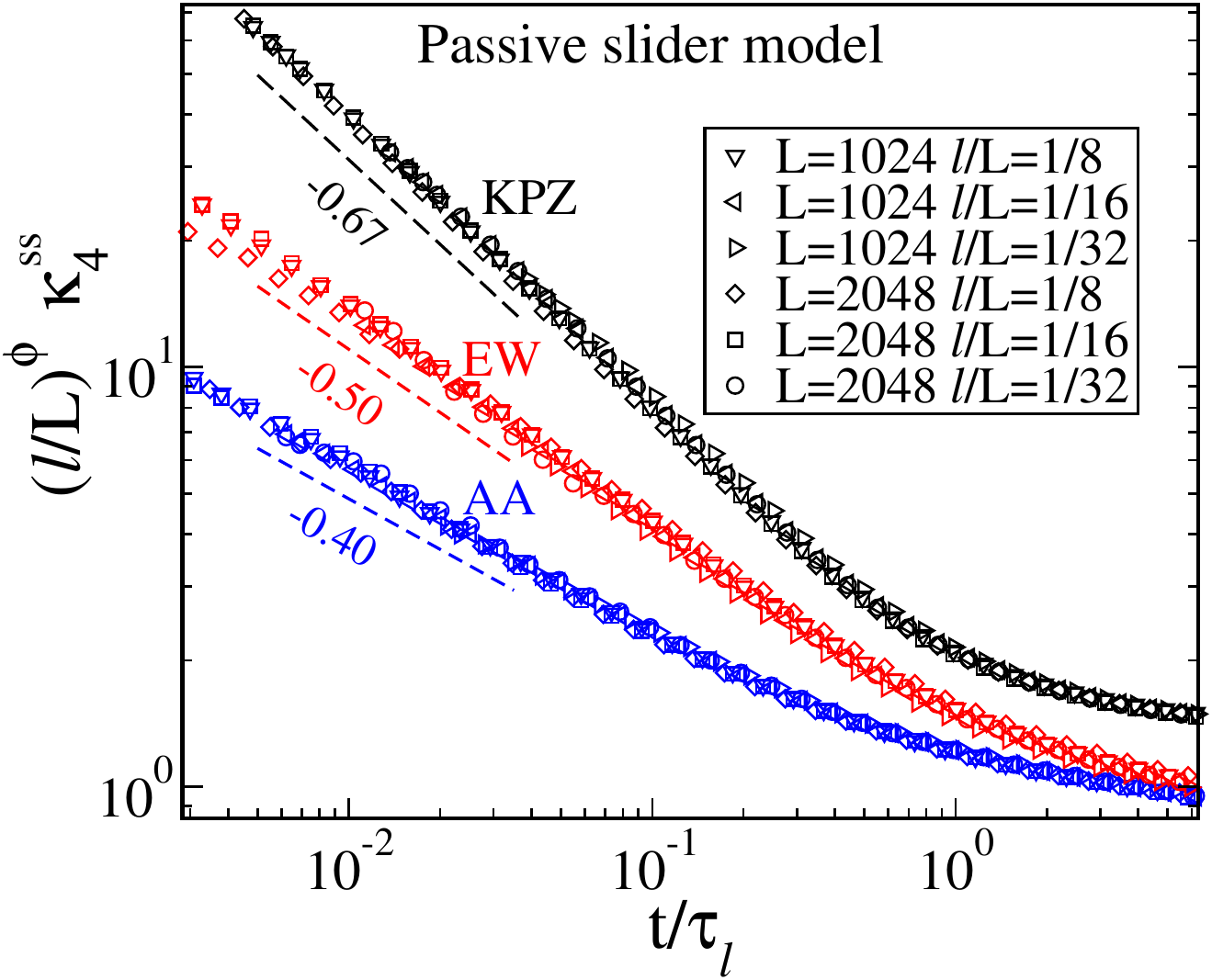}
\end{minipage}%
\hspace{1.3mm}
\begin{minipage}{0.45\textwidth}
\includegraphics[width=\textwidth,height=.25\textheight]{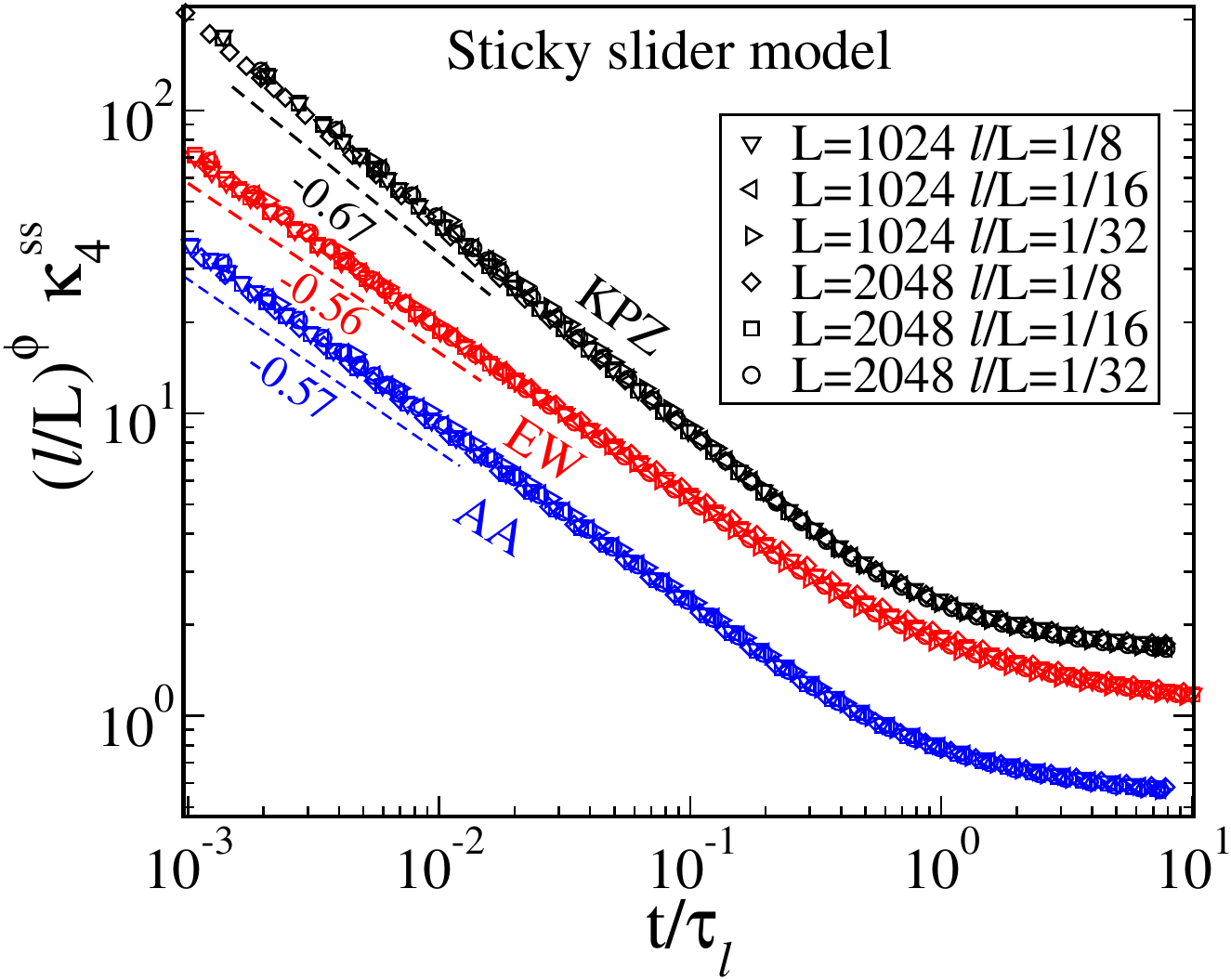}
\end{minipage}%
\caption{The divergence of flatness indicates temporal intermittency in steady state
for both PSM and SSM. For ease of display, $\kappa^{ss}_4$ is multiplied by constant factors, namely $1.5$, $2.5$, $2$ and $3$ for PSM-EW, PSM-KPZ, SSM-EW and SSM-KPZ, respectively.}
\label{Fig:Flatness_SS_PSM_SSM}
\end{figure}
\vspace{4mm}

\emph{Passive slider model:} We study the flatness for different values of system size $L$ and stretch length $l$, i.e., $l/L=1/8$, $1/16$ and $1/32$ for $L=1024$ and $2048$ for the PSM with KPZ, EW and KPZ-AA dynamics. In the scaling limit, the flatness diverges with a power law $t/\tau_l \rightarrow 0 $ and  saturates as $t/\tau_l \rightarrow \infty $. The saturation values depend on $L/l$, but collapse when time $t$ is scaled by $\tau_l$ and $\kappa_4^{ss}$ is scaled by $(L/l)^{\phi}$. 
Since particles are non-interacting, the time scale $\tau_l$ is determined by the time taken 
by a \emph{single} particle to cover a distance $l$, i.e., $\tau_l \sim l^z$, where the dynamic exponent $z$ depends on the surface driving as we have seen in Section 2.

This results in the compact scaling form 
\begin{eqnarray}
\kappa_4^{ss} \sim (L/l)^{\phi}  F_{\text{PSM}}\left(\frac{t}{\tau_{l}}\right) 
\label{eq:SS_scaling}
\end{eqnarray} 
 where $F_{\text{PSM}}(y) \sim y^{-\gamma} $ as $y \rightarrow 0$ and $F_{\text{PSM}}(y) \rightarrow \text{const}$ as $y\rightarrow \infty$. As seen in Fig. \ref{Fig:Flatness_SS_PSM_SSM}, our numerical simulations estimate $\phi \simeq 1$
for both KPZ and EW driving, and $\phi \simeq 0.75$ for KPZ-AA. Similarly, exponents corresponding  to
$F_{\text{PSM}}(y)$ have the values $\gamma \simeq 0.67$, $\simeq 0.50$ and $\simeq 0.40$ for KPZ, EW and KPZ-AA driving, respectively (see Table I).

\emph{Sticky slider model:} To get some insight into the occurrence of scaling, we study the SSM \cite{Singha2018} , defined by the rule that particles which find themselves on the same site undergo irreversible aggregation and do not separate once they are together.  Starting from a configuration with random placement of particles, the number of clusters decreases in time in the coarsening regime, finally reaching a single cluster which moves all over the system in steady state. By design, the SSM is a model of extreme clustering which is simple enough that one can understand the origin of scaling through analytic arguments. On the quantitative front, the SSM for KPZ driving resembles the corresponding PSM  fairly closely, whereas the SSM and PSM for both EW and KPZ-AA cases differ substantially from each other. This is borne out by the results shown in Table I.

In the steady state of the SSM, a single aggregate $A_{N}$ with $N$ particles slides stochastically on a 1D stochastically evolving surface of size $L$. Its motion is identical to that of a single walker, so in time $t$, its typical  displacement  $\mathcal{R}(t) \sim t^{1/z}$ implying that $A_N$ takes time $\tau_l \sim l^z$ to traverse the stretch length $l$. In order to estimate $S^{ss}_q$  given in Eq. \ref{struc} for SSMs, let us consider 
the location $R_0$ and $R$ of $A_{N}$ at times $t_0$ and $(t_0+t)$ respectively. The probability that $R_0$ is inside the stretch $l$ is $l/L$ in which case the probability of $R$ falling outside $l$ is of the order of $p_1 = \frac{\mathcal{R}(t)}{L}$. Likewise, when $R_0$ is outside $l$, the probability of $R$ falling inside $l$ is $p_2 = (1-\frac{l}{L}) \frac{\mathcal{R}(t)}{L}$. Hence, the $q$-th order structure function is given by
\begin{equation}
S_q^{ss} = p_1 N^q+ p_2 N^q . 
\label{SS_SecondMoment}
\end{equation}
Hence, by considering $N=L$ (which guarantees unit global density), we find that the flatness $\kappa_4^{ss}(t)= S^{ss}_4/(S^{ss}_2)^{2}$ is given by 
\begin{equation}
\kappa_4^{ss}  \simeq \frac{L}{\mathcal{R}(t)} \sim \frac{L}{t^{1/z}}. 
\label{Eq:SS_SSM}
\end{equation}
Thus for the SSM, the distinction between different surface drivings enters only through the values of $z$ for the different models. Similarly, the higher order normalized cumulants can be calculated straightforwardly.

Figure \ref{Fig:Flatness_SS_PSM_SSM}(b) shows that the flatness for the SSM for different surface drivings is given by 
\begin{equation}
\kappa_4^{ss} \sim (L/l)^{\phi}  F_{\text{SSM}}\left(\frac{t}{\tau_{l}}\right)
\end{equation}
as for the PSM, but with different exponents (Table I). The scaling function $F_{\text{SSM}}(y) \sim y^{-1/z}$ as $y \rightarrow 0$ and $F_{\text{SSM}}(y) \rightarrow \text{const}$ as $y\rightarrow \infty$ where the exponent $ 1/z \simeq 0.67$, $\simeq 0.56$ and $\simeq 0.57$ in the KPZ, EW and KPZ-AA cases respectively. Further, the scaling functions for $\kappa^{ss}$ are different for different drivings. 

For KPZ driving, the decay exponent  $\gamma$ of $F_{\text{PSM}}(y)$ in Eq.\  \ref{eq:SS_scaling} is the same as $1/z$ of $F_{\text{SSM}}(y)$, but it is substantially different for PSMs with EW and KPZ-AA drivings. Exponent values for PSMs and SSMs for the three types of driving are given in Table I.

\subsection{Adiabatic Approximation}

\vspace{4mm}
\begin{figure*}[ht!]
\begin{minipage}{0.315\textwidth}
\includegraphics[width=\textwidth,height=.20\textheight]{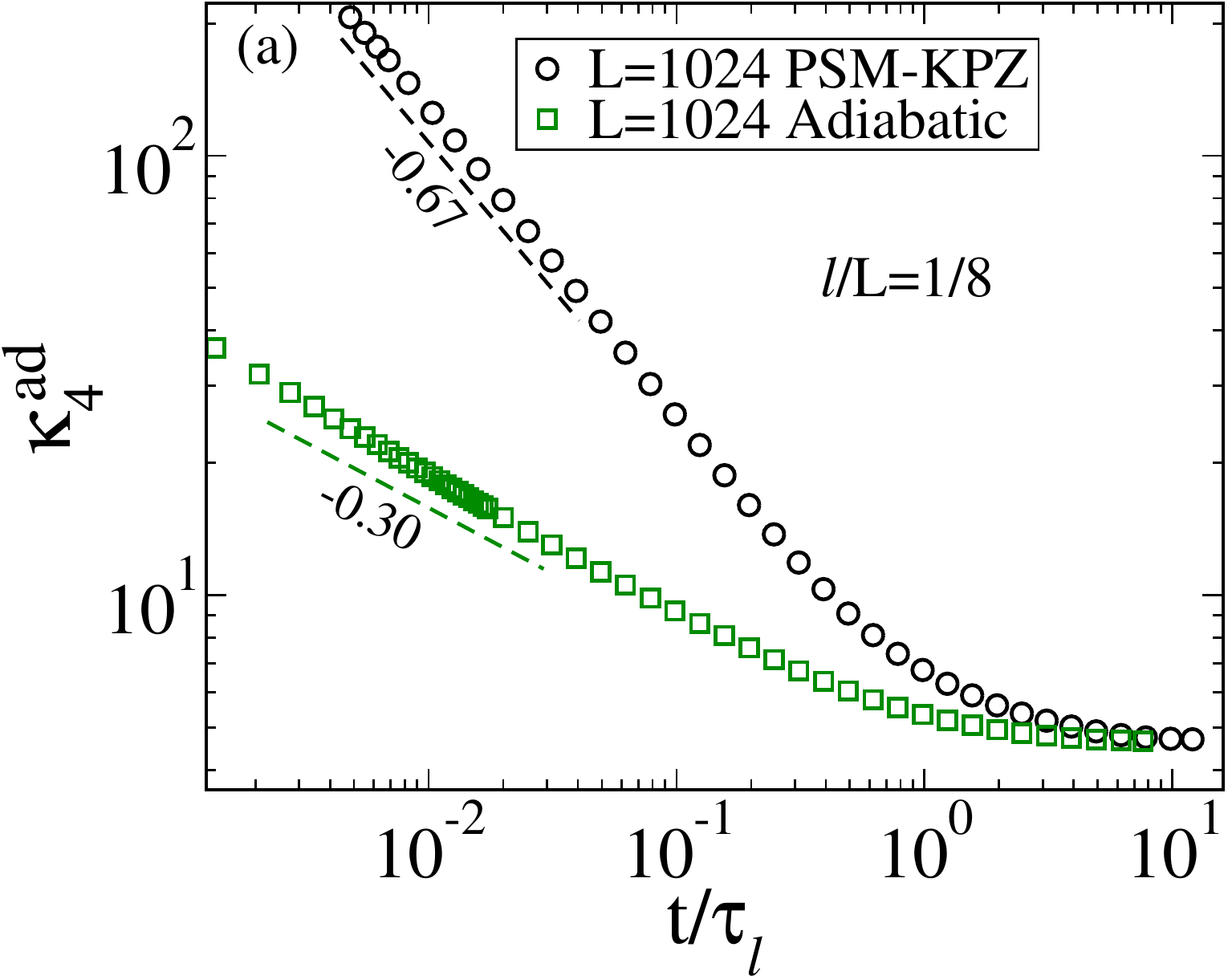}
\end{minipage}%
\hspace{1.6mm}
\begin{minipage}{0.315\textwidth}
\includegraphics[width=\textwidth,height=.20\textheight]{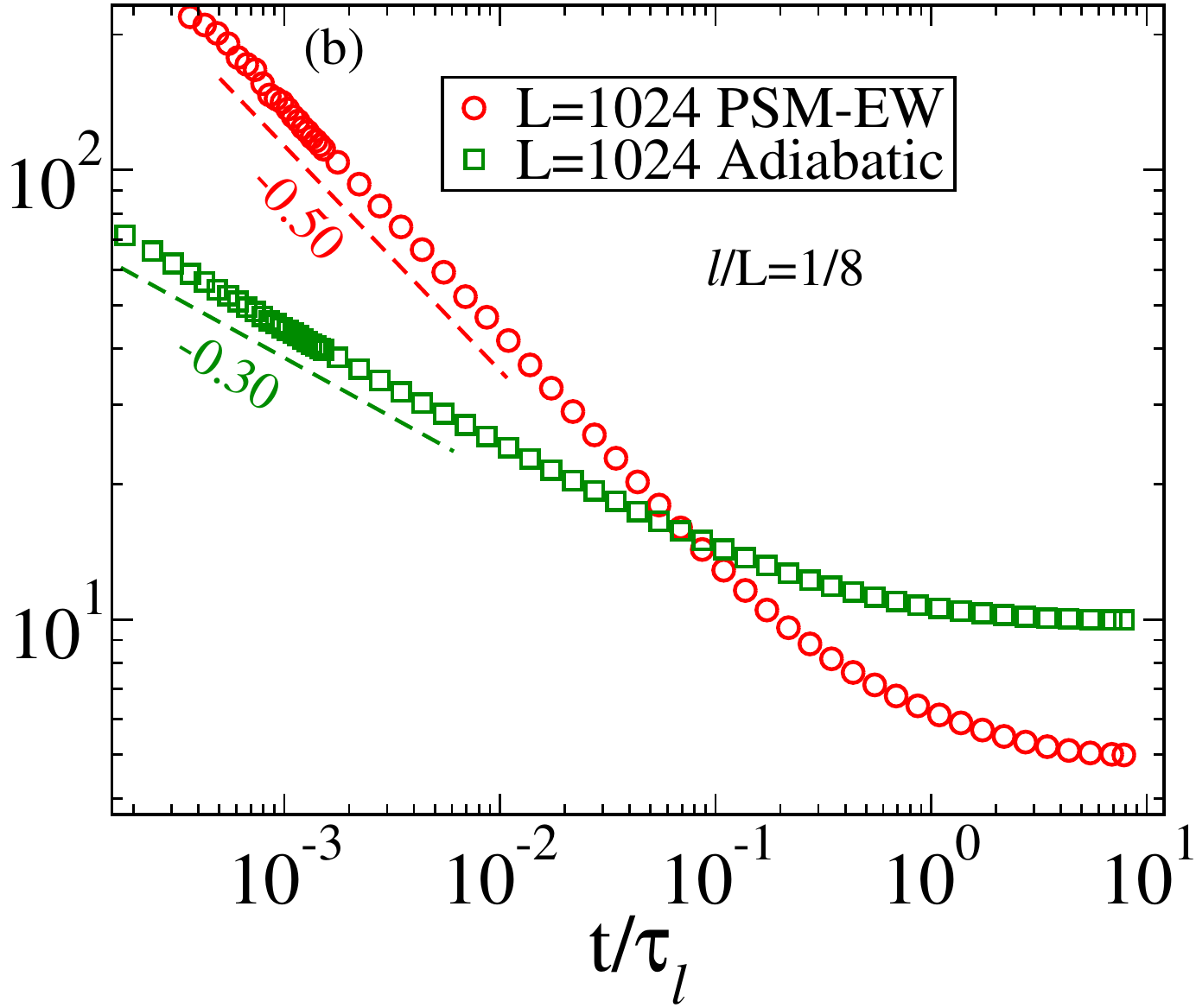}
\end{minipage}
\hspace{1.6mm}
\begin{minipage}{0.315\textwidth}
\includegraphics[width=\textwidth,height=.205\textheight]{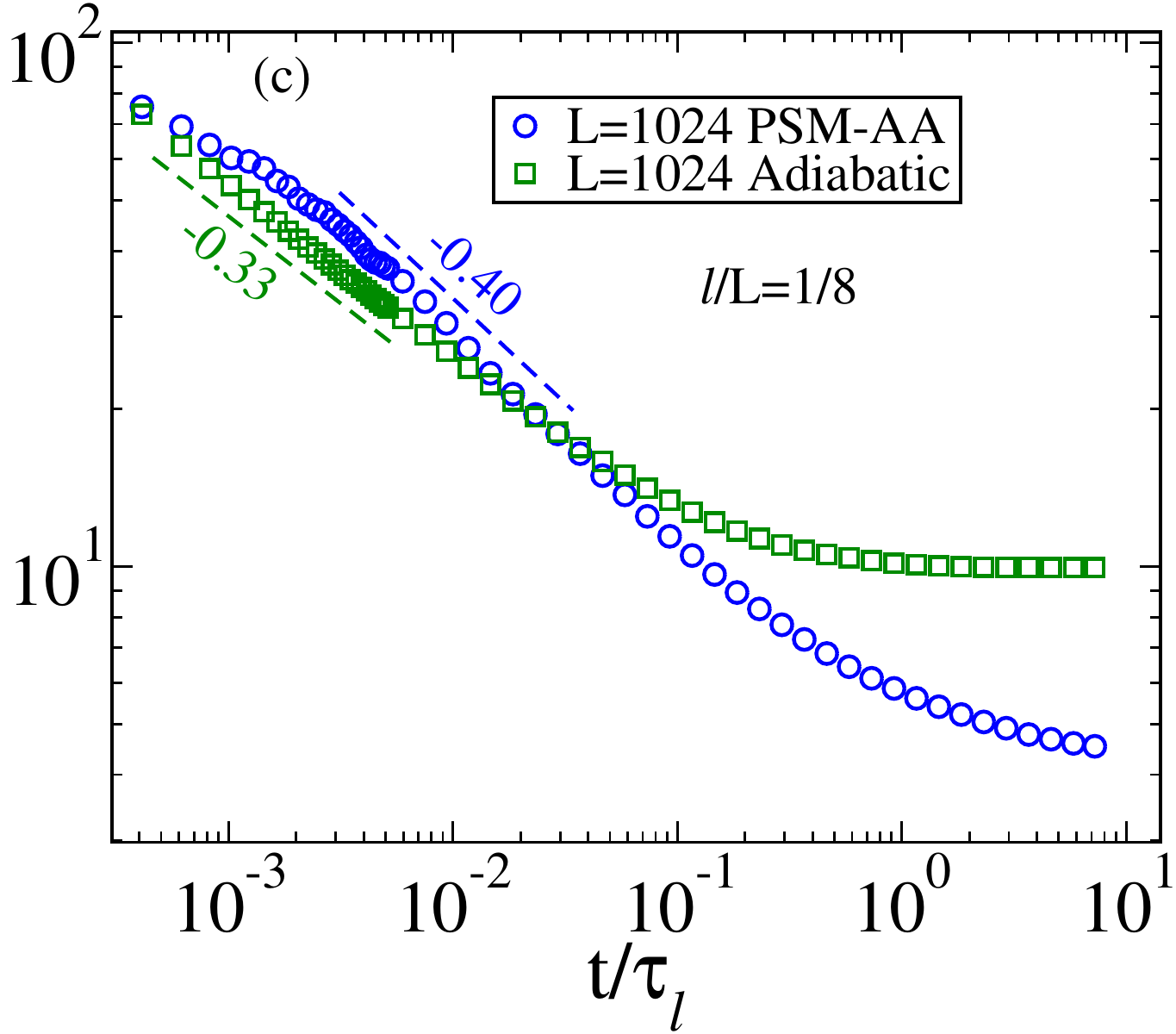}
\end{minipage}%
\caption{Time dependent flatness in the adiabatic limit with $\beta=\infty$, corresponding to the particles being at the deepest valley. The divergence of $\kappa^{ad}_4$ indicates intermittency. However, the corresponding exponents differ from those of the corresponding PSMs.}
\label{SS_Adiabatic}
\end{figure*}
\vspace{4mm}

It is often a good strategy to examine the extreme limits of a process in order to get a qualitative understanding. The adiabatic limit corresponds to the situation in which particles move infinitely faster than the surface \cite{Bouchaud1995}. The problem then reduces to the Sinai model of random walkers in a random potential \cite{Sinai1982}. By using a path-integral method, the two-point density-density correlation function and single-site probability distributions were calculated analytically in steady state  \cite{Comtet1998}. Surprisingly, static results agree very well with the numerical simulations \cite{NMB_2006}. Therefore, it is natural to ask how well the adiabatic approximation would work for the dynamics of the passive particles. 

In order to check this, we study second and fourth order structure functions $S^{ss}_q(t_0, t, l)$ and the corresponding flatness $\kappa_4^{ad} (t_0, t, l)$ in the adiabatic limit for the three drivings. In this limit, the particles reach thermal equilibrium, with the particle number  density $n_i(t)$ given by the Boltzmann-Gibbs form
\begin{equation}
n_i(t) = N \frac{e^{-\beta h_i(t)}}{Z}
\end{equation}
where $h_i(t)$ is the effective potential, $Z=\sum^{L}_{i=1} e^{-\beta h_i(t)}$ is the partition function, and $\beta$ is proportional to the inverse temperature. For simplicity, we consider the limit $\beta \rightarrow \infty$, in which case particles occupy only the sites where the height is minimum ($h_i(t)=h_{min}(t)$). Thus we numerically study the dynamics of global \emph{deepest valley} and monitor its visit to stretch $l$. To compare with the PSMs, we  present the variation of $\kappa_4^{ad}(t,l)$ in the adiabatic limit along with their corresponding PSMs in Fig.\ \ref{SS_Adiabatic}. We choose  $\omega=1$, i.e., equally fast surface and particle updates, for all the numerical simulations for PSMs. Figure \ref{SS_Adiabatic} shows that $\kappa_4^{ad}$ shows a divergence, $\sim (t/\tau_l)^{-\gamma_{ad}}$ with $\gamma_{ad} \simeq 0.33$, $\simeq 0.30$, and $\simeq 0.33$ for KPZ, EW, and KPZ-AA driving, respectively. These values are far from the numerically determined values for the PSMs, and we conclude that the adiabatic approximation does not work well for the dynamics.

\section{Coarsening regime}

In this section, we study the growth of the two-point density-density correlation function and show that it diverges in the limit of scaled separation going to zero. A comparison of the flatness with the three different drivings shows the difference in the degree of intermittency. Finally, numerical results for the sticky slider model are found to agree fairly well with those for passive scalars, for KPZ advection and EW driving but not for KPZ-AA.

\subsection{Correlation function}
 Initially, particles tend to move to the closest local minima of the co-evolving surface. As time $t$ passes, each passive particle typically move to a deeper valley a distance $\mathcal{R}(t)\sim t^{1/z}$ from its starting point, as discussed in Section 2. Therefore, in time $t$, particles from a catchment region (of length $\mathcal{L}(t)$ say) collect near the valley bottom. Evidently, $\mathcal{L}(t)$ is of order of $\mathcal{R}(t)$, and since we start with random placement of the particles, the typical number of particles in the catchment region is $ \rho \mathcal{L}(t)$. With unit density, this reduces to $\mathcal{L}(t)$. 
\vspace{4mm}
\begin{figure*}[ht!] 
\begin{minipage}{0.315\textwidth}
\includegraphics[width=\textwidth,height=0.20\textheight]{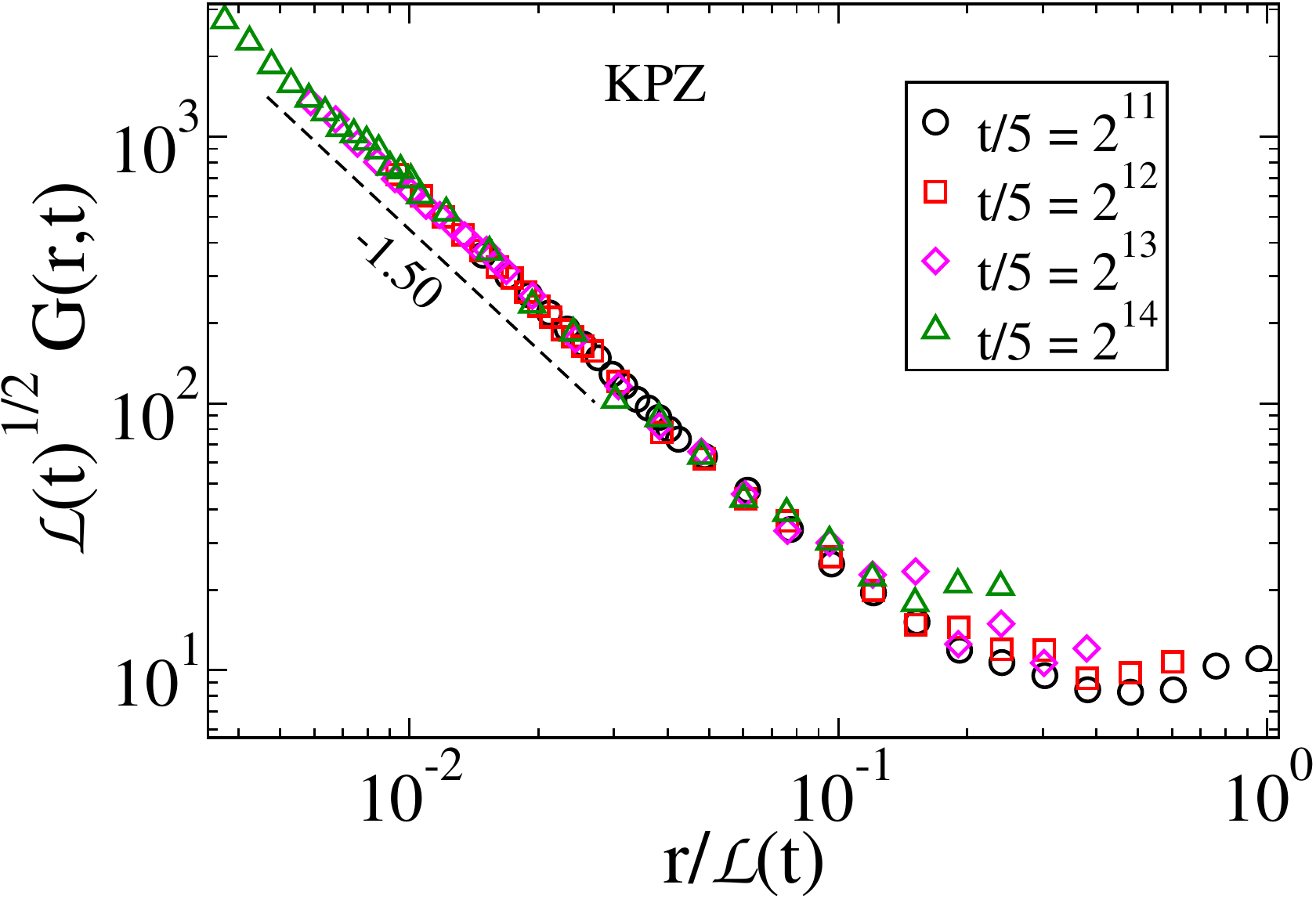}
\end{minipage}%
\hspace{1.4mm}
\begin{minipage}{0.315\textwidth}
\includegraphics[width=\textwidth,height=0.215\textheight]{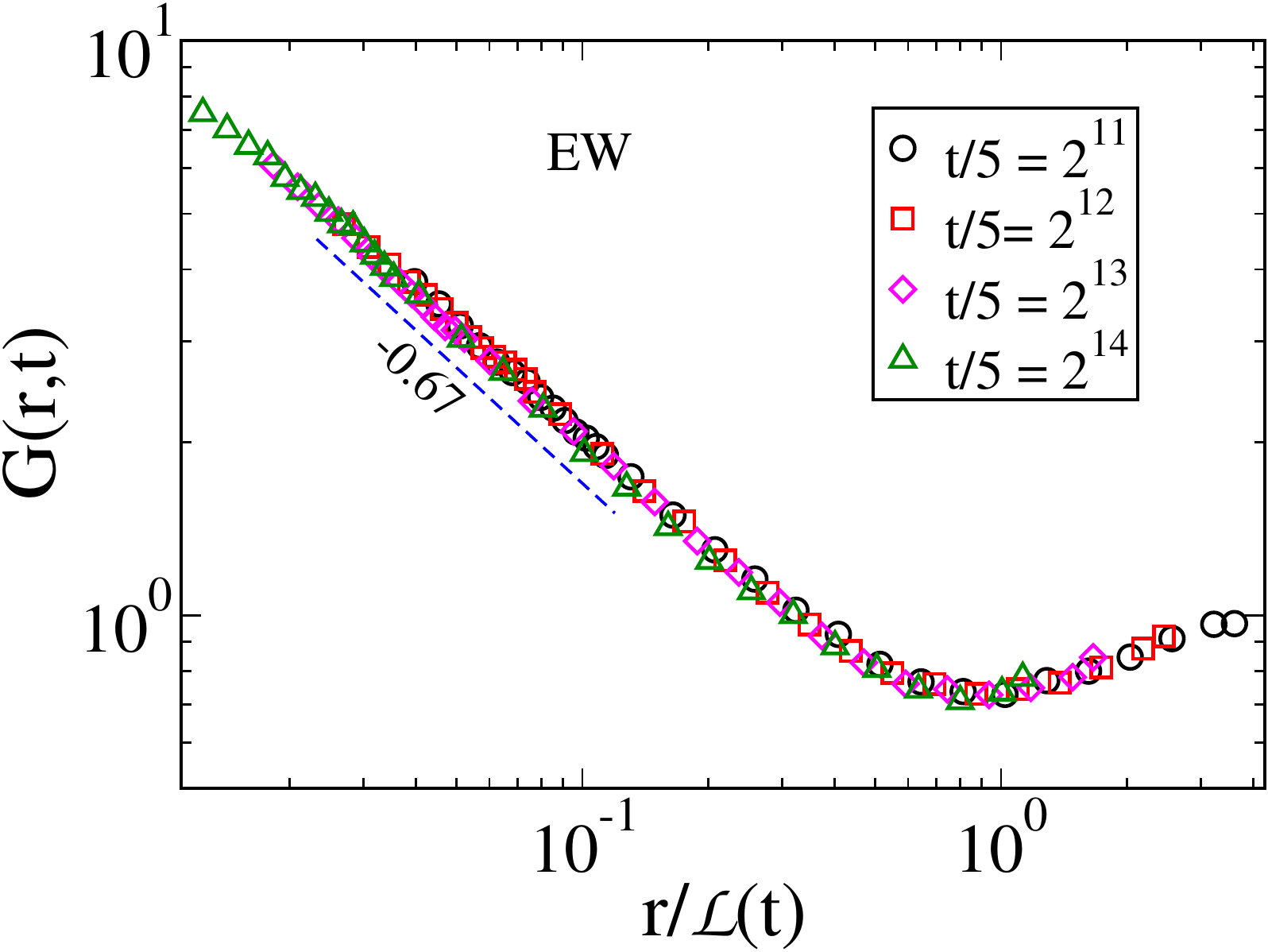}
\end{minipage}%
\hspace{1.5mm}
\begin{minipage}{0.315\textwidth}
\includegraphics[width=\textwidth,height=0.21\textheight]{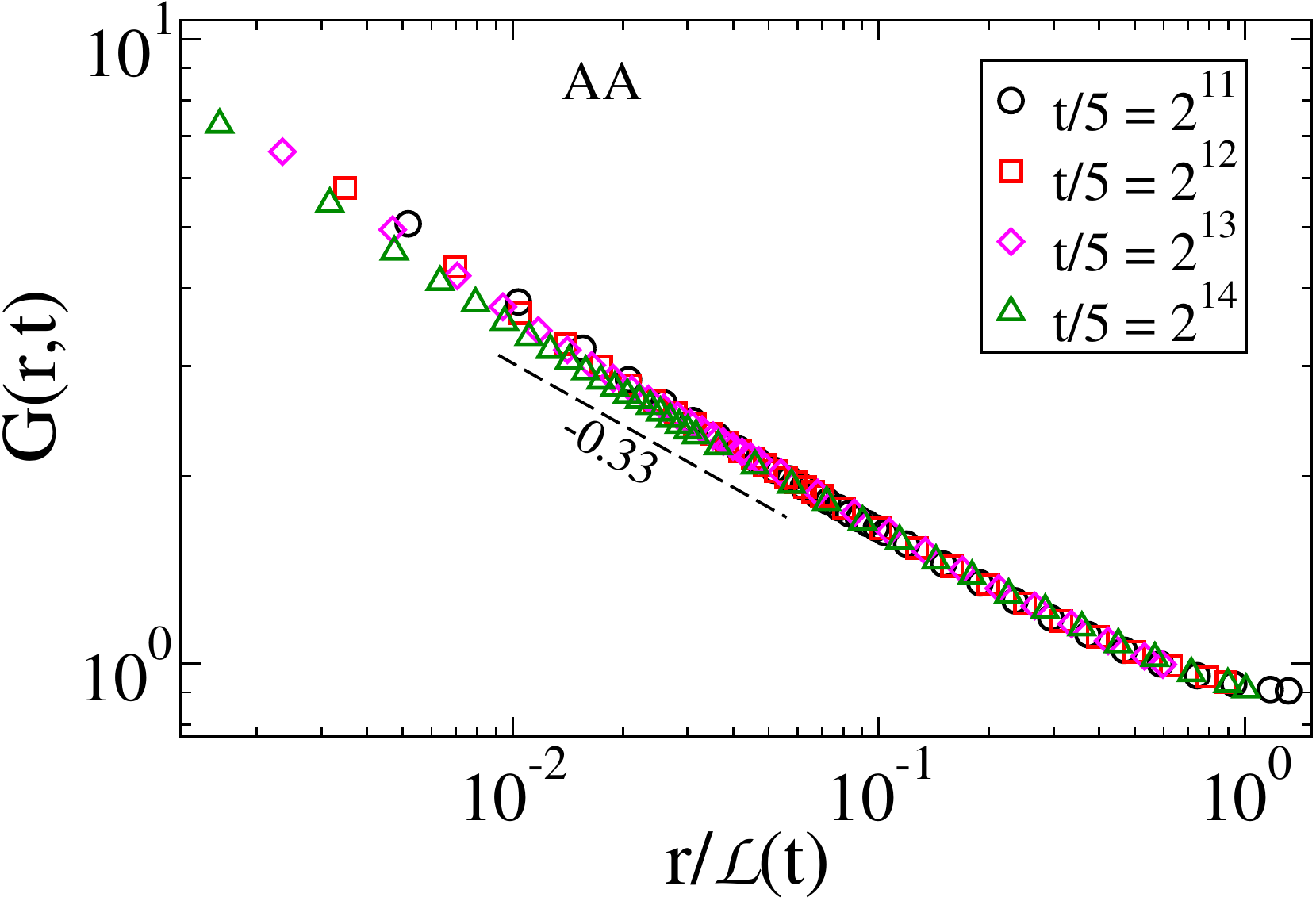}
\end{minipage}%
\caption{Correlation function in the coarsening regime. $G(r,t)$ diverges in the limit 
$r/\mathcal{L}(t) \rightarrow 0$ in all cases.}
\label{DenDenCorrKPZ_EW_AA}
\end{figure*}
\vspace{4mm}
\vspace{4mm}

\begin{figure}[ht!]
\begin{minipage}{0.45\textwidth}
\includegraphics[width=\textwidth,height=.25\textheight]{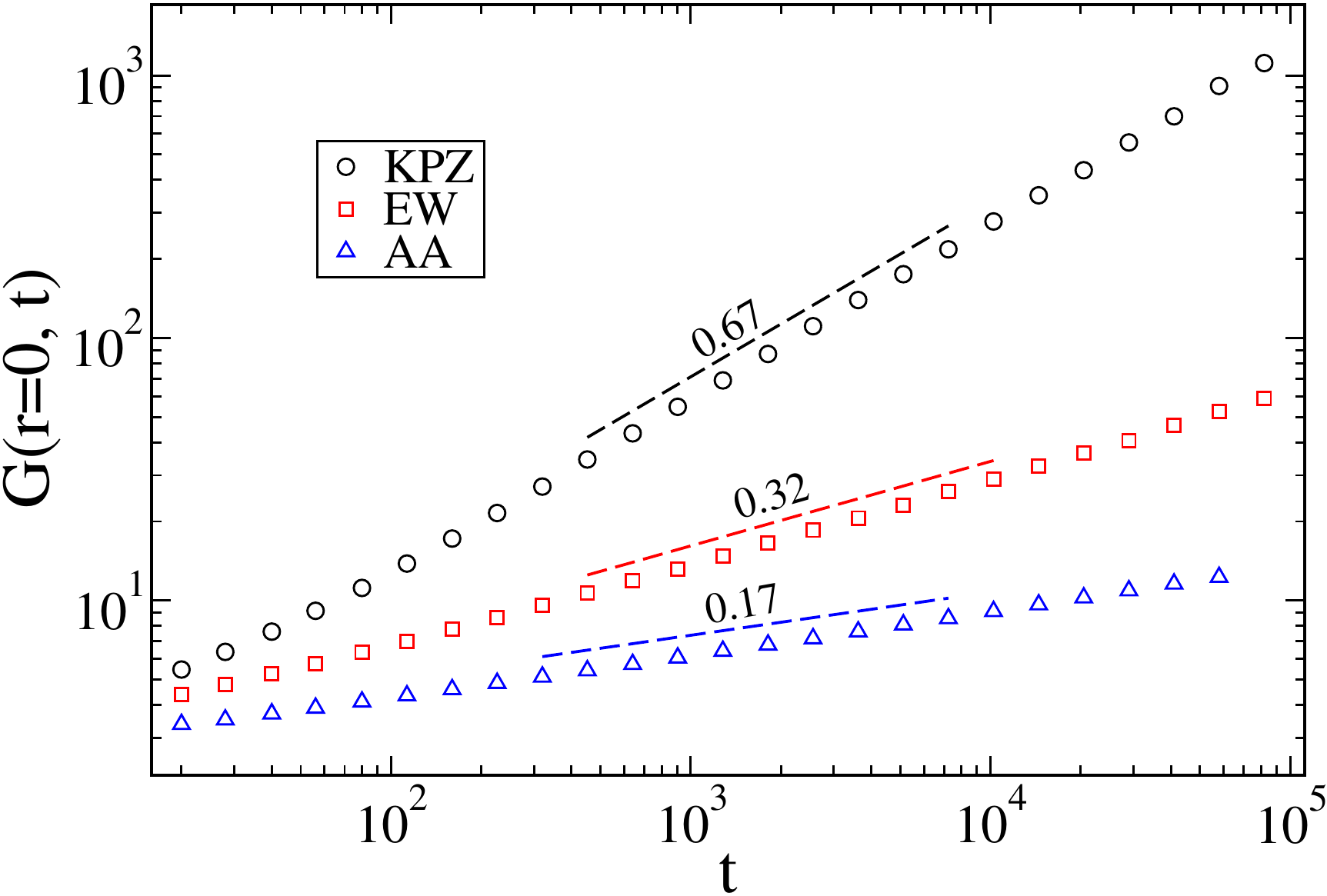}
\end{minipage}%
\caption{Time evolution of the density-density autocorrelation function in the coarsening regime.}  
\label{Auto_correlation}
\end{figure}
\vspace{4mm}

\begin{figure}[ht!] 
\begin{minipage}{0.45\textwidth}
\includegraphics[width=\textwidth,height=.25\textheight]{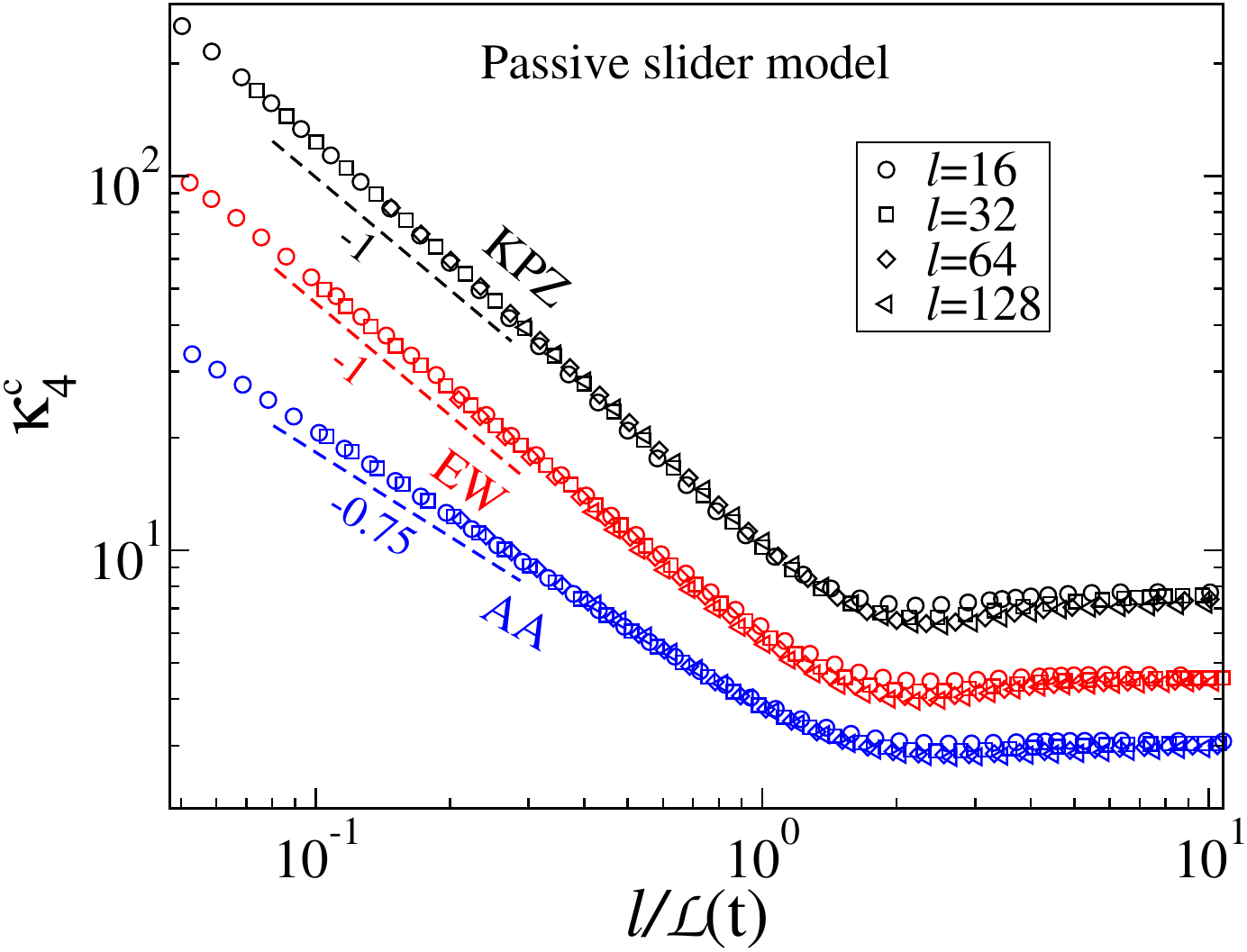}
\end{minipage}%
\hspace{1.5mm}
\begin{minipage}{0.45\textwidth}
\includegraphics[width=\textwidth,height=.25\textheight]{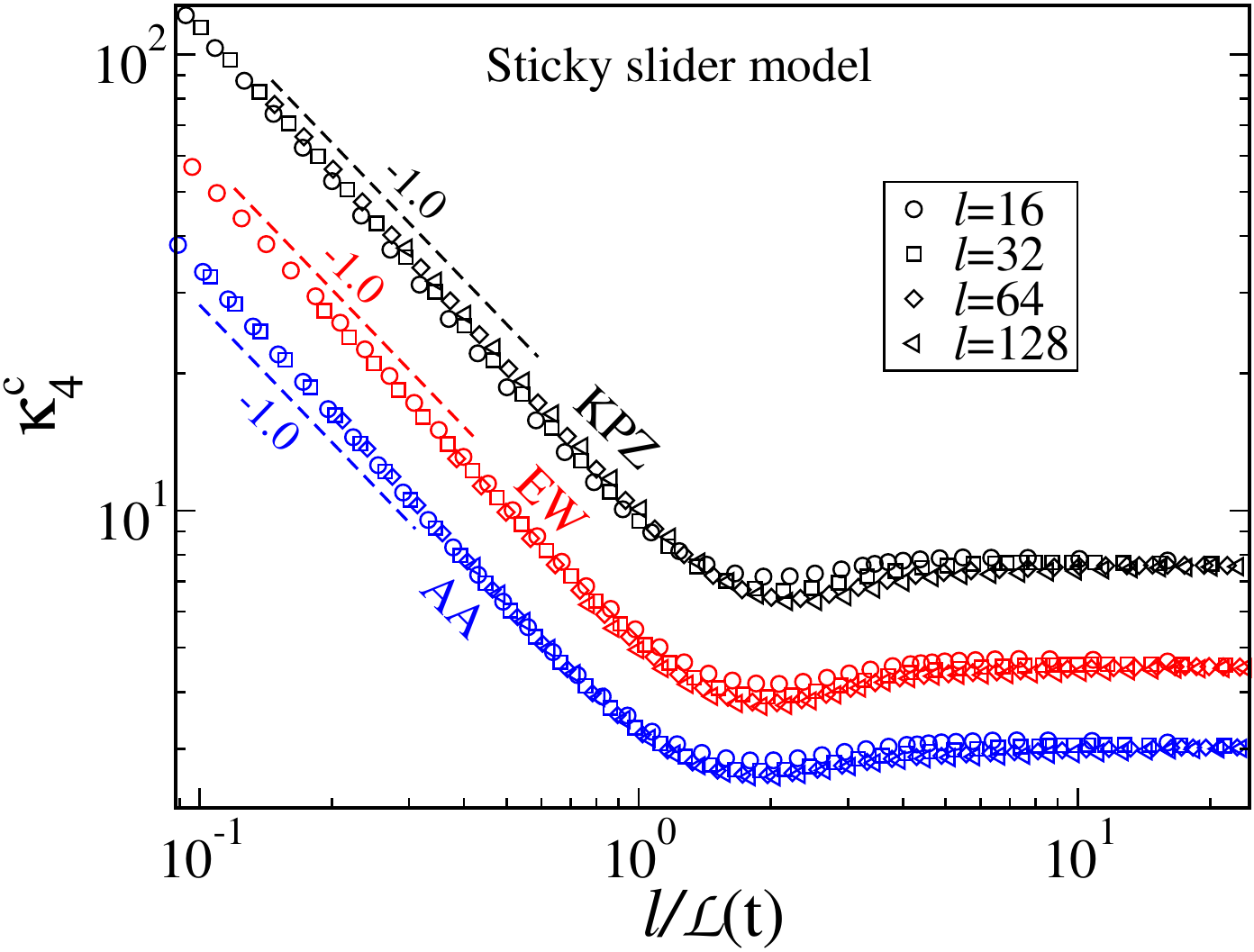}
\end{minipage}%
\caption{In the coarsening regime, flatness diverges as $t$ increases. For ease of display, $\kappa_4$ is multiplied by $1.5$ for PSM-EW and SSM-EW, and $2.5$ for PSM-KPZ and SSM-KPZ.}
\label{Coarsening_PSM_SSM} 
\end{figure}
\vspace{4mm}
The typical number of particles in a cluster increases with time, reminiscent of phase ordering dynamics, where ordered domains grow in time. We study the two-point density density correlation $G_c(r,t)= \langle n_i(t) n_{i+r}(t)\rangle$  where  $n_i(t)$ denotes the number of particles at $i$-th site at time $t$. Numerical simulations of $G_c(r,t)$ for KPZ, EW, and KPZ-AA drivings (fig.\ \ref{DenDenCorrKPZ_EW_AA}) show that data for different times exhibit a scaling collapse when the separation $r$ is scaled by the growing length scale $\mathcal{L}(t)$ and $G_c(r,t)$ is scaled with $\mathcal{L}^{\theta}(t)$. Our numerical simulations indicate $\theta \simeq 0.50$ for KPZ and $\simeq 0$ for both EW and KPZ-AA drivings.

\begin{table}
\begin{center}
\caption{ Exponents values corresponding to the correlation function} 
\begin{tabular}{ |p{1.78cm}||p{1.78cm}|p{1.78cm}|p{1.78cm}|}
\hline
System & $\theta$  & $\nu$ & $\delta/z$ \\
\hline
PSM-KPZ  & $ 0.50$ & $  1.50 $ & $0.67$   \\
\hline
PSM-EW  & $ 0$  & $  0.67 $ & $0.32$  \\
\hline
PSM-AA  & $  0$  & $  0.33 $ & $0.17$   \\
\hline
\end{tabular}
\label{Table_correlation}
\end{center}
\end{table}
The scaling form of the correlation can be written as 
\begin{equation}
G_c(r,t) \sim \frac{1}{\mathcal{L}^{\theta}(t)} Y_c\left(\frac{r}{\mathcal{L}(t)}\right)
\label{Den_DenCorr}
\end{equation}   
where $Y_c(y) \sim y^{-\nu}$ as $y \rightarrow 0$ which indicates the divergence of $G_c(r,t)$, with $\nu \simeq 1.50$, $ \simeq 0.67$, and $ \simeq 0.33$ for the PSM with KPZ, EW, and KPZ-AA drivings, respectively; the different exponent values quantify the spreading of the clusters for different drivings. KPZ-AA driving leads to a relatively large spread while KPZ shows the least and EW lies in between. 
 
The scaling form for two-point density correlation function in Eq.\ \ref{Den_DenCorr} is consistent with the steady state two-point density density correlation given in Eq.\ 15 of Ref. \cite{NMB_2006} when the system size $L$ in \cite{NMB_2006} is replaced by $\mathcal{L}(t)$. As for the steady state, we find $G_c(0,t) \sim \mathcal{L}^{\delta}(t)$ with $\delta/z \simeq 0.67$, $0.33$, and $0.17$ for KPZ, EW and KPZ-AA, respectively as shown in Fig.\ \ref{Auto_correlation}. In Table II we present the exponent values associated with $G_c(r,t)$. On comparing the values of steady state $\delta$ with our estimated $\delta/z$ (in the coarsening regime), we find good agreement for KPZ and for KPZ-AA cases, while 
$\delta$ for EW is smaller than its corresponding value in the steady state.

\subsection{Structure functions and Intermittency}

To track the intermittent signal as the system evolves from an initially random state towards the clustered steady state, we monitor the flatness of the distribution of particle number fluctuation $[N_l(t)-N_l(0)]$ in a stretch length $l$. We study the $q$'th order structure function $S^c_q(t,l) = \langle [N_l(t)-N_l(0)]^q \rangle$  for $q=2$ and $q=4$, and the flatness $\kappa_4$ for both the PSM and SSM models with KPZ, EW and KPZ-AA driving. 

 For the SSM, we argue that $S^c_q(t,l)$ obeys scaling, and compare with numerical simulations. Recalling that particles are drawn into basins of typical size $\mathcal{L}(t)$, let us assume that there is a single SSM cluster with $\mathcal{L}(t)$ particles in each such basin. At early time, when $\mathcal{L}(t) \ll l$, 
particle movements within $l$ do not affect $N_l$ and number fluctuations arise from  random motion of particles in and out of the edges. Thus the statistics of the number fluctuations are Gaussian (and the value of the flatness $\kappa_4^c=3$). This ceases to hold once $t$ is large enough that $\mathcal{L}(t) \sim l$. Once $\mathcal{L}(t) \gg l$, the probability that the SSM cluster falls within the $l$-stretch is $\frac{l}{\mathcal{L}(t)}$ which implies  
\begin{equation}
S^c_q \sim \frac{l}{\mathcal{L}(t)} \mathcal{L}(t)^q \    \    \  \text{for} \ \mathcal{L}(t) \gg l
\end{equation}      
and leads to the estimate 
\begin{equation}
\kappa_4^c (t,l) \sim \frac{\mathcal{L}(t)}{l} .
\label{Eq:Coarsening_SSM}
\end{equation}
Thus, in the `coarsening' regime, in contrast to the steady state, the flatness diverges in the limit $t/\tau_l \rightarrow \infty$. These two limits are captured by the scaling form  
\begin{equation}
\kappa_4^c \sim h\left(l/\mathcal{L}(t)\right)
\label{Eq_Coarsening_scaling}
\end{equation}
with $h(y)\sim y^{-\psi}$ as $y \rightarrow 0$ and $h(y) \rightarrow \text{const.}$
as $y \rightarrow \infty.$ In view of Eq.\ \ref{Eq:Coarsening_SSM}, we expect $\psi=1$. This analytical prediction is well confirmed by numerical simulations of $\kappa^c(t,l)$ for SSMs with all three drivings (see Fig.\ \ref{Coarsening_PSM_SSM} and Table II). 

Motivated by the success of scaling for the SSM, we performed simulations for the PSMs and SSMs for the three types of surface driving and have plotted the scaled data in Fig.\ \ref{Coarsening_PSM_SSM}. We see that scaling holds in all three cases. However, the exponent $\psi$ coincides with the corresponding SSM value only for KPZ and EW driving. It differs substantially from the SSM prediction for KPZ-AA driving, reflecting the smaller degree of clustering. Results are summarized in Table II.

\section{Aging}

During the process of coarsening, the system exhibits aging, namely changes in the pattern of dynamic evolution as time passes. Traditionally, these changes are studied by monitoring a two-time correlation function between an initial time $t_0$ and a final time $t_0+t$. This approach has been used in diverse contexts, e.g, phase ordering kinetics \cite{Bray1994}, and interface evolution \cite{Ramasco2006, Chatterjee2006, Bustingorry2007,Henkel2012}.

In this section, we consider the effect of aging on the flatness and show that it is a monotonic function of time if $t_0 < \tau_l$ for all three drivings. For the case $t_0 > \tau_l$, we find that a striking nonmonotonic behavior noted earlier for advection and EW driving, holds for KPZ-AA as well.

\subsection{$\tau_l < t_0$}
\begin{figure}[ht!]
\begin{minipage}{0.45\textwidth}
\includegraphics[width=\textwidth,height=.25\textheight]{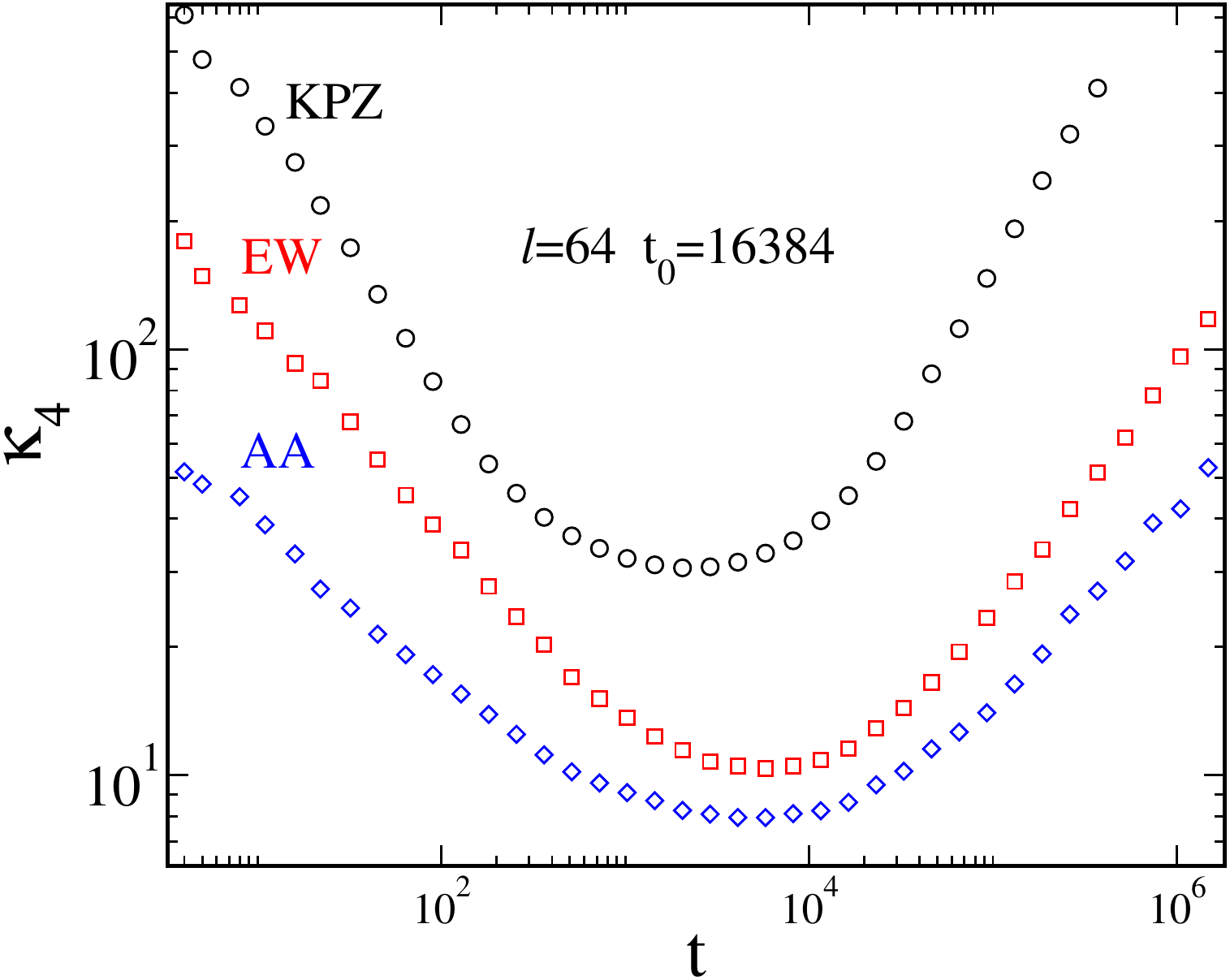}
\end{minipage}%
\caption{In the aging regime, $\kappa_4$ is a nonmonotonic function of time. } 
\label{Nonmono_Flatness}
\end{figure}

If $\tau_l \ll t_0$, the flatness shows two distinct wings (left and right) separated by an intermediate plateau regime shown in Fig. \ref{Nonmono_Flatness} as will be discussed below. The left wings of the curves in Fig. \ref{Nonmono_Flatness} corresponds to a quasi-steady state (QSS) regime, while the right wings correspond to long-time aging (LTA). To understand the nonmonotonicity of flatness, we first study $\kappa_4 (t_0,t,l)$ for SSMs via a probabilistic arguments. Within the SSM, when $t \ll \tau_l \ll t_0$, a typical catchment of size $\mathcal{L}(t_0)$ typically contains a single aggregate with $\mathcal{L}(t_0)$ particles. The position of the single aggregate can be anywhere within $\mathcal{L}(t_0)$ (reminiscent of the steady state where a single aggregate moves over the system size $L$) which implies that within  $\mathcal{L}(t_0)$, a local steady state is reached. Therefore, the structure functions and corresponding flatness in QSS can be estimated by replacing $L$ by $\mathcal{L}(t_0)$ in Eqs. \ref{SS_SecondMoment} and \ref{Eq:SS_SSM}. The flatness for SSM in QSS is thus
\begin{equation}
\kappa_4 \sim (t_0/t)^{1/z} . 
\label{QSS_flatness}
\end{equation} 

\vspace{4mm}
\vspace{4mm}

 On the other hand, when $t \gg t_0 $, the right wing of the nonmonotonic flatness where a typical catchment $\mathcal{L}(t_0+t)$ increases with $t$ and accordingly, number of particles in an aggregate increases because of the increasing basin size. This process continues until the difference time $t \ll L^{z}$. The $q$-th order structure function, defined by Eq.\ \ref{struc}, is estimated as 
\begin{equation}
S_q(t_0,t,l) \simeq \frac{l}{\mathcal{L}(t)} [\mathcal{L}(t_0+t)]^{q}
\label{LTA_Structure}
\end{equation}
where $t \gg t_0$. Therefore, Eq.\ \ref{LTA_Structure} can be approximated and the corresponding flatness is obtained as
\begin{equation}
\kappa_4(t_0,t,l) \simeq \left(\frac{t}{\tau_l}\right)^{1/z}
\label{LTA_flatness}
\end{equation}
which diverges as $t$ increases.
\vspace{4mm}
\begin{figure}[ht!]
\begin{minipage}{0.45\textwidth}
\includegraphics[width=\textwidth,height=.25\textheight]{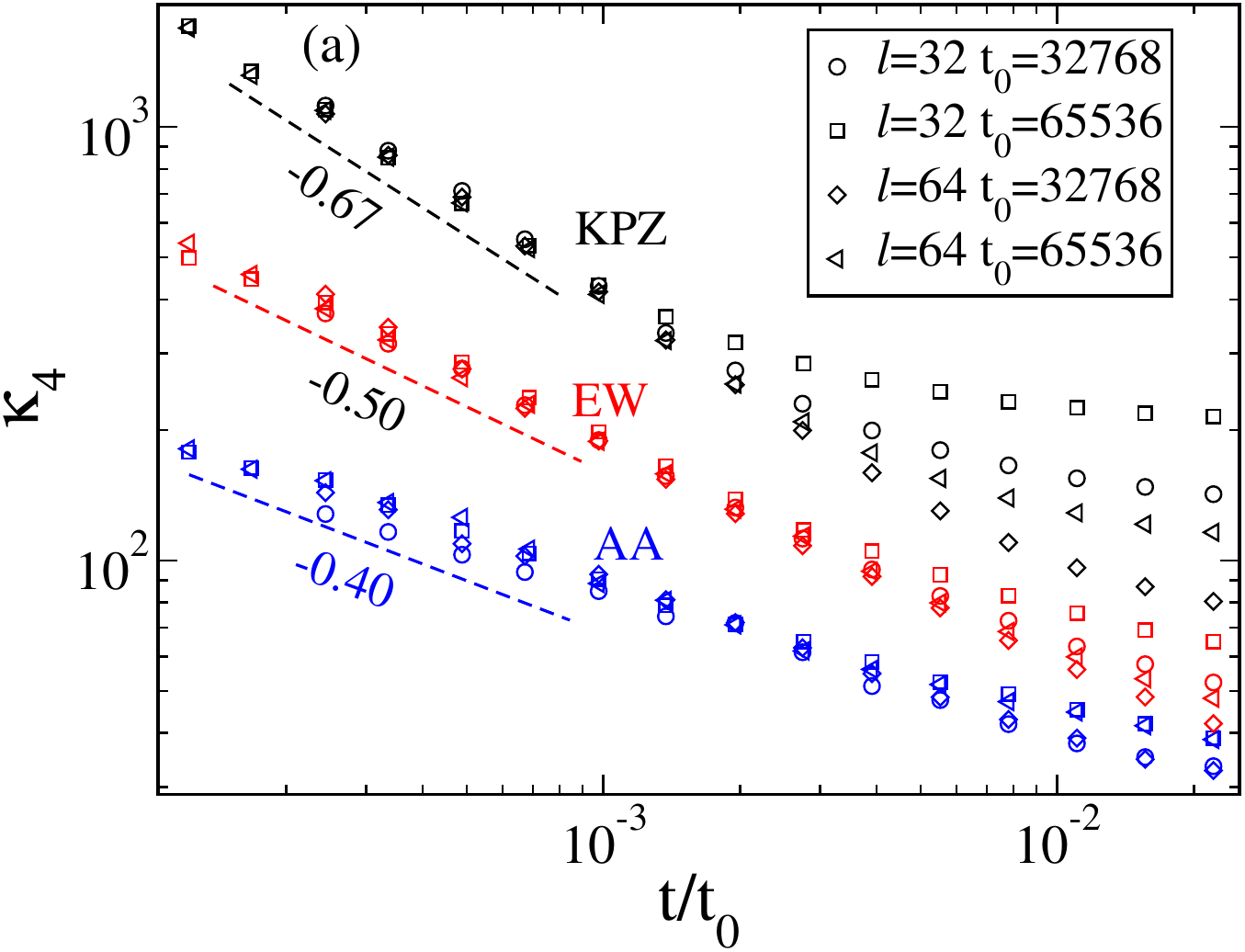}
\end{minipage}%
\hspace{1mm}
\begin{minipage}{0.45\textwidth}
\includegraphics[width=\textwidth,height=.25\textheight]{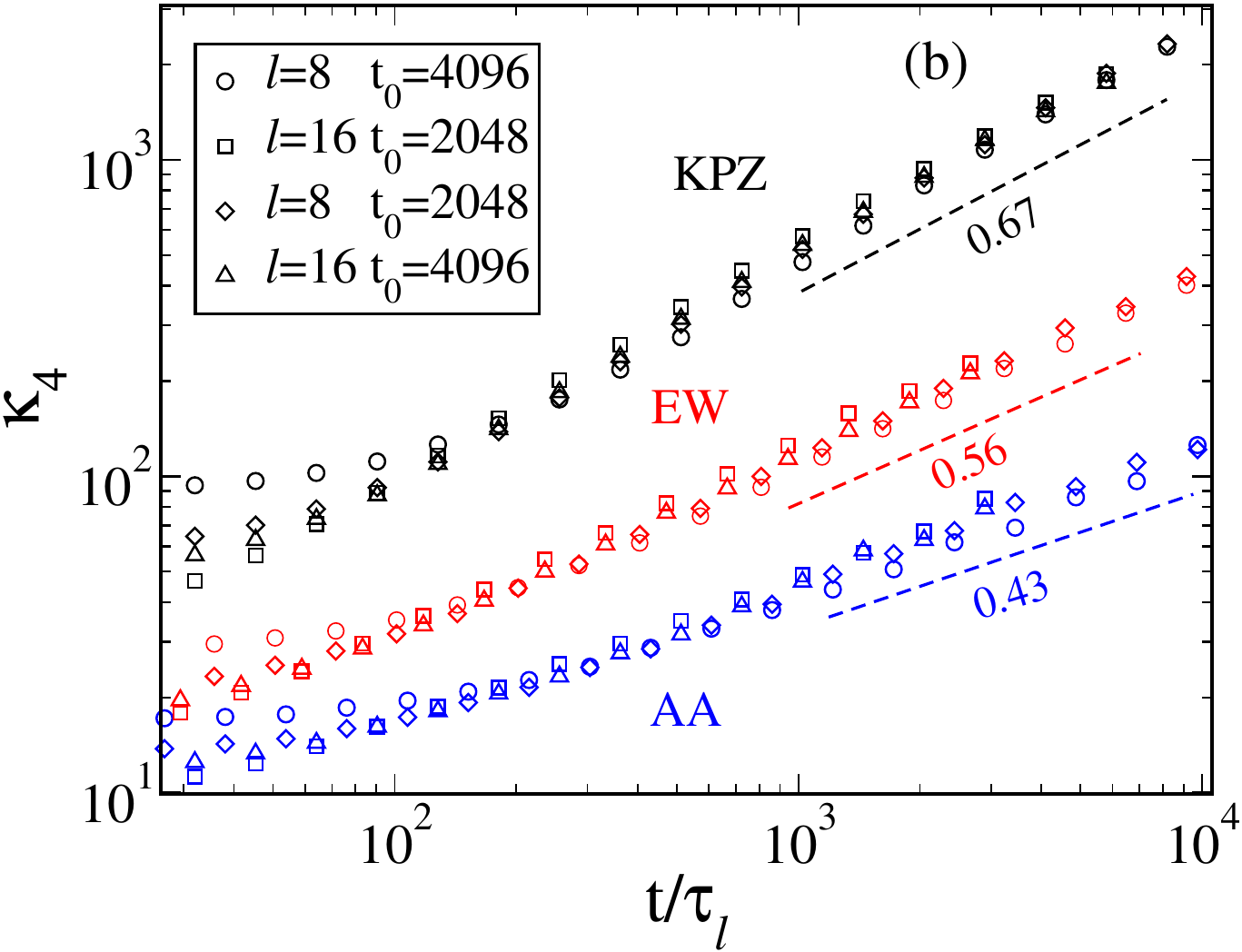}
\end{minipage}%
\caption{Scaling forms describe the two branches of the nonmonotonic flatness in the aging regime. (a) Quasi-steady state (QSS): $\kappa_4$ versus time for PSM with KPZ, EW, and KPZ-AA driving. (b) Long-time aging (LTA): $\kappa_4$ versus time for PSM with KPZ, EW, and KPZ-AA driving. For ease of display, flatness shown in Fig. \ref{Flatness_QSS}(a), is multiplied by $1.5$ for PSM-EW and PSM-KPZ.}
\label{Flatness_QSS}
\end{figure}
\vspace{4mm}
 The extent of the plateau regime is $\sim (t_0-\tau_l)$. Substituting $t=\tau_l$ in Eq.\ \ref{QSS_flatness}, we get $\kappa_4 \sim (t_0/\tau_l)^{1/z}$ which smoothly matches with the LTA regime by setting $t=t_0$ in Eq. \ref{LTA_flatness}.
 
Using the SSM results as a guide, we now discuss the numerical simulations of the PSMs. Figure \ref{Flatness_QSS} shows results of   the numerical simulation of PSMs for different values of $\tau_l$ and $t_0$. In the QSS, the data for several values of $\tau_l$ and $t_0$ collapse in the limit $t/t_0 \rightarrow 0$ when $t$ is scaled by $t_0$ shown in Fig.\ \ref{Flatness_QSS}(a). The flatness then can be estimated by replacing $L$ by $\mathcal{L}(t_0)$ in Eq.\ \ref{eq:SS_scaling}, leading to  
\begin{equation}
\kappa_4(t,t_0,\tau_l) \sim  \left(\frac{\mathcal{L}(t_0)}{l}\right)^{\phi}  
g\left(\frac{t}{\tau_l}\right) .
\label{PSM_QSS}
\end{equation}

For the LTA regime (corresponding to the right hand branch) numerical results for different values of $\tau_l$ and $t_0$ collapse when separation time $t$ is scaled by
$\tau_l$ shown in Fig. \ref{Flatness_QSS}(b). The flatness diverges in the limit $t/\tau_l \rightarrow \infty$ for an infinite system. The numerical results for PSM-KPZ follow the scaling form predicted by the SSM, whereas results for the PSM in the EW and KPZ-AA cases deviate from the corresponding SSMs.

\subsection{$\tau_l > t_0$}
\begin{figure}[ht!]
\centering
\begin{minipage}{0.45\textwidth}
\includegraphics[width=\textwidth,height=.25\textheight]{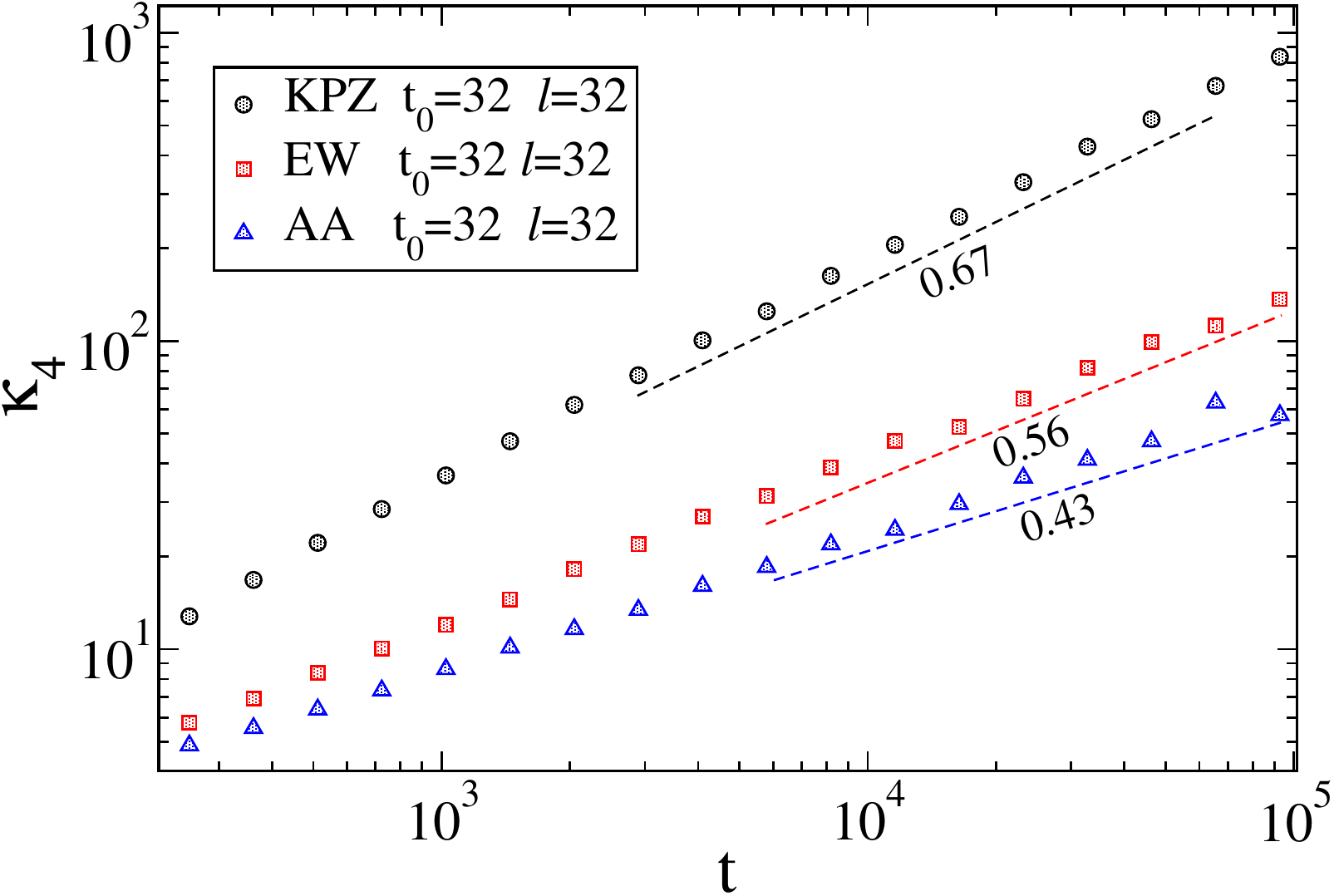}
\caption{In the regime $\tau_l > t_0$, $\kappa_4$ increases monotonically for the three drivings.}
\label{Fig_Flatness_Monotonic}
\end{minipage}%
\end{figure}

In the less interesting case $\tau_l > t_0$, the number fluctuations in $l$ increase with $t$ and consequently, $\kappa_4$ increases monotonically with $t$ as shown in Fig. \ref{Fig_Flatness_Monotonic}. This monotonic behavior of $\kappa_4$ can be identified with the LTA regime in Fig.\ref{Flatness_QSS}(b). 

\vspace{4mm}
\vspace{4mm}




\section{Conclusion}

In this work, we have characterized the intermittent steady state of passive particles driven by fluctuating surfaces, and the manner in which such a steady state is approached. We have given strong evidence that a scaling description holds for the three types of surface driving considered (KPZ advection, EW, and KPZ antiadvection), although the scaling functions differ considerably in the three cases, reflecting the different degree of clustering in both space and time.

It was known earlier that the single particle dynamic exponent $z$, defined through $\mathcal{R}(t) \sim t^{1/z}$, coincides with the surface dynamic exponent for KPZ advection \cite{BP1993, DrosselPRB2002, Chin2002, NMB_2006}, and differs from it for antiadvection \cite{DK2000, DrosselPRB2002, NMB_2006}, indicating that particles are slaved to the surface dynamics in the former case. For EW driving, logarithmic corrections to $z=z_s$ were indicated earlier \cite{BP1993, Huveneers}. Our study corroborates this finding and provides an estimate of the power of the logarithm. The exponent $z$ is significant for our study of the many-particle system, as it enters in scaling descriptions of correlations, both in the steady state and in the coarsening and aging regimes.

In order to understand the correlation between two passive particles, we studied the evolution of the probability distribution $P(r_s, t)$ of their separation for the three different drivings. In \cite{Ueda_Sasa2015} it was found numerically that with KPZ driving, $P(0,t)$ approaches a nonzero constant as $t \rightarrow \infty$ even though $ \langle r_s^2 \rangle \sim t$ for large $t$. We showed that these features follow from the fact that $P(r_s,t)$ is a function of  $r_s/\mathcal{L}(t)$ with $\mathcal{L}(t) \sim t^{1/z}$. Interestingly, we found that the time evolution of the average overlap of trajectory pairs, which enters into the discussion of replica symmetry breaking in trajectory space \cite{Ueda_Sasa2015}, is also obtained from the scaling form of $P(r_s,t)$.

In the many-particle system, clustering of particles leads to intermittency in space and time. Our numerical study of the phenomenon was supplemented by analytic arguments for a simplified sticky slider model, which suggested scaling forms for the passive particle model with all three drivings. Spatial multiscaling was demonstrated numerically in the steady state for all three drivings, with KPZ advection showing the strongest effect, EW driving being intermediate, and antiadvection displaying the weakest effect of the three, but still quite different from the usual scaling. Further, intermittency was also quantified by monitoring the divergence of flatness as a function of scaled distance or time, confirming the sequence of relative strengths.

We studied the approach to the steady state through the time evolution of the two-point density-density correlation function. It is a function of the separation $r$ scaled by $\mathcal{L}(t) \sim t^{1/z}$  where the growing length scale $\mathcal{L}(t)$ describes the spatial extent of the basin from which particles are drawn to form clusters. It also enters in the scaling properties of the time-dependent flatness. We also investigated aging by monitoring the flatness with different waiting times $t_0$ within the coarsening regime, and found that in a broad region, it is a nonmonotonic function, with two separate scaling regimes.

An interesting point that emerges from our study is that the intense clustering induced by KPZ advection differs qualitatively from that induced by EW or KPZ-AA driving. There are several pointers. For instance, single particle motion is slaved to the surface with KPZ advection, and not in the other cases. The probability that the separation of two particles lies within a specified finite range approaches a nonzero value for KPZ driving, but decays to zero as a power in the other two cases. This can be traced to the values of the critical exponents $\theta$ and $\nu$, which satisfies the relation $\nu-\theta=1$ in the KPZ case. However, $\theta=0$ for EW and KPZ-AA driving, and the relation fails to hold. In the many-particle system, the feature that stands out is the close similarity of KPZ advection and the SSM. This is apparent in spatial multiscaling plots (Fig.\ref{Exponents_Structure_factors}) which indicate that, for KPZ driving, a single exponent determines the structure functions for all orders, whereas a range of exponents is found for EW and KPZ-AA drivings. Both in the steady state and during the approach to it, temporal intermittency is characterized by exponents of the diverging flatness; it is strongest for KPZ advection and weakest for KPZ-AA.

We conclude by pointing out some open problems. For antiadvection, the variation of $a$,  the coupling of the driving surface to the passive particles, seems to induce nonuniversality, in that the dynamic exponent $z$ was found to depend on $a$ \cite{DrosselPRB2002}. It would be interesting to see how this variation affects the measures of intermittency studied here. Likewise, changing $\omega$, the ratio of particle to surface updates, may generate interesting effects, in view of the strong variation of the effective value of $z$ found for a single particle with EW driving \cite{Manoj2004}. Finally, we note that the scaling analysis is presented in this paper, is expected to be applicable to a broad set of problems. Thus it would be interesting to attempt such analyses for theoretical models which incorporate long-range correlated noise \cite{Deutsch1985}, as also for models which display real space condensation \cite{Evans2005}. The studies in Refs. \cite{SachdevaPRE2011,SachdevaSciRep2016} and Ref. \cite{DasPRL2016} of intermittency due to clustering at the cellular level, suggest that these methods may work well in the biological context as well.

\subsection*{Acknowledgments}

We acknowledge useful interactions with P. Perlekar, F. Huveneers, and N. Rana.


\end{document}